\documentclass[]{aastex}

\usepackage{emulateapj5,ifthen}
\usepackage[dvips,bookmarks=false]{hyperref}
\shorttitle{Radio Galaxies in Clusters}
\shortauthors{Lin \& Mohr}

\newcommand\xray{\hbox{X--ray} }
\newcommand\kev{{\rm\ keV}}
\newcommand\eg{e.g.~}
\newcommand\ie{i.e.~}

\newcommand\beq{\begin{equation}}
\newcommand\eeq{\end{equation}}

\newcommand{\figtype}{EPS}
\def\myputfigure#1#2#3#4#5%
{\vskip#5pt\makebox[0pt]{\hskip#2in
\includegraphics[width=#3\textwidth]{#1}}\vskip#4pt\hfill}
\newenvironment{inlinefigure}{
\def\@captype{figure}
\noindent\begin{minipage}{0.999\linewidth}\begin{center}}
{\end{center}\end{minipage}\smallskip}

\begin{document}

\submitted{to appear in ApJS}

\title{Radio Sources in Galaxy Clusters: Radial Distribution, and
  1.4 GH\lowercase{z} and $K$-band Bivariate Luminosity Function}

\author{Yen-Ting Lin\altaffilmark{1,2} and Joseph J.~Mohr\altaffilmark{2,3}}
\altaffiltext{1}{Department of Astrophysical Sciences,
Princeton University, Princeton, NJ 08544; Departamento 
de Astronom\'{i}a y Astrof\'{i}sica, Pontificia Universidad Cat\'{o}lica de
Chile; ytlin@astro.princeton.edu}
\altaffiltext{2}{Department of Astronomy, University of Illinois,
Urbana, IL 61801}
\altaffiltext{3}{Department of Physics, University of Illinois,
Urbana, IL 61801; jmohr@uiuc.edu}

\begin{abstract}

We present a statistical study of several fundamental properties of radio
sources in nearby clusters ($z\le 0.2$), including the radial distribution
within clusters, the radio luminosity function (RLF), and the fraction of
galaxies that is radio-active (radio active fraction, RAF).  The analysis is
carried out for a sample of 573 clusters detected in the X--ray and also
observed at 1.4 GHz in the {\it NRAO VLA Sky Survey}.  The X--ray data are used
to locate the cluster center and estimate cluster mass.  In addition, near-IR
data from the {\it Two Micron All-Sky Survey} are used to identify the
brightest cluster galaxies (BCGs), and to construct the $K$-band luminosity
function. Our main results include: (1) The surface density profile of
radio-loud active galactic nuclei (AGNs) is much more concentrated than that of
all galaxies, and can be described by the \citeauthor*{navarro97} profile with
concentration $\sim 25$.  More powerful radio sources are more concentrated
than the weaker ones.  (2) A comparison of the RLFs in the clusters and in the
field shows that the cluster AGN number density is about 5,700 times higher,
corresponding to a factor of 6.8 higher probability of a galaxy being radio
active in the cluster than in the field.  We suggest that about $40-50\%$ of
radio-loud AGNs in clusters may reside in low mass galaxies ($M_K\gtrsim -23$);
if an equivalent population exists in the field, the RLFs in the two
environments can be brought into better agreement.  The strongest support for
the existence of this low mass population comes from their spatial distribution
and RLF.  (3) The RAFs of cluster galaxies of different stellar mass are
estimated directly. About $5\%$ of galaxies more luminous than the characteristic
luminosity ($M_K\le M_{K*} \approx -24$) host radio-loud AGNs.  The RAF for
BCGs is $>30\%$, and depends on the cluster mass.  Compared to the field
population, cluster galaxies have $5-10$ times higher RAF.

Combining the AGN RLF and spatial distribution within clusters, we estimate
that they may inject an energy of $\sim 0.13$ keV per particle to the
intracluster medium near the cluster center.  We also investigate the degree of
contamination by cluster radio sources on the yields of Sunyaev-Zel'dovich
effect (SZE) cluster surveys.  Under our simple model extrapolating to higher
frequency, we estimate that as many as 10\% of clusters detected at 150 GHz may
host AGNs whose flux is comparable to the cluster SZE signal.  The fraction is
expected to be much higher at lower frequency.

\end{abstract}


\section{Introduction}
\label{sec:ra_intro}

Studying the radio-loud active galactic nuclei (AGNs) is one of the primary
methods to understand the central supermassive black hole (BH) population that
is believed to be present in all galaxies \citep{magorrian98}. With the
discovery that the intracluster medium (ICM) in central regions of clusters
does not cool to the extent predicted by the cooling flow theory
\citep[e.g.][]{tamura01,fabian01,peterson03}, more and more attention has been
paid to the energy sources that may heat up the ICM. Among the interesting
candidates are the AGNs (in the form of radio outflows; \eg
\citealt[][]{binney95,churazov01}). Analyzing the radio properties of the AGNs
in galaxy clusters thus should provide guidance in assessing the feasibility of
the AGN feedback scenario
\citep[e.g.][]{eilek04,birzan04,mcnamara05,sanderson05,best06}.

Another motivation for a comprehensive study of radio sources in clusters is
from cluster surveys at cm and mm wavelengths. The Sunyaev-Zel'dovich effect
\citep[SZE;][]{sunyaev70} is a striking cluster signature that allows clusters
to be selected using high angular resolution cosmic microwave background (CMB)
observations.
In a nutshell, the hot electrons in the ICM inverse-Compton scatter off the CMB
photons, causing a distortion in the CMB spectrum in the direction of a
cluster. The effect is independent of cluster redshifts and therefore is an
efficient way of detecting high redshift clusters \citep[e.g.][]{carlstrom02}.
We note, however, radio point sources that are often found at the centers of
clusters are a potential problem for SZE surveys. The point sources can easily
be so powerful that they significantly contaminate the SZE signature from the
cluster. With many SZE surveys planned or underway (e.g.~AMI,
\citealt{barker06}; SZA, \citealt{muchovej06}; ACT, \citealt{kosowsky03}; SPT,
\citealt{ruhl04}; AMiBA, \citealt{lo05}), there are several published studies
that address the point source problem
\citep[e.g.][]{loeb97,cooray98,hlin02,holder02,massardi04,aghanim05,
mwhite04,knox04,pierpaoli04,melin05,dezotti05}; in this paper we calculate the
effects the cluster radio sources have on the SZE based on their luminosity
function.

We present here a study of several fundamental properties of cluster radio
galaxies, based on the public data archive of the NRAO VLA Sky Survey (NVSS,
\citealt{condon98}). It has long been recognized that the two main galaxy
populations that are active at radio wavelengths, the galaxies with an AGN and
the star forming galaxies, can be roughly separated by their radio luminosity,
with a division at $\sim 10^{23}$ W$\,$Hz$^{-1}$ \citep[e.g.][]{condon92}. 
In this paper we will mainly study the properties of higher luminosity radio
sources, the radio-loud AGNs.  Hereafter we loosely use the term ``radio
source''  to refer to the whole population, and ``AGN'' to refer to the high
luminosity sources ($P\ge10^{23}$ W$\,$Hz$^{-1}$).

We study the luminosity function (LF), the radial distribution (in terms of
surface density profile), and the fraction of cluster galaxies that are
luminous in the radio. These ``classical'' properties have been subject to
extensive investigations
\citep[e.g.][]{ledlow95,ledlow96,miller02,morrison03,rizza03,best04,best05,best06}.
In the current treatment, we study the radio sources within a large sample of
clusters with \xray detections and measured redshifts. In addition to ensuring
the reality of the clusters, the \xray data also make some important auxiliary
cluster properties available, such as the cluster center and estimated mass;
these are used to define the cluster region and a fundamental length scale for
the calculation of the radial distribution of sources. With the knowledge of
the radial distribution, we can correct for the projection effect (\ie from a
cylindrical to a spherical volume) and calculate the LF.  Finally, by
cross-matching the NVSS source catalog with the source catalog of
the Two Micron All-Sky Survey (2MASS, \citealt{jarrett00}), we probe the
radio and near-IR (NIR) $K$-band bivariate luminosity function, which enables
estimation of the ``radio active fraction'' of cluster galaxies.  These
novelties and advantages of our study over previous ones will help theoretical
modeling of accretion onto black holes in cluster galaxies and estimating the
selection function for SZE surveys.

Even though both NVSS and 2MASS surveys do not deliver any redshift
information, it is possible to construct ensemble cluster properties, such as
the LF. Effectively, we are examining the properties of the excess of sources
toward clusters relative to a typical location on the sky.  The key is to know
the relation between the surface number density and the flux (or magnitude),
known as the $\log N$--$\log S$ relation, derived over a large enough area so
that it is an accurate description of the mean sky. Once we have this tool, we
can look for cluster signals by subtracting the background contributions.
Moreover, we can use the angular clustering properties of these sources to
estimate the variance in this background estimate.  The background-subtracted
or residual population averaged over many clusters provides the input for our
study of the cluster radio galaxies.  This residual population is at the
cluster redshift.  In the current study the signal is not strong because radio
sources in clusters are not intrinsically abundant.  One novelty of our
``stacking'' method over previous ones is that we have included the information
about the volume surveyed for each cluster, and thus we are able to present the
luminosity function of sources, not just the ``luminosity distribution'' (the
mean number of sources per luminosity interval per cluster).  We have applied
this method previously to study the galaxy properties in ensembles of clusters
\citep{lin03b,lin04,lin04b}.

Given the correlation between the masses of the central supermassive BH and the
bulge of galaxies \citep{magorrian98}, one naturally expects that low mass
galaxies will exhibit weaker nuclear activity. We present evidence suggesting
the existence of a population of low mass cluster galaxies that host moderately
powerful AGNs. By comparison of the radio LFs from all NVSS sources and from
those that exist in both NVSS and 2MASS catalogs, we can infer the number
density of AGNs that are hosted by galaxies which are too faint (not massive
enough) to be detected by the 2MASS.

In \S\ref{sec:ra_analysis} we give an overview of our analysis, describing the
cluster sample selection and mass estimation, the radio and NIR data used, and
some details necessary to carry out the calculations presented in later
sections. 
The radial distribution of the radio sources, as well as a comparison with that
of cluster galaxies as a whole, are presented in \S\ref{sec:sdp}.  In
\S\ref{sec:ra_blf} we examine the radio and $K$-band LFs in detail, calculate
the radio active fractions, and estimate the duty cycle of AGNs.  In
\S\ref{sec:lowmass} we present evidence supporting the existence of a
population of low mass galaxies that host moderately powerful AGNs.  The
implications of our findings on the heating of the ICM and on SZE cluster
surveys are discussed in \S\S\ref{sec:ra_heating} and \ref{sec:ra_sz},
respectively.  Our main results are summarized in \S\ref{sec:ra_summary}.
Unless noted, we assume the density parameters for the matter and the
cosmological constant to be $\Omega_M = 0.3$, $\Omega_\Lambda = 0.7$,
respectively, and the Hubble constant to be
$H_0=70\,h_{70}$~km~s$^{-1}$~Mpc$^{-1}$.

\section{Analysis Overview}
\label{sec:ra_analysis}

Below we first describe our cluster sample selection and cluster mass
estimation (\S\ref{sec:ra_sample}). In \S\ref{sec:ra_data} we give brief
accounts of the radio and $K$-band data from the NVSS and 2MASS surveys, and
the way cross-matching is done. As has been noticed by earlier studies
\citep[e.g.][]{edge91a,lin04b,gonzalez05}, the brightest cluster galaxies
(BCGs) play an important role in cluster formation and evolution, both in the
optical/near-IR (OIR) and radio wavelengths. We identify the BCGs in clusters
in a statistical sense, which is described in \S\ref{sec:ra_bcg}. Finally in
\S\ref{sec:ra_index} we examine the distribution of spectral index of radio
sources for a subset of the sample where multi-frequency detections are
available. This motivates our choice of the spectral index, which is required
when one converts from the observed specific radio flux to specific luminosity
in the rest frame 1.4~GHz.

\subsection{Cluster Sample and Mass Estimator}
\label{sec:ra_sample}

We construct our cluster sample from two large \xray cluster catalogs drawn
from the ROSAT All-Sky Survey (RASS), {\it NORAS} \citep{boehringer00} and
{\it REFLEX} \citep{boehringer04b}. For brevity, hereafter we refer to the
two samples together as the {\it RASS} sample. These catalogs provide the
cluster center (determined from the \xray emission) and the cluster restframe
luminosity $L_X$. We select only the clusters with galactic latitude $|b|\ge
20$ deg.  Using the observed relation between the \xray luminosity and cluster
binding mass \citep{reiprich02} 
\beq 
\label{eq:ra_xlm} 
\log \left[{1.4^2L_X(0.1-2.4\kev) \over h_{70}^{-2} 10^{40} {\rm ergs\ s}^{-1}}
\right] = A + \alpha \log \left( {1.4 M_{200} \over h_{70}^{-1} M_\odot} \right), 
\eeq 
where $A=-18.857$ and $\alpha=1.571$, we estimate the cluster
virial mass $M_{200}\equiv (4\pi/3)r_{200}^3\times 200\rho_c$, which is defined
as the mass enclosed by $r_{200}$, a radius within which the mean overdensity
is 200 times the critical density of the universe, $\rho_c$. We note that this
scaling relation provides a mass estimate accurate to $\lesssim 50\%$
(\citealt{reiprich02}; see also \citealt{reiprich06}) and a virial radius $r_{200}$ estimate accurate to 15\%. We
first evolve the restframe $L_X$ to $z=0$ (dividing $L_X$ by $E^2(z)$, where
$E(z) \equiv H(z)/H_0$ and $H(z)$ is the Hubble parameter), then use
Eqn.~\ref{eq:ra_xlm} to obtain the mass and calculate the virial radius of the
cluster. We assume any additional evolution of the scaling relation is
negligible, as our clusters are fairly nearby (see below).

After removing the clusters that are included in both the {\it NORAS} and {\it
REFLEX} catalogs, those that have no redshift measurements, and those that lie
too close to each other in projection (i.e.~the distance between the \xray
centers is less than the sum of the $r_{200}$'s), there are more than 700
clusters in the sample. 
We further exclude systems at $z>0.2$ or with $M_{200}<10^{13} M_\odot$. 
These cuts leave us with 577 clusters.

\subsection{Radio and Near-IR Data}
\label{sec:ra_data}

To study the radio sources in clusters, we use the 1.4 GHz data from the NVSS.
The survey covers the sky north of $-40$ deg.~in declination, with $45\arcsec$
FWHM resolution, and is complete to 2.5 mJy \citep{condon98}.
We use the survey catalog from SIMBAD to select radio sources within the
cluster fields. Because the NVSS is carried out in the ``snapshot'' mode,
sidelobes from strong sources may not be completely removed during the cleaning
process \citep{condon98}. We eliminate 4 clusters with very strong sources
(flux density $S>1000$ Jy) from the sample. {\it Our final sample contains 573
clusters.}

In the absence of redshift information about individual radio sources, we can
only estimate the expected cluster signal statistically, which is done by
subtracting the averaged background contribution from the observed data. In
\S\ref{sec:sdp} when we study the radial distribution of radio sources, we
determine the background level from the radial profile extending to large
clustercentric distance. For the luminosity function calculations (\S\ref{sec:ra_blf}), 
the
background contribution is estimated statistically from the $\log N$--$\log S$.
With the statistical background subtraction, the signal-to-noise ratio is
fairly low for most of the clusters. We therefore ``stack'' many clusters
together, and we defer the description of our stacking method to
\S\S\ref{sec:sdp} and \ref{sec:ra_blf}.

The background number is obtained by integrating the differential $\log
N$--$\log S$ above a flux threshold $S_{lim}$ over the cluster region, where
the $\log N$--$\log S$ is derived from a region of 60 deg.~radius centered at
the north Galactic pole (NGP). In Fig.~\ref{fig:ra_lnls} we show the cumulative
source surface density $\sigma(\ge S)$. For our analysis, a
relatively high flux threshold $S_{lim}=10$ mJy is chosen, which follows the
suggestion of \citet{blake02}, who find that near the NVSS completeness limit
(2.5 mJy), the mean surface density of radio sources varies as a function of
declination. Such a fluctuation becomes insignificant at $S\approx 15$ mJy. Our
choice of flux threshold is a compromise between the uniformity of the
background counts and the number of cluster sources available. For reference,
we find that $\sigma(\ge 10\,{\rm mJy})=16.91$ deg$^{-2}$, which is consistent
with that found by \citet{blake04}, 16.9 deg$^{-2}$.

\begin{inlinefigure}
   \ifthenelse{\equal{\figtype}{EPS}}{
   \begin{center} 
   \epsfxsize=8.cm
   \begin{minipage}{\epsfxsize}\epsffile{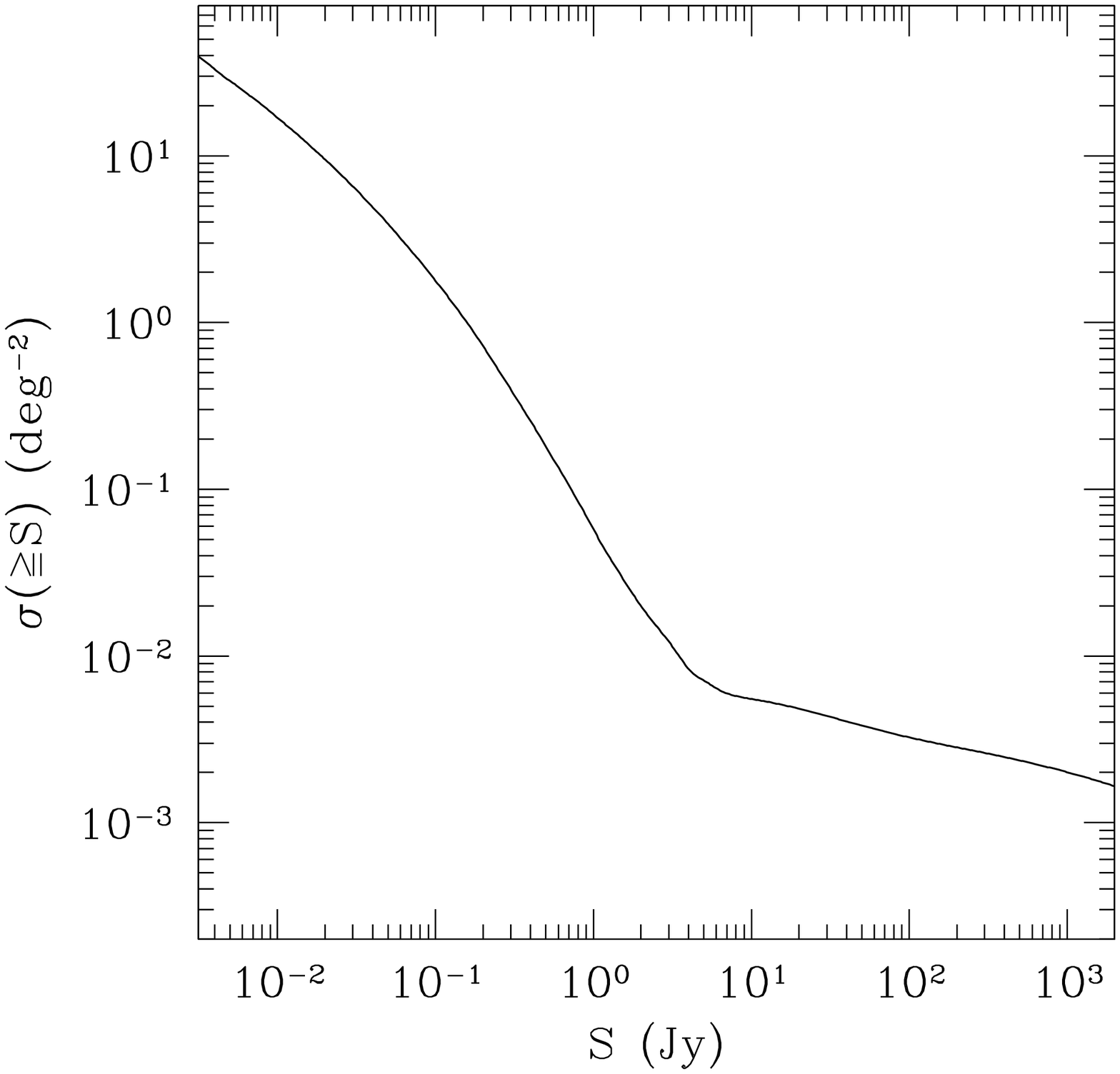}\end{minipage}
   \end{center}}
   {\myputfigure{f1.pdf}{0.1}{1.}{-20}{-5}} 
   \figcaption{\label{fig:ra_lnls}
    The cumulative $\log N$--$\log S$ from all NVSS sources within 60 deg.~from
    the North Galactic Pole. The NVSS survey completeness limit is 2.5 mJy 
    \citep{condon98}; however, we analyze the sample brighter than
    10 mJy throughout this paper. 
     }
\end{inlinefigure}

In some parts of our study we wish to connect the properties of radio sources
with the host galaxies (e.g.~identifying the brightest cluster galaxy), which
in practice is done by cross-matching the NVSS sources with the 2MASS $K$-band
point and extended source catalogs\footnote{The abundance of radio stars and
pulsars is very small compared to the extragalactic radio sources
\citep{helfand99}. Given that our clusters lie at 20 deg.~away from the
Galactic plane, inclusion of the point sources will increase the number of
NVSS-2MASS cross-matched sources (by $\sim 20\%$) without contaminating our
results with Galactic objects.}, for the $K$-band light is a good tracer of
stellar mass.  Preference is given to extended sources when there are both
point and extended sources present within a certain angular separation to an
NVSS object.  In doing so we need to determine an optimal maximum separation
$d_{max}$ between the central positions between the radio and 2MASS sources. 
We cross-correlate NVSS sources with $S\ge 10$ mJy within a region of 60
deg.~radius centered at the NGP with all objects in the 2MASS catalogs and
identify those whose positions in the two catalogs differ by less than
$1\arcmin$. At 10 mJy, the positional uncertainty for NVSS sources is about
$1-2\arcsec$ \citep{condon98}.  Examining the distribution of separation $d$
for sources that are matched (Fig.~\ref{fig:sep}), we find that the
distribution for NVSS-2MASS {\it point} source pairs shows a peak at small
angular separation ($d<3\arcsec$) and a rising trend with increasing
separation, which is due to chance projection. We therefore set
$d_{max}=5\arcsec$ when cross-correlating radio and $K$-band point sources. As
for the NVSS-2MASS {\it extended} source pairs, the distribution peaks sharply
at $d=1-2\arcsec$, and has a roughly constant tail beyond $d=20\arcsec$.
Because our clusters span a wide range in mass and in redshift, we follow the
approach of \citet{miller01b} and choose $d_{max}$ for each cluster so that the
probability of chance projections is expected to be less than $5\times
10^{-3}$.  In short, given a cluster, for every radio source within the virial
radius, we calculate the surface density of 2MASS extended sources around it
(within $30\arcmin$). The average of the surface densities, $\bar{\sigma}_{K}$,
is then used to solve for $d_{max}$ via $p(<d_{max}) = 0.005 = 1-\exp(-\pi
\bar{\sigma}_K d_{max}^2)$ \citep{condon98}.  The resulting $d_{max}$ lies in
the range of $9-30\arcsec$, which is similar to that used by \citet{miller01b}.

\begin{inlinefigure}
   \ifthenelse{\equal{\figtype}{EPS}}{
   \begin{center}
   \epsfxsize=8.cm
   \begin{minipage}{\epsfxsize}\epsffile{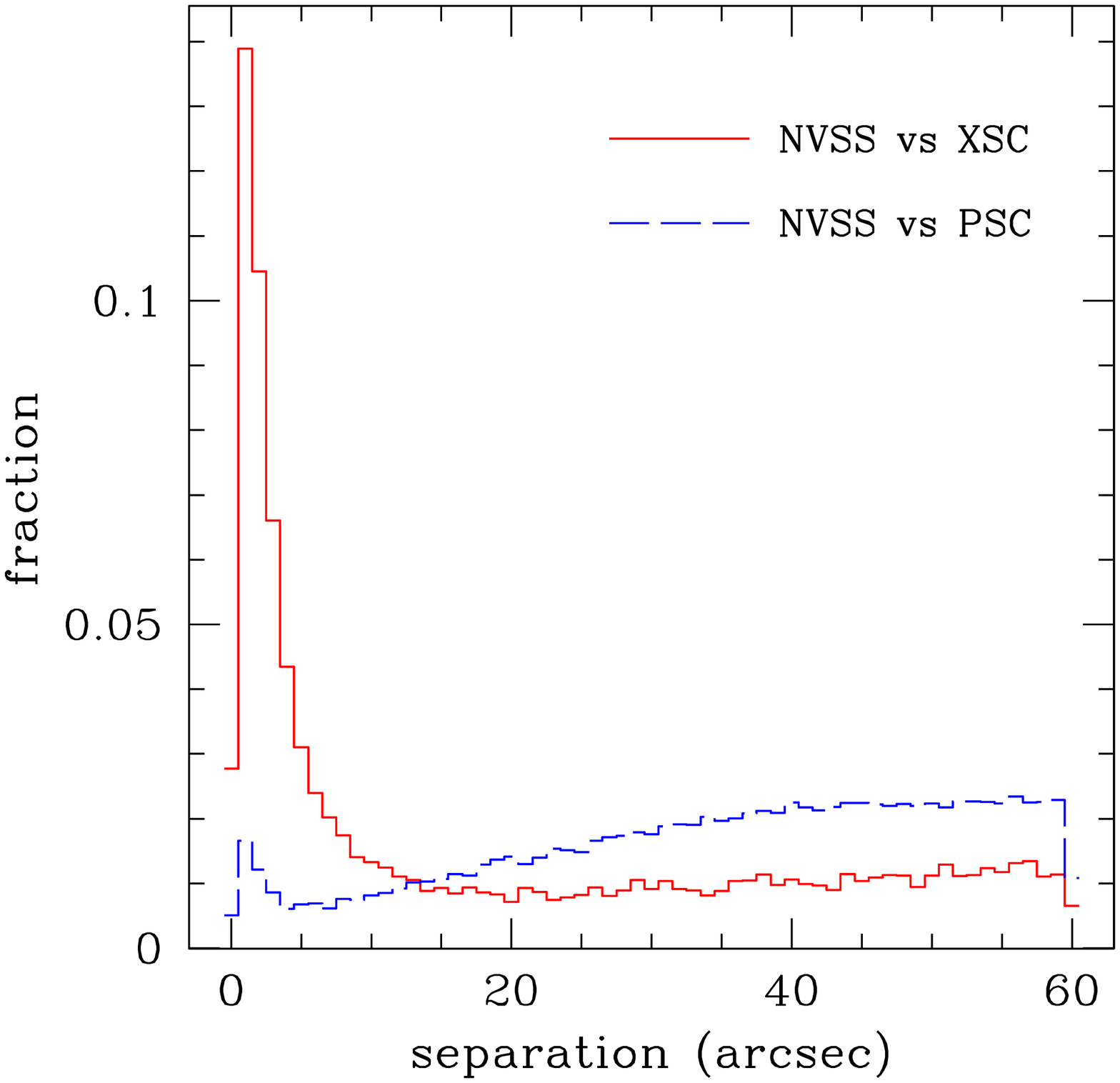}\end{minipage} 
   \end{center}}
   {\myputfigure{f1.pdf}{0.1}{1.}{-20}{-5}} 
   \figcaption{\label{fig:sep}
      Using a maximum matching radius of $1\arcmin$, we first match NVSS sources
stronger than 10 mJy with 2MASS extended sources. The unmatched NVSS sources
are further matched with 2MASS point sources.  The resulting distributions of
the separation are shown as the solid (NVSS with extended sources) and dashed
(NVSS with point sources) histograms, respectively.  In our analysis we match
NVSS sources with 2MASS point sources with a maximum separation
$d_{max}=5\arcsec$. See text for how $d_{max}$ for matching NVSS with 2MASS
extended sources is determined. 
     }
\end{inlinefigure}

Our results are robust against the way cross-matching radius is determined; for
example, using a constant $d_{max}=20\arcsec$ will not change the main
conclusions in this paper.

%


In the following, we refer to the NVSS-2MASS cross-correlated sources as the
``XC'' sources.

\subsection{Selection of Brightest Cluster Galaxies}
\label{sec:ra_bcg}

The most luminous cluster galaxy detected in the OIR wavelengths, the so-called
BCG, is a special class of object (see,
e.g.~\citealt{oemler76,schombert86,dubinski98,lin04b,gonzalez05}).  Their
dominance over the cluster OIR luminosity (especially so for low mass clusters,
\citealt{lin04b}) encodes clues to the merger history of clusters.  Outside the
realm of OIR wavelengths, the OIR-selected BCGs are often found to be dominant
as well (\eg in \xray or/and radio), and may be actively involved in a range of
thermodynamical processes in clusters
\citep[e.g.][]{tucker83,fabian03,mcnamara05,croton06}.

We identify the BCGs statistically from the 2MASS catalog for each cluster
using two selection criteria: a BCG must lie within $0.1 r_{200}$ of the
cluster center, and its $K$-band magnitude must be consistent with the BCG
luminosity-cluster mass ($L_b$--$M$) correlation found from a sample of 93 BCGs
\citep{lin04b}. We evolve the local $L_b$--$M$ relation using a population
synthesis model \citep{bruzual03}, assuming passive evolution in luminosity for
the BCGs \citep[e.g.][]{stanford98}. As we have shown previously
\citep{lin04b}, approximately 80\% of BCGs lie within 10\% of the cluster
center as defined by the peak in X-ray emission.

With these criteria, we are able to identify the BCG for 342 clusters; out of
these, 122 BCGs also have a radio counterpart with $S\ge 10$ mJy. This
corresponds to a detection rate of 36\%, which is consistent with the
measurements of the fraction of radio active BCGs found in earlier studies
\citep[e.g.][]{burns90,ball93}. We discuss more about the radio activity of
BCGs in \S\ref{sec:ra_klf}.

To see if there is any systematic difference between the full RASS sample and
the radio-loud BCG subsample,
we perform a 2D Kolmogorov-Smirnov test on the $z$ v.s.~$L_X$ distribution
for these cluster samples. The result indicates that the samples are similar. 

\subsection{Choice of Spectral Index}
\label{sec:ra_index}

For each cluster with known redshift, we can convert the radio flux to
luminosity.  In doing so, we need to account for the fact that what we observe
at 1.4 GHz corresponds to different wavelength in the restframe of the sources
(i.e.~the need to do a ``$k$-correction''). This requires some knowledge of the
average shape of the galaxy spectrum in the radio wavelengths, which is
typically described by a power-law $S_\nu \propto \nu^\alpha$, where $\nu$ is
the frequency.

By cross-correlating the NVSS catalog with two catalogs at 4.85 GHz (from the
GB6 and PMN surveys; \citealt[][and associated papars]{gregory96,griffith93}),
we can determine the distribution of spectral index $\alpha(1.4,4.85) \equiv
\log (S_{4.85}/S_{1.4})/\log (4.85/1.4)$ for radio sources within the cluster
fields. For 4036 NVSS sources projected within the virial radius of all {\it
RASS} clusters with $S\ge 10$ mJy, we search for their 4.85 GHz counterparts
within $1\arcmin$ (note that the FWHM of the GB6 and PMN surveys is about
$3.5\arcmin$ and $4.2\arcmin$, respectively; \citealt{gregory96,griffith93}).
In total there are 622 identifications, corresponding to a fraction of $15\%$
of all the 1.4 GHz sources considered.  We caution that these sources may be
background sources that happen to be projected within the cluster fields.  The
spectral indices are calculated using the fluxes given by the surveys; we do
not consider effects of the differences in the beam size. However, we do
include the uncertainties in the flux measurements in the determination of the
mean spectral index.

Because of the extraordinary properties of the BCGs, we also pay attention to
their spectral index distribution (SID). Out of the 622 matches between NVSS
and 4.85 GHz catalogs, 54 are the BCGs. Using data from the literature
\citep{burns90,ball93}, we calculate the BCG SID based on 73 galaxies.

In Fig.~\ref{fig:ra_alpha} we show the SID between 1.4 and 4.85 GHz for 568
non-BCG sources within cluster fields as the solid histogram. The SID has a
mean of $\bar{\alpha}(1.4,4.85)=-0.47\pm 0.15$, with a large root-mean-square
(RMS) scatter of 0.52. About 17\% of these sources have $\alpha\ge 0$. 
The SID for the BCGs is shown as the dashed histogram in
Fig.~\ref{fig:ra_alpha}.  Compared to the general population of (cluster) radio
sources, the BCGs seem to have a steeper spectrum; the mean is
$\bar{\alpha}(1.4,4.85)=-0.74\pm 0.05$, with a RMS scatter of 0.60. We note
that $14\%$ of the BCGs have a rising spectrum ($\alpha\ge 0$). 
Finally, the mean and scatter of the spectral indices from all 641 sources
considered here are $-0.51\pm 0.15$ and 0.54, respectively.

Because of the higher flux limits of the 4.85 GHz surveys, the SIDs deduced
here may be biased against large negative values of $\alpha$. This effect may
be stronger for non-BCG sources as they are more abundant and potentially less
powerful than BCGs.
Partly because of this bias and also to remain generally consistent with
previous literature \citep[e.g.][]{condon92,cooray98,coble06}, we adopt
$\bar{\alpha}=-0.8$ when applying the $k$-correction to obtain the rest frame
1.4 GHz luminosity (in \S\S 3-6).
In \S\ref{sec:ra_sz}, however, when we extrapolate our results at 1.4 GHz to
higher frequency, we will use the SID constructed from all the sources
considered in this section.
Bearing in mind that a 1D Kolmogorov-Smirnov test suggests that the probability
of the SIDs of BCGs and non-BCGs being drawn from the same distribution is
0.003, we will discuss the effect on our results due to this simplifying
assumption.

\begin{inlinefigure}
   \ifthenelse{\equal{\figtype}{EPS}}{
   \begin{center}
   \epsfxsize=8.cm
   \begin{minipage}{\epsfxsize}\epsffile{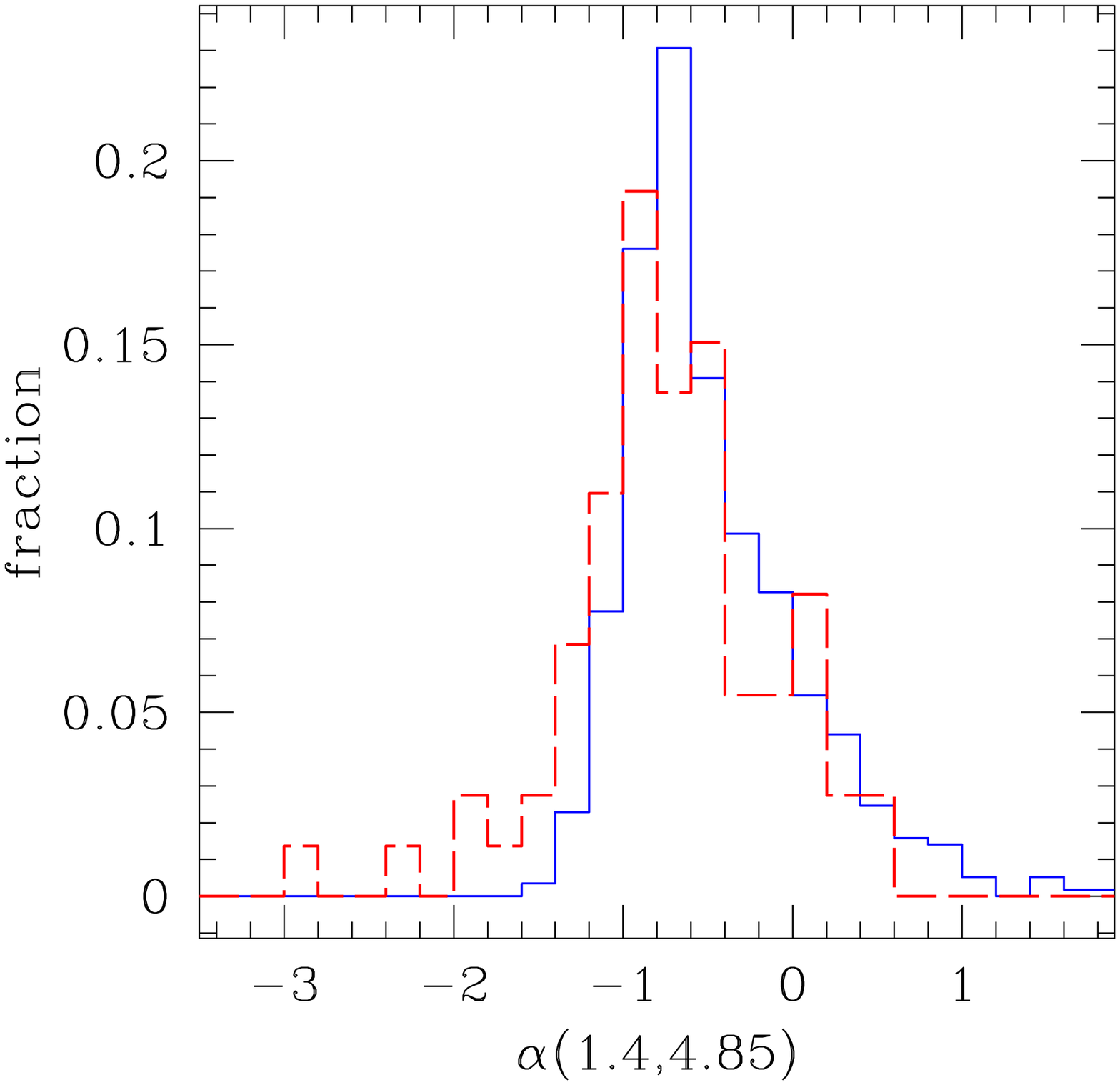}\end{minipage}
   \end{center}}
   {\myputfigure{f1.pdf}{0.1}{1.}{-20}{-5}}
   \figcaption{\label{fig:ra_alpha}
    The spectral index $\alpha$ for galaxies within cluster fields where
    $S_\nu \propto \nu^\alpha$.  We have matched the NVSS catalog with the GB6
    and the PMN catalogs within the {\it RASS} cluster fields. The distribution
    of $\alpha$ for 568 sources that are not associated with BCGs is shown as
    the solid histogram; the dashed histogram describes the SID for 73 BCGs
    (including 19 sources from the literature). The BCGs have a
    steeper spectral index.
     }
\end{inlinefigure}

\section{Ensemble Properties of Cluster Radio Sources: Surface Density Profile}
\label{sec:sdp}

There have been numerous investigations of the surface density profile of the
cluster radio sources
\citep[e.g.][]{ledlow95,slee98,miller02,morrison03b,reddy04,best04,branchesi06}.
Most of these quantify the surface density as a function of {\it metric}
distances from the cluster center. However, clusters are not of the same size
(our clusters span a factor of $\sim 200$ in mass); the cluster mass defines a
fundamental size scale, namely the virial radius \citep[e.g.][]{evrard96}. Here
we examine the radial distribution relative to $r_{200}$, which is a proxy of
the virial radius, and we use the  \xray luminosity to estimate this quantity
(see Eqn.~\ref{eq:ra_xlm}).

\begin{inlinefigure}
   \ifthenelse{\equal{\figtype}{EPS}}{
   \begin{center}
   \epsfxsize=8.cm
   \begin{minipage}{\epsfxsize}\epsffile{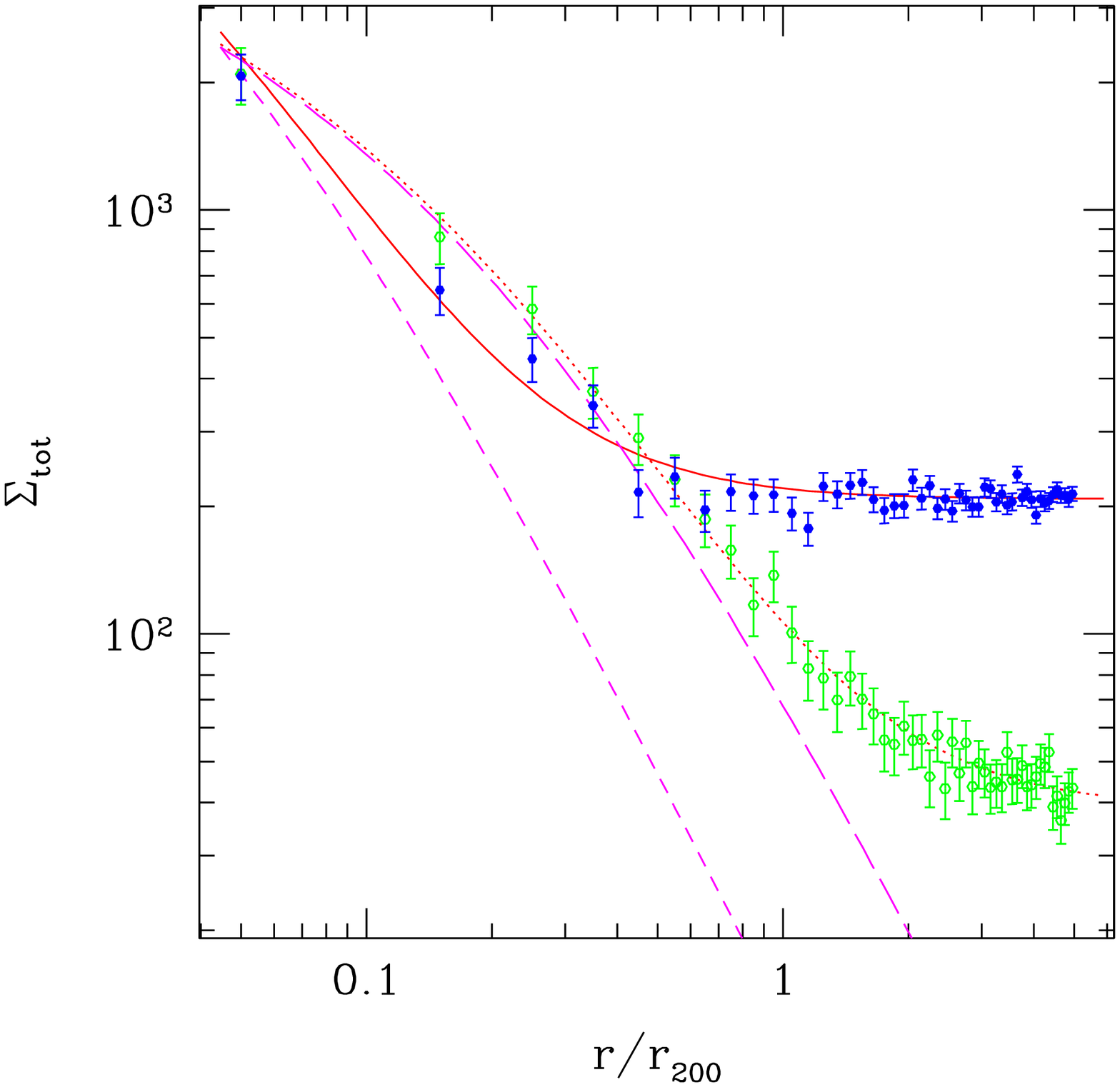}\end{minipage}
   \end{center}}
   {\myputfigure{f1.pdf}{0.1}{1.}{-20}{-5}}
   \figcaption{\label{fig:ra_rad}
       Radial distribution of radio sources with luminosity
    $P\ge 10^{23}{\rm W\,Hz}^{-1}$ in clusters (solid points).  The best-fit
    total profile (cluster plus background) is shown as the solid line, while
    the best-fit cluster model (an NFW
    profile with concentration $c=25\pm 7$) is shown as the short
    dashed line. No BCGs (which, by our definition, all lie within
    $0.1 r_{200}$) are included in the profile .  Inclusion of
    the BCGs results in a profile with $c=52^{+25}_{-14}$.
    The normalization of the profile is arbitrary.
    As a comparison we also show the surface density profile of all galaxies
    more luminous than $M_K = -24$ as the hollow points. The dotted and long
    dashed lines are the best-fit models for the total and cluster profiles.
    Apparently the galaxies have a much broader distribution than the AGNs.
    More details of the comparison are presented in \S\ref{sec:ra_2massprof}.
     }
\end{inlinefigure}

The surface density of all sources within a cluster field can be decomposed
into a cluster and a (constant) background component. To reduce the statistical
noise, we construct the total surface density by summing up contributions from
many cluster fields. For each cluster, the projected distance of every source
to the center (defined as the \xray emission peak) is normalized to
$r_{200}$ and then binned. We include sources out to $5r_{200}$ from the
cluster center, to better determine the background value.  We use the
``universal'' profile of \citet[][NFW]{navarro97} as the model for the spatial
distribution of radio sources in clusters, where 
the normalization of the projected profile 
and the concentration $c \equiv r_{200}/r_s$ are the
parameters to be fit. Here $r_s$ is the characteristic scale of the NFW
profile.  The best fit model and background value are obtained by minimizing
the $\chi^2$.
The uncertainty in each radial bin is estimated assuming Poisson statistics,
following the prescription given by \citet{gehrels86}.

In Fig.~\ref{fig:ra_rad} we show the total surface density for sources with
radio luminosity $P \ge 10^{23}\, {\rm W\,Hz}^{-1}$ out to $5 r_{200}$. 
For a given lower limit of luminosity, there is a
corresponding upper limit in cluster redshift $z_{up}$ so that all the sources
included have fluxes greater than our adopted NVSS completeness limit (10 mJy).
The clusters used in constructing the surface density profile shown here are at
$z\le z_{up}=0.0647$.
In total,
16,646 objects from 188 clusters contribute to the total profile, of which 836
lie within $r_{200}$.  
At large radii, the profile is quite flat, which allows an accurate
determination of the background level.  Within $r_{200}$, we estimate the number
of background sources is 621.
The best-fit cluster profile (dashed
line) has $c=25\pm 7$.
Mock observations are used to estimate the uncertainties in the best-fit
concentration parameter and evaluate the systematics in the fitting procedure.
We describe our method in \S\ref{sec:appen}.

We note that no BCGs are included in the total
profile. By our definition, the BCGs all lie within $0.1 r_{200}$.
Furthermore, because there are typically only a few sources per cluster, the
BCGs stand as a relatively large proportion in the overall source sample
(there are 31 BCGs that have $\log P\ge 23$ in this 188 cluster 
subsample; the number of non-BCG cluster radio sources within $r_{200}$ 
is estimated to be 183).  The
inclusion of the BCGs thus makes the resulting profile much peakier.
Fitting an NFW profile to it gives $c=52^{+25}_{-14}$.

\subsection{Comparison with Profile from Spectroscopically Confirmed
  Members}
\label{sec:ra_morrison}

The surface density profile derived from our statistical method can be compared
to that from a targeted spectroscopic survey \citep[][hereafter M03a,b,
respectively]{morrison03,morrison03b}.  M03a observe a sample of 30 clusters
detected in the RASS out to $z\le 0.25$ with the VLA at 1.4 GHz in the A and C
configurations.  The sample is complete to $2\times 10^{22}$ W$\,$Hz$^{-1}$.
Optical observations are carried out for identifying and confirming cluster
radio sources. M03b obtain the radio source surface density by stacking the
clusters in the metric distance space, and fit the resulting profile with a
King model. We apply our method to the data from M03a. Of the 30 clusters in
the M03a sample, 14 are also in our sample. For these clusters we use the \xray
luminosity and \xray center provided by the {\it RASS} catalogs, and construct
the surface density profile based on the location of spectroscopically
confirmed members listed in Table 4 of M03a. We use only the galaxies with
redshifts within 3000 km$\,$s$^{-1}$ from the cluster restframe. Within
$r_{200}$, there are 56 confirmed member galaxies in the ensemble cluster. The
resulting surface density can be described by an NFW profile with
$c=14^{+18}_{-6}$. 
Note, however, we do not distinguish and exclude the BCGs in M03's data. A fair
comparison therefore would be to include BCGs and apply a radio luminosity cut
at $2\times 10^{22}$ W$\,$Hz$^{-1}$ in our NVSS source sample as well. 
To reach down to $P_{lim}=2\times 10^{22}$ W$\,$Hz$^{-1}$, $z_{up}=0.0296$. For
50 clusters that are within the redshift limit, we find an NFW model with
$c=21^{+19}_{-9}$ is a fair description. 
Finally, if we build the surface density from the M03 sample with $P_{lim}=
10^{23}$ W$\,$Hz$^{-1}$ (28 objects), the best-fit has $c=48^{+18}_{-12}$, also
in good agreement with our result (with BCGs).  We therefore conclude the
profile derived from our data is consistent with that from the M03 survey.

\subsection{Dependencies on Radio Luminosity and Cluster Mass}
\label{sec:ra_dep}

We list in Table \ref{ra_1} the concentration parameters of the best-fit NFW
profiles for the radial distribution of radio sources selected at various
luminosities, with and without BCGs.  
In the Table for fits with $c>80$, we do not list the fit values, as these are
practically power-law, and the NFW profile is no longer a useful description.

\begin{table*}[htb]
\begin{center}
\begin{minipage}{0.65\textwidth}
\begin{center}
\caption{Radio Luminosity Dependence of Radial Distribution}
\label{ra_1}
\vspace{1mm}
\begin{tabular}{clcll}
\hline \hline
 & & & \multicolumn{2}{c}{concentration}\\
\cline{4-5}
$\log P_{lim}^a$ & $z_{up}^b$ & $N_{cl}^c$ & w/o BCGs & w/ BCGs \\
\hline
22.5 & 0.03705 & 84 & $25\pm 12$ & $43^{+32}_{-15}$\\
23.0 & 0.06470 & 188 & $25\pm 7$ & $52^{+25}_{-14}$\\
23.5 & 0.1118 & 364 & $21\pm 7$ & $48^{+18}_{-12}$\\
24.0 & 0.1899 & 554 & $22\pm 6$ & $66^{+24}_{-15}$\\
24.5 & 0.20   & 573 & $59\pm 11$  & $>80$ \\
25.0 & 0.20   & 573 & $>80$  & $>80$\\
\hline
\end{tabular}
\end{center}
{\small
Note.-- Surface density constructed from NVSS-only sources, using the
full {\it RASS} cluster sample.

$^a$ Minimum radio luminosity.

$^b$ Maximum redshift for a cluster to be included in each subsample.

$^c$ Number of clusters.
}
\end{minipage}
\end{center} 
\end{table*}

From Table \ref{ra_1} it seems that, except for very luminous AGNs (those with
$\log P \ge 24.5$), there is no suggestion of luminosity-dependence of the
concentration (\ie the shape of the profiles). The fact that very luminous AGNs
are more centrally located has long been noticed
\citep[e.g.][M03b]{ledlow95,miller02,rizza03}. This can result from the combined
effects of (1) more luminous galaxies are more strongly clustered, and (2) more
luminous galaxies have larger probability of hosting more powerful AGNs. We
will discuss these in \S\ref{sec:ra_2massprof}.  A third possibility is the
confining pressure due to the dense ICM near the cluster center. In
\S\ref{sec:disc_faint} we will discuss this issue in more detail.

Earlier studies of the radio source surface density have usually compared the
data to a King profile, and found evidence of an excess of sources in the
central bins. This suggests that the distribution is more concentrated than the
model. As we have shown here, a cuspy profile like the NFW (or even a
power-law) provides a reasonable description of the composite surface density
profile of cluster radio galaxies.

We also check if there are any cluster mass-related trends in the radial
distribution. We consider two limiting radio luminosities to examine the issue
($\log P_{lim}=23.5$ and 24), and arbitrarily set the division between the high
and low mass clusters at $\log M_{200}=14.2$. The resulting best-fit
concentration parameters are listed in Table \ref{ra_mdep}. When BCGs are
included, the profiles are very different for low and high mass clusters,
although the difference is much smaller when BCGs are excluded from the fit.
This probably mainly reflects the different radio active fraction for BCGs in
high and low mass clusters (\ie BCGs in high mass clusters tend to be more
active in the radio; see \S\ref{sec:ra_klf}). We note in passing that, with
$\log P_{lim}=24$, we can include clusters with redshift up to $z_{up}=0.1899$,
which allows us to check if there is any redshift-dependent trend. Separating
the sample at $z=0.1$, we find that the radial distributions at the two
redshift bins are similar.

\begin{table*}[htb]
\begin{center}
\begin{minipage}{0.65\textwidth}
\begin{center}
\caption{Cluster Mass Dependence of Radial Distribution}
\label{ra_mdep}
\vspace{1mm}
\begin{tabular}{clcll}
\hline \hline
& & & \multicolumn{2}{c}{concentration}\\
\cline{4-5}
$\log P_{lim}$ & $\log M_{200}$ & $N_{cl}$ & w/o BCGs & w/ BCGs\\
\hline
23.5 & $13-14.2$   & 106 & $15^{+12}_{-6}$ & $17^{+15}_{-7}$\\
     & $14.2-15.5$ & 258 & $23^{+10}_{-6}$ & $61^{+28}_{-17}$\\
\cline{1-5}
24.0 & $13-14.2$   & 106 & $13^{+17}_{-6}$ & $14^{+18}_{-7}$\\
     & $14.2-15.5$ & 448 & $24^{+8}_{-5}$ & $79^{+33}_{-21}$\\
\hline
\end{tabular}
\end{center}
{\small
Note.-- Surface density constructed from NVSS-only sources, using the
full {\it RASS} cluster sample.
}
\end{minipage}
\end{center}
\end{table*}

\subsection{Comparison with $K$-band Galaxy Radial Profile}
\label{sec:ra_2massprof}

Following the same procedure outlined in \S\ref{sec:sdp} (also see \citealt{lin04},
hereafter LMS04), we find the best-fit cluster
profile from the total surface density for 2MASS galaxies. 
The $k$-correction of the form $k(z)=-6\log(1+z)$ is used \citep{kochanek01}. 
The galaxies are selected from a cluster subsample which is constructed 
so that no cluster lies in projection within six times the virial radius of any 
other cluster. We refer to
these clusters as the ``isolated'' sample, and will explain below the reason for
using this sample to study the profile. 
A comparison between the galaxy total surface density
profile ($M_K\le -24$; hollow points) with that of the radio sources ($\log P\ge 23$; 
solid points) is shown in Fig.~\ref{fig:ra_rad}. 
The best-fit NFW profile has $c=4.2^{+0.5}_{-0.4}$.  
The normalization of the profiles is arbitrary, and we have shifted the radio
source profile so that the best-fit cluster models (long and short dashed
lines) meet in the innermost point in the figure. It is apparent from the
figure that the two populations have very different radial distributions.
We note that by fitting a King model to the projected surface density profiles,
M03b find that the radio-loud AGNs (HRLGs in their terminology) has a smaller
core radius compared to the red galaxies, implying that powerful radio galaxies
is more centrally distributed.

Similar to our treatment of radio sources, 
we use mock observations to estimate the uncertainties in the best-fit
concentration parameter; the details of the procedure
are provided in \S\ref{sec:appen2}.

Compared to the galaxy concentration $c=2.9\pm 0.2$ found in LMS04 (\S3.1
therein), the
value obtained here ($c=4.2^{+0.5}_{-0.4}$) seems to be quite high. This
discrepancy is caused by three factors: (1) the use of a luminosity cut in selecting
the galaxies in the present analysis (as opposed to including all galaxies brighter than
the 2MASS completeness limit), (2) the use of a different cluster mass indicator
(\xray luminosity as opposed to \xray temperature), and (3) the sample selection and
background estimation. We explain their effects in turn.\\
%
%
{\it i. Luminosity Cut}:
As we will show below, more luminous galaxies are more centrally distributed in clusters.
Therefore, mixing faint and luminous galaxies effectively reduces the concentration
of the surface density profile. Furthermore, as the faint galaxies are more numerous than
the luminous ones, they have more weight in the resulting profile.\\
%
{\it ii. Cluster Mass Estimators}:
It has been found that for clusters hotter than $\sim5$
keV, the mass derived from $L_X-M$ relation is about 30\% larger than that from
the $T_X-M$ relation \citep{popesso05}. For the same galaxy distribution (in
metric space), a larger virial radius ($r_{200}$) corresponds to a more
centrally concentrated profile, namely a larger $c$. With the difference in
mass caused by the different scaling relations, we estimate the change in
concentration is at $10-20\%$ level (see e.g.~\S6 of LMS04 for discussion on
the effects of cluster mass uncertainty in the estimation of concentration).
To further quantify this effect, we compile a sample of 50 clusters which are
present in both the {\it RASS} sample and the 93 cluster sample used in LMS04
(and therefore both $L_X$ and $T_X$ are available).
We construct two surface density profiles with this sample, one using $L_X$
to infer $r_{200}$ and the other using $T_X$.  It is found that the concentration
value of the profile using $L_X$ as the mass indicator is higher than its $T_X$
counterpart by 15\%, in good agreement with the estimate above.\\
%
{\it iii. Cluster Sample Selection and Background Estimation}:
Another difference of the two analyses is the way background level is estimated.
In LMS04 the number of background galaxies is estimated based on the global
2MASS number count ($\log N$--$\log S$). In the current analysis, we fit the cluster
and background
surface densities simultaneously, using galaxies extending to $5 r_{200}$. While
the former method ignores the presence of large scale structure in the foreground or
background of the clusters, the latter (which we refer to as the ``local'' estimate) 
may be affected by it strongly. Averaging over
many different line-of-sights, one might hope to find an agreement between the
two estimates. In \S\ref{sec:appen} we show that this is indeed the case for the radio
sources with the {\it RASS} cluster sample (which is the reason we use the full cluster
sample to study the profile of the radio sources in \S\ref{sec:sdp}).
Most likely because of their extended spatial distribution, for galaxies, the comparison 
of the global and local background estimates shows a surprising difference.
For $M_K\le -24$ galaxies selected from the full {\it RASS} cluster sample, we find that
within three, four, and five times the virial radius, the number of background sources
derived locally is 14\%, 23\%, and 24\% higher than that based on the global estimates.
To better understand the origin of the discrepancy, we have selected a subsample 
from all the {\it RASS} clusters which are not closer to each other (in projection) 
than $8 r_{200}$, and
found that the differences between the locally and globally derived background estimates 
become $\lesssim 4\%$ beyond $3 r_{200}$. Based on the expectation that the two
methods of background estimation should converge at some large clustercentric distance,
we regard such a cluster sample as suitable for the study of the surface density profile of
cluster galaxies. The drawback of demanding clusters to be isolated out to
$8 r_{200}$ is that the sample size becomes small ($\sim 50$) and the measurement of the profile
is noisy. We find that a subsample isolated out to $6 r_{200}$ is a good compromise
between the accuracy of background determination ($\lesssim 5\%$) and the number of 
clusters available ($\sim 100$, for studying $M_K\le -24$ galaxies).


In short, we attribute the differences in the value of concentration found in 
the current analysis and our previous study to the three factors described above.
With a better understanding of the systematics of the procedure to fit the
surface density profile, the results presented here should be more representative
of the nature of cluster galaxies.

To conclude this section, we examine the dependencies of the galaxy concentration on
the magnitudes of the galaxies.  The results are shown in Table
\ref{ra_3}. The last row in Table \ref{ra_3} suggest a trend of luminosity
segregation; the most luminous galaxies (e.g.~$M_K\le -25$) tend to distribute with larger
concentrations. In LMS04 (\S5.2 therein) we examine a sample of 93 clusters and find that there
is no difference in concentration for luminous or fainter galaxies (with an
arbitrary division set at $M_K=-23.5$). We repeat the analysis by examining the concentration for
galaxies with $M_K\le -23.5$ and $-23.5<M_K\le -21$. We find that $c=4.2^{+0.5}_{-0.4}$ 
and $3.9^{+1.3}_{-0.9}$, respectively.
Essentially, the luminosity segregation is only strongly apparent with the most
luminous galaxies, and the larger cluster sample used here allows analysis of more finely
binned gradations in galaxy luminosity.

\begin{table*}[htb]
\begin{center}
\begin{minipage}{0.65\textwidth}
\begin{center}
\caption{$K$-band Luminosity Dependence of Radial Distribution}
\label{ra_3}
\vspace{1mm}
\begin{tabular}{llcll}
\hline \hline
& & & \multicolumn{2}{c}{concentration}\\
\cline{4-5}
$M_{K,lim}$ & $z_{up}$ & $N_{cl}$ & w/o BCGs & w/ BCGs\\
\hline
$-22.0$ & 0.02710 & 15 &  $5.0^{+1.3}_{-1.0}$ & $6.4^{+1.7}_{-1.2}$\\
$-23.0$ & 0.04325 & 49 &  $3.6^{+0.5}_{-0.4}$ & $4.7^{+0.6}_{-0.5}$\\
$-24.0$ & 0.06930 & 106 &  $4.2^{+0.5}_{-0.4}$ & $6.4^{+0.7}_{-0.6}$\\
$-25.0$ & 0.1117 & 207 &  $7.8^{+1.4}_{-1.1}$ & $28.5^{+5.3}_{-4.1}$\\
\hline
\end{tabular}
\end{center}
{\small
Note.-- Surface density constructed from 2MASS sources, using the ``isolated''
cluster subsample (see \S\ref{sec:ra_2massprof}).
}
\end{minipage}
\end{center}
\end{table*}

\section{Ensemble Properties of Cluster Radio Sources: 
  Radio and K-band Bivariate Luminosity Function}
\label{sec:ra_blf}

Consider the $K$-band and radio bivariate luminosity function of galaxies
$\psi$, where $\psi(L_K, P) dL_K dP$ is the spatial
number density of galaxies having $K$-band and radio luminosities within the
intervals [$L_K,L_K+dL_K$] and [$P,P+dP$], respectively. 
When this function is projected onto
one of the luminosity axes, one obtains a univariate LF
$\phi(L_K)$ or $\phi(P)$.

The bivariate LF $\psi(L_K, P)$ is extremely useful, because it can be thought
of as the conditional probability of a galaxy of a given $K$-band luminosity (a
mass surrogate) to host an AGN of radio luminosity $P$. With it we can begin to
explore how the central engines are fueled (whether galaxy environment is 
relevant?) and how long outbursts or periods of activity last (\ie the ``duty
cycle''), which in turn should provide important insights into galaxy
evolution \citep[e.g.][hereafter LO96]{ledlow96}.

Without accurate redshift determinations, it is not possible to map out
the bivariate LF (BLF) fully. Using the NVSS-2MASS XC sources, however, helps
significantly as they allow us to navigate through the $L_K$--$P$ space.
By specifying a cutoff in either the radio or $K$-band fluxes, we can build
semi-BLFs such as $\psi(L_K, \ge P)$ and $\psi(\ge L_K, P)$.  Furthermore,
comparisons between these semi-BLFs and the univariate LFs yield clues about
the populations that have either $K$-band or radio luminosities too faint to be
included in the surveys. 
For example, the ratio between $\psi(L_K, \ge
P)$ and $\phi(L_K)$ is the fraction of galaxies with a given $K$-band luminosity that
have radio luminosity greater than $P$, which we refer to as the 
radio active fraction (RAF).

The pioneering study of LO96 (see also \citealt{ledlow95b,ledlow95}) examines the
BLF based on a large sample of Abell clusters.  Our results build upon
their results for several reasons. First, \citeauthor{ledlow96}'s
analysis, despite its large cluster sample and the spectroscopic redshift
confirmation of the cluster memberships, focuses only on sources within 0.3
Abell radius, and thus may not reflect the full cluster population.  Second,
they adopt the cluster optical LF from other studies, while we construct the
$K$-band LF from the same sets of clusters. Third, our cluster sample is \xray
selected, thus significantly reducing contamination from spurious clusters that
are due to chance projection\footnote{Contaminations due to point sources
also happen in \xray surveys. However, as the {\it NORAS} and {\it REFLEX}
surveys have been complemented with spectroscopic confirmation, as well
as correlations with galaxy distribution, the contamination rate in the final
catalogs should be very small \citep{boehringer00,boehringer04b}.
}.
In addition, \xray data provide an estimate of
the cluster virial region, which sets a physically motivated region for the
examination of cluster properties. Finally, we pay particular attention to any
possible trends with cluster mass, and to the effects of the BCGs, neither of
which have been addressed in previous studies.

\subsection{Radio Luminosity Function}
\label{sec:ra_rlf}


We construct the radio source LF based on a method developed in LMS04.
Essentially, we use the cluster radio galaxy excess toward each cluster to
build up a luminosity function. Specifically, for each cluster, we assume every
source is at the cluster redshift $z_c$ and transform the flux density into the
luminosity (in units of W$\,$Hz$^{-1}$) with the standard $k$-correction (e.g.
M03a), assuming an average spectral index of $\alpha=-0.8$ (see
\S\ref{sec:ra_index}). We bin the radio sources projected within $r_{200}$ in
the logarithmic luminosity space. The faintest bin included is the bin with a
faint edge that is just greater than the luminosity that the limiting flux
density (i.e.~10~mJy) corresponds to at $z_c$. For each bin we estimate the
background count from the $\log N$--$\log S$ (by transforming the luminosity
interval into flux densities). Once every cluster field has been examined, for
each bin we subtract the cumulative background contribution from the total
observed number of sources, and that difference is then divided by the sum of
volumes from all the clusters that contribute to the bin. For each
cluster we adjust the cluster volume to account for the expected difference
between the number of cluster sources identified in the cylindrical volume
surveyed and the number expected in the spherical volume we are trying to
study. We do this using an NFW profile for the cluster radio galaxies with
$c=25$, consistent with the observed radial profile (see \eg Table~\ref{ra_1}).

The uncertainty in each bin includes the Poisson noise,  the
cosmic variance in the background count, and the uncertainty in
cluster mass. We follow the standard
treatment \citep[e.g.][]{mwhite95} for the variance in the mean
surface density of the background
\begin{eqnarray}
\nonumber
\sigma_{cv}^2 & = & {\sigma^2 \over 16\pi^3 (1-\cos\theta_{200})^2}
  \sum_{l=1}^{l_{max}} (2l+1) C_l B_l^2, \\
B_l & = & 2\pi \int_{\cos \theta_{200}}^1 P_l(x) dx,
\end{eqnarray}
where $\sigma$ is the mean background surface density (from the
$\log N$--$\log S$), $\theta_{200}$ is the cluster angular size, $C_l$
is the radio source angular power spectrum (from \citealt{blake04}),
$P_l(x)$ is the Legendre polynomial, and $l_{max}$ is the maximum
angular frequency corresponding to the angular resolution of the NVSS. The
cosmic variance in each luminosity bin is calculated by summing up
in quadrature the $\sigma_{cv}$ terms from every cluster.

To account for the effect of uncertainties in cluster mass, given a cluster
sample, for every cluster we perturb its mass from the nominal value (inferred
from $L_X$) in the logarithmic space by a Gaussian random number with
a standard deviation of 0.2 (appropriate for a fractional uncertainty in mass
of $\sim 50\%$; \S\ref{sec:ra_sample}), and use the resulting $r_{200}$ to
include radio sources and construct the LF. Repeating this process ten times,
for each luminosity bin, we calculate the dispersion of the ten LFs, which
is then summed with the uncertainties due to Poisson statistics and the cosmic
variance.


We show in Fig.~\ref{fig:ra_lf} the composite radio LF (RLF) for all 573
clusters (solid points). 
To conveniently describe
our data, we take the LFs for AGN and star forming (SF) galaxies
determined by \citet[][hereafter C02]{condon02} for the {\it field}
population and fit the sum of the two to the data, allowing the
amplitudes of the RLFs to vary, while fixing the shape of the RLFs. In
doing so we are assuming that the shape of the RLFs are the same both
within and outside clusters. The reason we only fit for the
amplitudes is because our data only allow us to fit for the combined
contribution from the AGN and SF populations. The functional form
that C02 adopt is
\begin{equation}
\label{eq:ra_lffit} \log \phi = y - \left( b^2 + \left( \frac{\log
P-x}{w} \right)^2 \right)^{1/2} - 1.5 \log P.
\end{equation}
For the AGN RLF, C02 find that $(b,x,w)=(2.4,25.8,0.78)$; the same
set of parameters for the SF RLF is $(b,x,w)=(1.9,22.35,0.67)$.
These parameters, which control the shape of the RLFs, are kept fixed
during fitting. We find that RLFs with the amplitude parameters
$y=37.97$ and $35.00$ best describe the cluster AGN and star forming
galaxy populations, respectively. The dotted line in
Fig.~\ref{fig:ra_lf} shows the sum of our best-fit cluster AGN and
SF RLFs.

In Fig.~\ref{fig:ra_lf} we also plot the field RLF from C02 (AGN and SF
galaxies combined), multiplied by a factor of $200/\Omega_M (\approx
855^{+181}_{-101}$, where we use the value favored by the three-year WMAP
result, $\Omega_M=0.238^{+0.027}_{-0.045}$, \citealt{spergel06}) to account for
the difference in the mean overdensities in the clusters and the field (dashed
line).  Compared to the (scaled) field RLF, there is an overabundance of radio
sources in clusters. As a way to quantify the difference, we integrate both the
cluster and (unscaled) field AGN RLFs over the same range in the radio
luminosity.  Within the interval $\log P = 24-27$, the ratio between the
resulting number densities is $\approx 5690\pm 1000$, or $6.8\pm 1.7$ times the
difference in overdensity ($200/\Omega_M$).  Using the RLF of \citet{sadler02}
results in a larger difference ($\approx 8690\pm 1520$).  In these calculations
we have assumed a 5\% uncertainty of the field RLFs.
Our RLF is consistent with that obtained by \citet{massardi04}; integrating
their RLF (as tabulated in Table 2 therein), we find the number density in
clusters is $\approx 5540\pm 1090$ times higher than that in the field
(corresponding to $6.6\pm 1.8$ times the difference in overdensity).

The comparison is robust against our choice of the mean spectral index.
Using $\bar{\alpha}=-0.5$, which is the mean derived from all the sources for which
we have both 1.4 and 4.85 GHz fluxes (\S\ref{sec:ra_index}), we find the ratio
of number densities is $6.7\pm 1.7$ times higher than $200/\Omega_M$. 

\begin{inlinefigure}
   \ifthenelse{\equal{\figtype}{EPS}}{
   \begin{center}
   \epsfxsize=8.cm
   \begin{minipage}{\epsfxsize}\epsffile{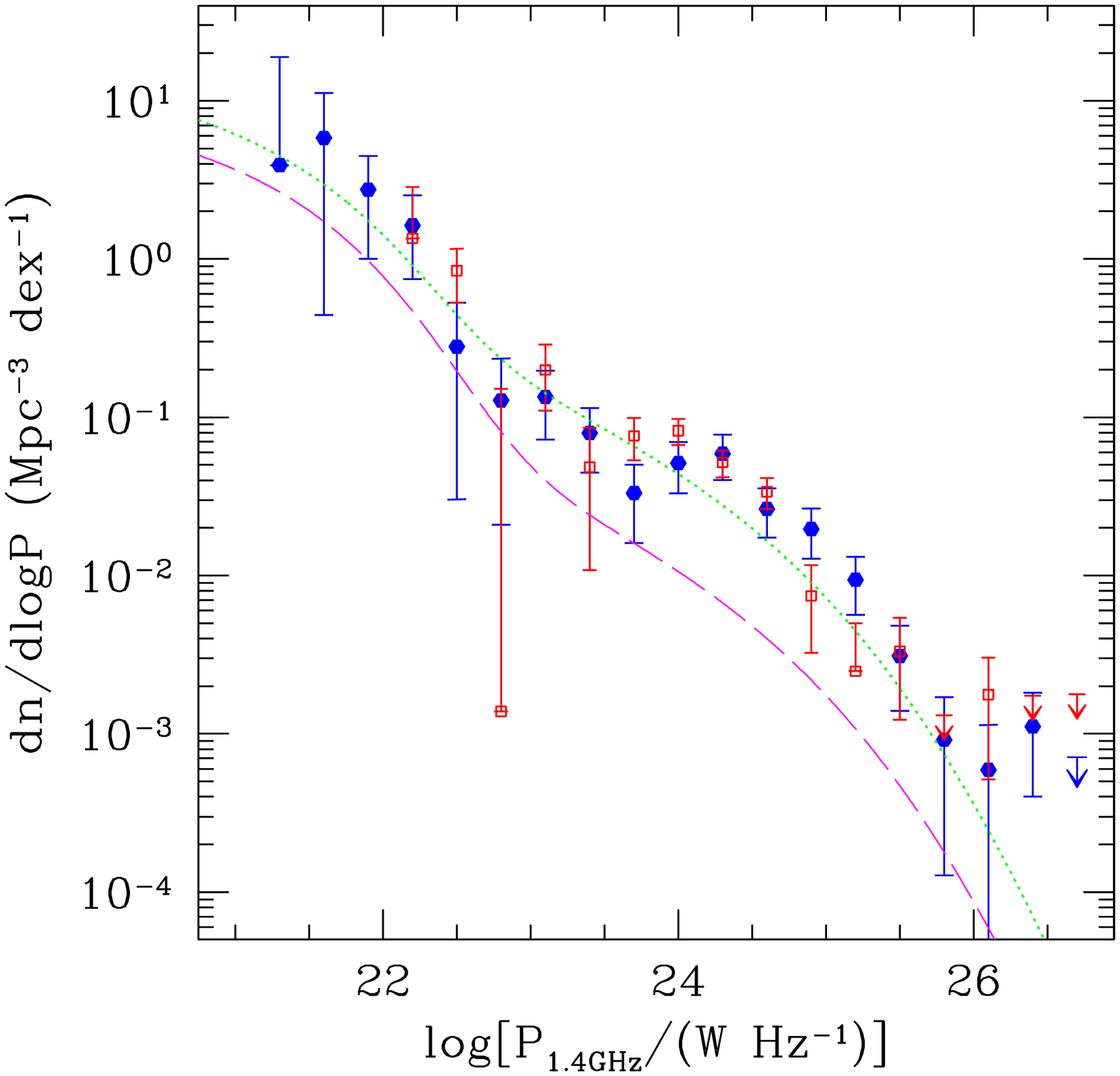}\end{minipage}
   \end{center}}
   {\myputfigure{f1.pdf}{0.1}{1.}{-20}{-5}}
   \figcaption{\label{fig:ra_lf}
The composite radio luminosity function (RLF) within $r_{200}$.  We denote
``per $\log_{10}$ interval of radio luminosity'' by ``dex$^{-1}$''.  We
construct the RLF based on two data sets: the NVSS (solid points) and FIRST
(hollow points) surveys. The two RLFs agree with each other (see
\S\ref{sec:robust}).  The dotted line is the fit to the NVSS RLF (see text for
more details).  The dashed line is the field RLF \citep{condon02}, shifted in
amplitude to account for different overdensities in the field and cluster
environments.  There is an overabundance of radio sources in clusters, compared
to the field environment.  We do not show the lower-side errorbars for bins
whose $1\sigma$ value is consistent with zero.
     }
\end{inlinefigure}

Our finding that radio galaxies are over-abundant in clusters is at odds with
the conclusion of LO96, who find no differences between the fraction of all
galaxies that is active in the radio in and outside clusters.  While our result
is based on the sources solely identified in the radio (irrespective of their
optical property), in the analysis of LO96, there is an explicit requirement
that the radio sources must have an optical counterpart. This is one of the
main reasons that lead to the different conclusions. A fair comparison to
LO96's analysis, therefore, may be done by using the sources that exist in both
NVSS and 2MASS catalogs (our ``XC'' sources; \S\ref{sec:ra_data}).
We will compare our results with theirs and comment on this issue in later
sections (see \S\ref{sec:ra_lo96}).

Before concluding the section, we check for any cluster mass dependencies of
the RLF. We separate the clusters into high ($\log M_{200}\ge 14.2$) and low
mass samples. The most luminous AGNs ($\log P\ge 25.5$) are only found in high
mass clusters. On the other hand, for less luminous bins, the amplitude of the
RLF for low mass clusters seems to be slightly larger. Summing up the RLF for
luminosity bins with $\log P\ge 23$ gives the number density of AGNs; we find
that the AGN number density in low mass clusters is $2.0\pm 0.7$ times higher
than that in high mass ones,  
that is, the data suggest that
low mass clusters has a higher amplitude of the RLF.

\subsection{$K$-band Luminosity Function and Radio Active Fraction}
\label{sec:ra_klf}

Here we study the $K$-band LF (KLF) of the radio sources, and compare it to
that from all 2MASS galaxies. The statistical background subtraction method is
similar in spirit to the one adopted in the previous section; see LMS04 for
more details.
This analysis will provide insights into the probability of a galaxy
hosting a radio-loud AGN.

We can only obtain KLF for NVSS sources that have counterparts in the 2MASS
catalog, namely the XC sources.  We show in Fig.~\ref{fig:ra_klf} the KLFs for
2MASS and XC sources, constructed from 364 clusters\footnote{The cluster sample
used in this section is the full {\it RASS} sample; we have checked that using
the ``isolated'' subsample (see \S\ref{sec:ra_2massprof}) the KLF is
essentially unchanged.} at $z\le 0.1118$.  The upper limit of redshift is set
so that all the XC sources have $\log P\ge 23.5$.  The solid symbols are the
2MASS-only sources, and the hollow points are the XC sources. To correct for
the background in the XC sources, at each cluster redshift we calculate the
flux density $S_l$ that corresponds to the limiting radio luminosity, and use
the $K$-band $\log N$--$\log S$ constructed from the XC sources whose radio
flux density $S\ge S_l$.

\begin{inlinefigure}
   \ifthenelse{\equal{\figtype}{EPS}}{
   \begin{center}
   \epsfxsize=8.cm
   \begin{minipage}{\epsfxsize}\epsffile{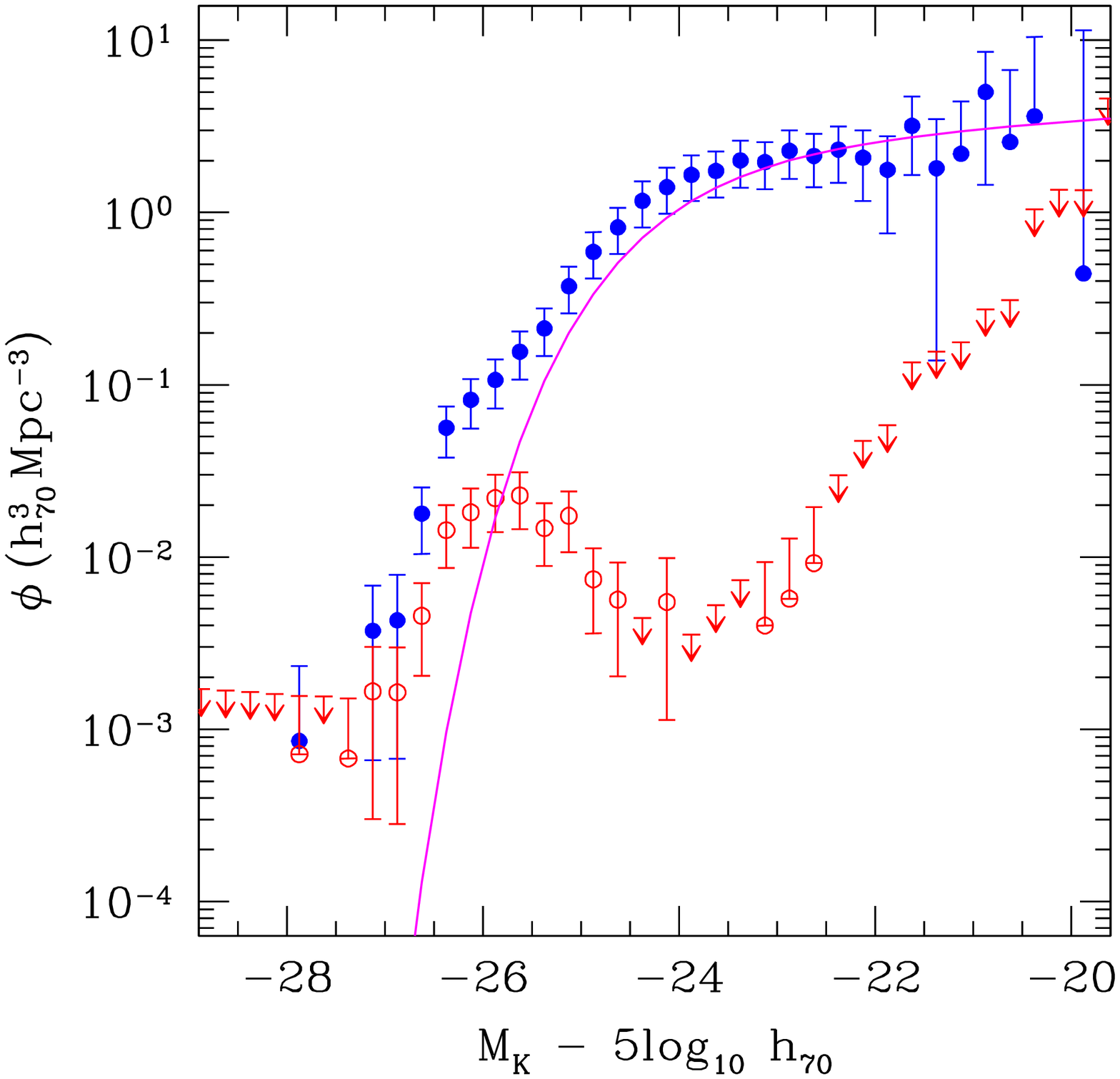}\end{minipage}
   \end{center}}
   {\myputfigure{f1.pdf}{0.1}{1.}{-20}{-5}}
   \figcaption{\label{fig:ra_klf}
    The composite $K$-band luminosity function within $r_{200}$ from
    364 clusters at $z\le 0.1118$ (solid points). The cross-correlated
    sources (with $\log P\ge 23.5$) are shown as hollow symbols.
    Also shown as the solid line is the field LF \citep{kochanek01}, shifted
    vertically to account for different overdensities in the field and
    cluster environments. We do not plot the errorbars for points
    whose 1-$\sigma$ uncertainty is consistent with zero.
    The inclusion of BCGs in the LF makes it deviate from the Schechter function
    at the bright end.
     }
\end{inlinefigure}

Also shown is the field LF
from \citet{kochanek01}, scaled up by $200/\Omega_M$ to account for different
overdensities in the field and cluster environments. Using the KLF
determined in LMS04, which is a good description of the data here, we find
the cluster-to-scaled field number density ratio is $1.02\pm0.22$ down to
$M_K=-21$.

By integrating the 2MASS-only and XC KLFs to a given $K$-band
magnitude $M_{lim}$, we can
calculate the fraction of galaxies more luminous than $M_{lim}$ that are hosts
of $P\ge P_{lim}$ radio sources.  As is clear from the Figure, the radio active
fraction (RAF) is a strong function of $K$-band luminosity or galactic stellar
mass.  We find that about $11.7\%$ of massive galaxies ($M_K\le -25$) have
radio luminosity $\log P\ge 23.5$, but only $2.8\%$ ($0.74\%$) of galaxies more
luminous than $-24$ ($-22$) mag host AGNs of the same radio luminosity range
($\log P\ge 23.5$).  Table~\ref{ra_5} contains the RAFs for $22\le \log P_{lim}
\le 25$ and $-22 \ge M_{K,lim}\ge -26$.  In the Table we include RAFs
calculated when the BCGs are included or excluded (shown in parenthesis).  Note
that at a given $P_{lim}$, a corresponding maximum redshift $z_{up}$ defines
the cluster subsample used to calculate the RAF, and therefore different
entries within a $M_{K,lim}$ column are not calculated based on the same
subsample.

\begin{table*}[htb] 
\begin{center} 
\caption{Radio Active Fraction (\%)}
\label{ra_5}
\vspace{1mm}
\begin{tabular}{clllll}
\hline \hline
$\log P_{lim}$ & $M_K\le -22$ & $M_{K}\le -23$ & $M_K\le -24$ &
$M_K\le -25$ & $M_K\le -26$ \\
\hline
22 & 8.1\phn\phn\phn(7.2)  & 12.5\phn\phn(11.0)    & 19.9\phn\phn(16.7)       & 45.7\phn(35.7)   &
  37.6\phn(18.3)\\
23 & 1.5\phn\phn\phn(1.1) & \phn2.1\phn\phn(\phn1.4)& \phn4.9\phn\phn(\phn3.3)& 19.1\phn(11.6)    &
  45.3\phn(16.4)\\
24 & 0.42\phn\phn(0.26)    & \phn0.66\phn(\phn0.39)   & \phn1.6\phn\phn(\phn0.95)   & \phn7.5\phn(\phn4.0) &
  19.1\phn(\phn5.9)\\
25 & 0.093\phn(0.048)      & \phn0.16\phn(\phn0.082)  & \phn0.40\phn(\phn0.20)   & \phn2.0\phn(\phn1.0) &
  \phn6.6\phn(\phn2.3)\\ 
\hline 
\end{tabular}
\end{center}
{\small
Note.-- All values calculated within $r_{200}$.
  Values in parentheses are calculated without BCGs.
  See Table~\ref{ra_bcgraf} for BCG-only RAFs.
}
\end{table*}

A couple of trends are apparent in Table~\ref{ra_5}.  
First, for galaxies of the same mass class (i.e.~same $M_{K,lim}$), fewer
galaxies are radio-active when $P_{lim}$ increases.  Second, the cumulative RAF
increases dramatically for a given $P_{lim}$ when considering more $K$-band
luminous (i.e.~more massive) galaxies.  For example, $4.9\%$ of the galaxies of
similar or higher luminosity than $M_*$ host AGNs ($\log P \ge 23$). For
galaxies with $M_K\le -26$, the fraction becomes 45.3\%.  If we exclude the
BCGs from the calculation, the luminous, non-BCG galaxies still have $\approx
5$ times higher probability (RAF $=16.4\%$, compared to the 3.3\% RAF for
$M_K\le -24$ non-BCG galaxies) to host radio-loud AGNs.

\begin{table*}[htb]
\begin{center}
\caption{BCG Radio Active Fraction (\%)}
\label{ra_bcgraf}
\vspace{1mm}
\begin{tabular}{cll}
\hline \hline
$\log P_{lim}$ & \multicolumn{1}{c}{all} & \multicolumn{1}{c}{$M_K\le -26$} \\
\hline
23 & 32.7 & 35.2 \\
24 & 19.9 & 21.0 \\
25 & \phn5.84 & \phn6.85 \\
26 & \phn0.585 & \phn0.913 \\
\hline
\end{tabular}
\end{center}
\end{table*}

We can calculate the RAF specifically for the BCGs, as the ratio between the
number of BCGs having radio luminosity $P\ge P_{lim}$ and all the BCGs that can
be identified statistically (342) in the {\it RASS} cluster sample. For $\log
P_{lim}=23$, 24, 25, and 26, we calculate the BCG RAFs and record them in Table
\ref{ra_bcgraf}.
Compared to non-BCG galaxies of $M_K\le M_*$, the BCGs are $\approx 10$
times more likely to host an AGN with $\log P \ge 23$, and $\approx 29$ times
more probable to harbor $\log P \ge 25$ AGNs!
To find out how much more likely it is for BCGs to host radio-loud AGNs
compared to galaxies of comparable luminosity/mass, we compare the RAFs for
BCGs more luminous than $M_K=-26$ (Table~\ref{ra_bcgraf}) to those listed in
parentheses under the $M_K\le -26$ column in Table~\ref{ra_5}. The BCGs are
about $2-3.6$ times more likely to be active in the radio than the similarly
massive, non-BCG cluster galaxies.  These findings are in agreement
with those presented in \citet{best06}. This boost in radio activity is most
likely due to the special location the BCGs have compared to other galaxies,
namely at the center of the cluster.

We note in passing that our results are not affected by the choice of the
spectral index ($\bar{\alpha}=-0.8$); using $\bar{\alpha}=-0.5$ lowers the RAF
very slightly.  For the most powerful sources (e.g.~$\log P>25$) for which the
effect is strongest, the change in RAF is only at few percent level.

\subsubsection{Dependencies on Cluster Mass}
\label{sec:ra_rafdep}

Next we 
examine the RAF as a function of cluster mass (when the limiting radio and
$K$-band luminosities are kept the same), both for the whole galaxy population
and for the BCGs. 
%
In Fig.~\ref{fig:raf} we show the comparison of RAFs for two limiting radio luminosities
($\log P_{lim}=23$, open symbols, and $\log P_{lim}=24$, solid symbols) as a function
of limiting $K$-band magnitude $M_{K,lim}$. 
The division of high and low mass is arbitrarily set to be $\log M_{200}=14.2$
The RAFs of high (low) mass clusters are shown as circles (triangles). 
For $M_{K,lim}>-24$, the RAFs of high and low mass clusters
are similar; for galaxies more luminous than $M_K=-24$, those in high mass 
clusters appear to be more
active in the radio than their counterparts in low mass clusters.
While the RAFs in high and low mass cluster differ at
$\lesssim 2\sigma$ level for the case of $\log P_{lim}=23$, the differences become
more significant for the case of $\log P_{lim}=24$
($>3\sigma$; e.g.~compare filled circles with filled triangles).

\begin{inlinefigure}
   \ifthenelse{\equal{\figtype}{EPS}}{
   \begin{center}
   \epsfxsize=8.cm
   \begin{minipage}{\epsfxsize}\epsffile{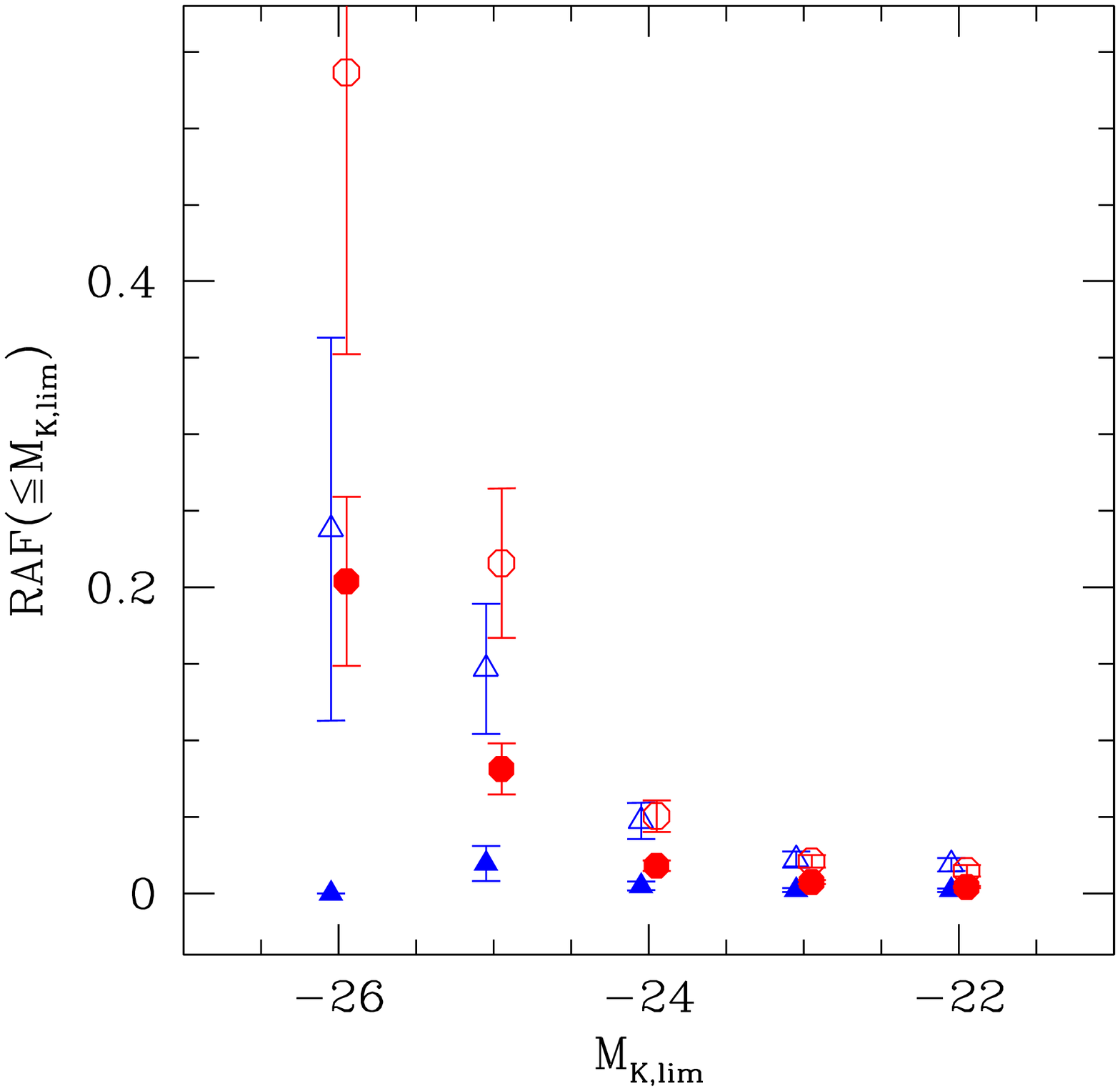}\end{minipage}
   \end{center}}
   {\myputfigure{f1.pdf}{0.1}{1.}{-20}{-5}}
   \figcaption{\label{fig:raf}
  Comparison of radio active fractions (RAFs) in high and low mass clusters,
where the division in mass is set at $\log M_{200}=14.2$. The RAF is defined as
the fraction of galaxies more luminous than $M_{K,lim}$ which also host radio
sources more powerful than $P_{lim}$. The RAF as a function of $M_{K,lim}$ of
high (low) mass clusters is shown  as circles (triangles). The open and solid
symbols denote the cases of $\log P_{lim}=23$ and 24, respectively. For both
cases the high mass clusters have higher RAFs when $M_{K,lim}\le-24$.
     }
\end{inlinefigure}

That RAF weakly correlates with cluster mass seems to be counter-intuitive to
the finding that the RLF of the low mass clusters has a slightly higher
amplitude than that of the high mass clusters (\S\ref{sec:ra_rlf}).  
This is because (1) for massive clusters the abundances of both luminous
galaxies and radio-loud AGNs at the bright-end of the KLF are higher than those
in low mass clusters, 
and (2) because of the shape and amplitude of $K$-band and radio LFs, there are
many more galaxies ($M_K \gtrsim M_{K*}$) than AGNs in low mass clusters.
Interestingly, these reflect the main differences between the KLFs in high and
low mass clusters (the lack of very luminous galaxies and the higher amplitude
of the KLF near $M_*$ in low mass clusters; see LMS04).

The trend for a higher RAF in more massive clusters is stronger for the BCGs.
Out of the 342 clusters for which we can identify a BCG statistically, 288 of
them are more massive than $\log M_{200}=14.2$, and 54 are lower mass clusters.
Among the BCGs in high mass clusters, $36\pm 3$\% have radio luminosity $\log
P\ge 23$, while only $13\pm 5$\% of the BCGs in low mass clusters are as
powerful in the radio. The average radio luminosity for the BCGs in high mass
clusters is much higher than that for the low mass cluster BCGs (geometric
means are $10^{24.30\pm0.07} {\rm W} {\rm Hz}^{-1}$ and $10^{22.88\pm 0.18}
{\rm W} {\rm Hz}^{-1}$, respectively). This enhanced RAF for BCGs in more
massive clusters may provide a way to assess the relative importance of gas
supply, local ICM pressure, and interactions/mergers between galaxies in
clusters of different masses.

Finally, we compare our results with the findings of \citet{best04}, who
suggests that the radio-loud AGNs are preferentially located in groups and low
mass clusters. By cross-matching NVSS with 2dFGRS, \citeauthor{best04} examines
the RAFs of galaxies more luminous than $M_B=-19$ (roughly corresponding to
$M_K=-23.2$) in groups of different richness ($4-8$, $9-20$, $21-50$, and
$>50$).  We translate these richness classes into mass bins ($\log
M_{200}=13.78-13.90$, $13.90-14.36$, $14.36-14.83$, $>14.83$), using the
correlation between the number of galaxies and cluster mass presented in LMS04
(see Eq.~\ref{eq:appen}).  Such a binning in mass is much finer than our simple
low v.s.~high mass separation, and reduces the size of cluster subsamples.
Nevertheless, for $\log P_{lim}=23$, we find the RAFs in these subsamples are
2.1, 3.1, 2.6, and 1.5\%, respectively. Although the trend with respect to
cluster mass is in general consistent with that presented in Fig.~7 of
\citet{best04}, we do not find a strong suppression of RAF in the most massive
clusters. Based on various mass binning and limiting radio and $K$-band
luminosities, we conclude that the RAF of the most massive clusters (e.g.~$\log
M_{200}>14.5$) is slightly smaller than that in intermediate-to-massive
clusters (e.g.~$\log M_{200}=14-14.5$), but the two are consistent within the
uncertainties. In most of the cases, the RAF of the most massive clusters is
greater than that of less massive clusters (e.g.~$\log M_{200}<14$).

A recent study of the optically-selected AGN population in clusters finds that
the AGN fraction is anti-correlated with the cluster velocity dispersion
\citep{popesso06}. As the velocity dispersion is positively correlated with
cluster mass, this finding appears to be an opposite trend to our results.
However, as the radio and optical (emission-line) AGN activities may be
independent of each other (see discussion in \S\ref{sec:disc_faint}), the
distinct trends with cluster mass of the radio and optical AGNs probably point
out to the different physical mechanisms that are responsible (e.g.~mergers or
interactions between member galaxies v.s.~cooling instability).

\subsubsection{Comparison with the Radio Active Fraction in the Field Environment}
\label{sec:ra_cffield}

Combining the fraction $\tilde{f}$ of galaxies that are radio-active as a function
of their stellar mass $\mathcal{M}$ with the stellar mass function $\tilde{\phi}$, one can estimate the
RAF in the field environment; that is,
\[
{\rm RAF} = { \int \tilde{f}(\mathcal{M}) \tilde{\phi}(\mathcal{M}) d\log \mathcal{M} \over  \int \tilde{\phi}(\mathcal{M})  d\log \mathcal{M}}.
\]
Based on Fig.~2 presented in \citet{best05}, we estimate that, for hosts of $\log P \ge 23$ AGNs,
$
\tilde{f}(\mathcal{M}) = 10^{-28.79} (\mathcal{M}/\mathcal{M}_\odot)^{2.43}.
$
Using the stellar mass function determined by
\citet{bell03}, and the stellar mass-to-light ratio of 0.7 solar unit ($K$-band; see \citealt{lin03b}), 
we find that the RAF in the field is $1.3\pm 0.5\%$ for $M_K\le -24$ galaxies.
For $\log P\ge 25$, we approximate the result of \citet{best05} by 
$\tilde{f}(\mathcal{M}) = 10^{-43.96} (\mathcal{M}/\mathcal{M}_\odot)^{3.63}$,
and find the fraction is $0.21\pm 0.08\%$ for $M_K\le -25$ galaxies.
These values are about $10-20\%$ of the RAFs found in clusters ($4.9\pm 0.5\%$ and 
$2.0\pm 0.3\%$, respectively; see Table~\ref{ra_5}). Together with the finding that the cluster
RLF has higher amplitude than the (scaled) RLF in the field (\S\ref{sec:ra_rlf}), these results 
suggest that cluster environment promotes AGN activity.
%

\subsection{AGN Duty Cycle}
\label{sec:duty}

The simplest application for the radio active fractions presented in the
previous section is to estimate the duty cycle of radio-loud AGNs. If every
galaxy has a central supermassive black hole (SMBH), the RAF is equivalent to
the probability that the central SMBH is accreting for a galaxy of given
luminosity range.  In addition, if the AGN activity is intermittent and
the RAF $f$ remains about constant over the lifetime of the galaxies, 
the lifetime of a radio
source is simply $t_r \sim f t_g$, where $t_g$ is the typical age of a galaxy 
\citep[e.g.][]{schmidt66}. For a $M_*$ galaxy
formed at $z=3$, $t_g=11.4$ Gyr; then $f\approx 0.05$ gives $t_r \approx
5.7\times 10^8$ yr. Based on bias and mass information from the spatial 
clustering of quasars, limits can be
derived for the lifetime $t_Q$ of quasars \citep[e.g.][]{martini01,haiman01c};
the recent determination from the 2dF QSO survey finds that $4\times 10^6< t_Q
< 6\times 10^8$ yr.~at $z\sim 2$ \citep{croom05}, consistent with our estimate.
At lower redshifts ($0.05\le z\le 0.95$), the 
fraction of galaxies which exhibit AGN signature in their optical spectra may be
as high as 40\%, suggesting a duty cycle of $\approx 2\times 10^8$ yr.~\citep{miller03},
which is within the uncertainties from our value.

The higher RAF of BCGs strongly suggests that their duty cycle is longer than
the other galaxies, which may be closely related to their location in clusters
and the ample supply of gas due to radiative cooling of the ICM. We explore the
implication of our results on the ICM energy injection from the radio-loud AGNs
in \S\ref{sec:ra_heating}.

\subsection{Comparison with Results of \citeauthor{ledlow96}}
\label{sec:ra_lo96}

In \S\ref{sec:ra_rlf} we compare the number densities obtained from the cluster
and field RLFs, and find that the clusters contain more radio-loud AGNs than
the field does, even after the difference in the overdensities has been
accounted for. On the other hand, LO96 cast their results in terms of
fractional luminosity function (meaning that the RLF is further normalized to
the total number of galaxies expected in each radio luminosity bin), which can
be compared to the RAFs presented in \S\ref{sec:ra_klf}.

LO96 consider both the univariate and bivariate LFs (in our notation,
$\psi(P,M_R\le -20.5)$ and $\psi(P,M_1\le M_R \le M_2)$, respectively). Summing
up the univariate LF (presented in their Fig.~6), they find that RAF=14\% (for
$\log P_{lim}=22$) for
early type galaxies with $M_R\le -20.5$, which can be compared to our value of
$12.5\pm 3.0\%$ (for $M_{K,lim}=-23.1$, where $R-K=2.6$ for early type galaxies
is used), although we do not distinguish between early and late type galaxies.

Next we compare our data with their bivariate LF for galaxies with $R$-band
magnitudes within $-23.5\le M_R\le -22.7$ (roughly corresponding to $-26.1\le
M_K\le -25.3$). At $\log P=23.8$, $24.2$, $24.6$, and $25$, the RAFs (in
percentage) based on their Table 2 and our data (in parantheses) are: $2.2\pm
0.9$ ($4.0\pm0.9$), $4.0\pm1.2$ ($4.1\pm0.8$), $5.8\pm1.5$ ($1.7\pm0.5$), and
$1.8\pm0.8$ ($1.5\pm0.5$).

There are several differences that make the comparison between the two analyses
challenging, including: (1) the distinction of galaxy morphological types, (2)
the portion of the cluster surveyed, (3) the deprojection of the RAF from a
cylindrical to a spherical volume. Given the relative abundance of the early
and late type galaxies in clusters, and the fact that radio-loud AGNs are
mainly hosted by early types, exclusion of the late type galaxies should not be
a problem. However, the second and third factors do affect the calculation of
the RAF.
The RAF will be radius dependent because of the rather different radial
distribution of radio sources and galaxies in clusters (see
Fig.~\ref{fig:ra_rad}). Let $f_\delta$ denote the RAF measured at a radius
corresponding to an overdensity $\delta$, one expects that $f_{\delta_1} >
f_{\delta_2}$ for $\delta_1 > \delta_2$.
LO96 calculate the RAF for galaxies within 0.3 Abell radius (about 0.6 Mpc).
This fixed metric radius corresponds to different overdensities for clusters of
different masses, and the RAF from LO96 should be regarded as an averaged value;
without specific knowledge of the mass distribution of their cluster sample, it
is not possible for a direct comparison. However, unless the majority of their
clusters are low mass systems, we would expect their RAF to be higher than
ours, given that 0.6 Mpc is within $r_{500}$ for a moderately massive cluster
(\eg $M_{200}\sim 5\times 10^{14} M_\odot$).

Furthermore, we expect their RAFs to be increased by about 30\%.  This is
because they do not explicitly take into account the projection of galaxies
along the line of sight. Knowing the radial profiles for galaxies and radio
sources enables one to calculate the correction factor to transform the RAF
measured within a cylindrical volume into one that is within a sphere of the
same radius.

LO96 compare their RLF with that obtained by \citet{sadler89} and find
no significant difference between the RLFs measured in the field and in
clusters. However, because of the reasons mentioned above, it may not be a fair
comparison; if their RAFs were to be increased (i.e.~due to deprojection),
then the cluster RAF would be higher than the field value. 
Furthermore, as the two measurements are made at different
frequencies (1.4 and 5 GHz), the effect of spread in the spectral index should
be taken into account (\S\ref{sec:ra_index}), which makes it harder to precisely
compare the results.

To summarize, a more direct comparison with the results of LO96 is made
possible with our XC sources. Although agreement is found for both univariate
and bivariate LFs between the two groups, we point out differences in the analyses
(the fraction of the cluster region surveyed and the deprojection of the RAF) that
may make their RAFs systematically higher than ours. A possible explanation
of the agreement is that the cluster sample used by LO96 is on average less massive 
than ours, so that the effects from the two factors mentioned above (portion of 
the clusters surveyed and the deprojection) roughly cancel out.

\section{Active Radio Sources in Low-Mass Cluster Galaxies}
\label{sec:lowmass}

The higher amplitude of the RLF in the cluster versus the field is an exciting
result that deserves further consideration.  We first discuss the possibility
that the lower amplitude in the field RLF is due to a different selection
technique, then examine the robustness of our determination of the cluster RLF.

To determine a field RLF, a usual procedure is to first define a sample of
galaxies limited by, say, an apparent magnitude $m_{lim}$, then conduct the
radio survey for these galaxies, and follow up with optical spectroscopy for
the sources that are cross-matched \citep[e.g.][]{sadler89,condon02,sadler02}.
We note that in doing so, the radio galaxy sample is only complete to the
absolute (optical) magnitude that the highest-redshift object in the sample corresponds
to. Such a survey will therefore miss the radio sources in low luminosity
galaxies, and the LF thus constructed will underestimate the true amplitude and
shape of the RLF. In some sense, the field RLF is built from objects like
our XC sources.  The comparison, therefore, would be more meaningful
between the field RLF that is based on sources selected with a fixed OIR absolute
magnitude limit and the RLF from our XC sample. 

Of course, if powerful radio sources (\eg $\log P\ge 24$) only reside in
luminous galaxies, then our concern for the field RLF will be a non-issue. We
will come back to this point below.

We show in Fig.~\ref{fig:ra_lf2} as the hollow symbols the cluster RLF based on
the XC sources that have $K$-band magnitude $M_K\le -24$ ($\sim M_{K*}$; \eg
LMS04) in 202 clusters. The highest redshift of clusters that we can include so that the
galaxies are complete to $M_K=-24$ is $z=0.0693$. The background contribution is
accounted for as follows: given a limiting $K$-band absolute magnitude, for
each cluster we calculate the corresponding apparent magnitude $K_{f}$, and use
the (radio) $\log N$--$\log S$ from the XC sources with $K\le K_f$.

The RLF for all NVSS sources from all our clusters is shown as the
solid points in Fig.~\ref{fig:ra_lf2}.  The difference in amplitudes between
the NVSS-only and the XC RLFs encodes the fraction of all radio sources that
are hosted by luminous, presumably massive, galaxies, as the $K$-band light
traces stellar mass well.
%
For example, down to $P=10^{23}$ W$\,$Hz$^{-1}$, $55\pm 14\%$ of all radio
sources are within galaxies more luminous than $M_K=-24$.  Repeating the
exercise, we find that $43\pm 11\%$ ($52\pm 15\%$) are in galaxies with $M_K\le
-25$ ($-23$).
From these comparisons we find that the fraction of cluster radio
sources associated with luminous galaxies is $\sim 1/2$.

\begin{inlinefigure}
   \ifthenelse{\equal{\figtype}{EPS}}{
   \begin{center}
   \epsfxsize=8.cm
   \begin{minipage}{\epsfxsize}\epsffile{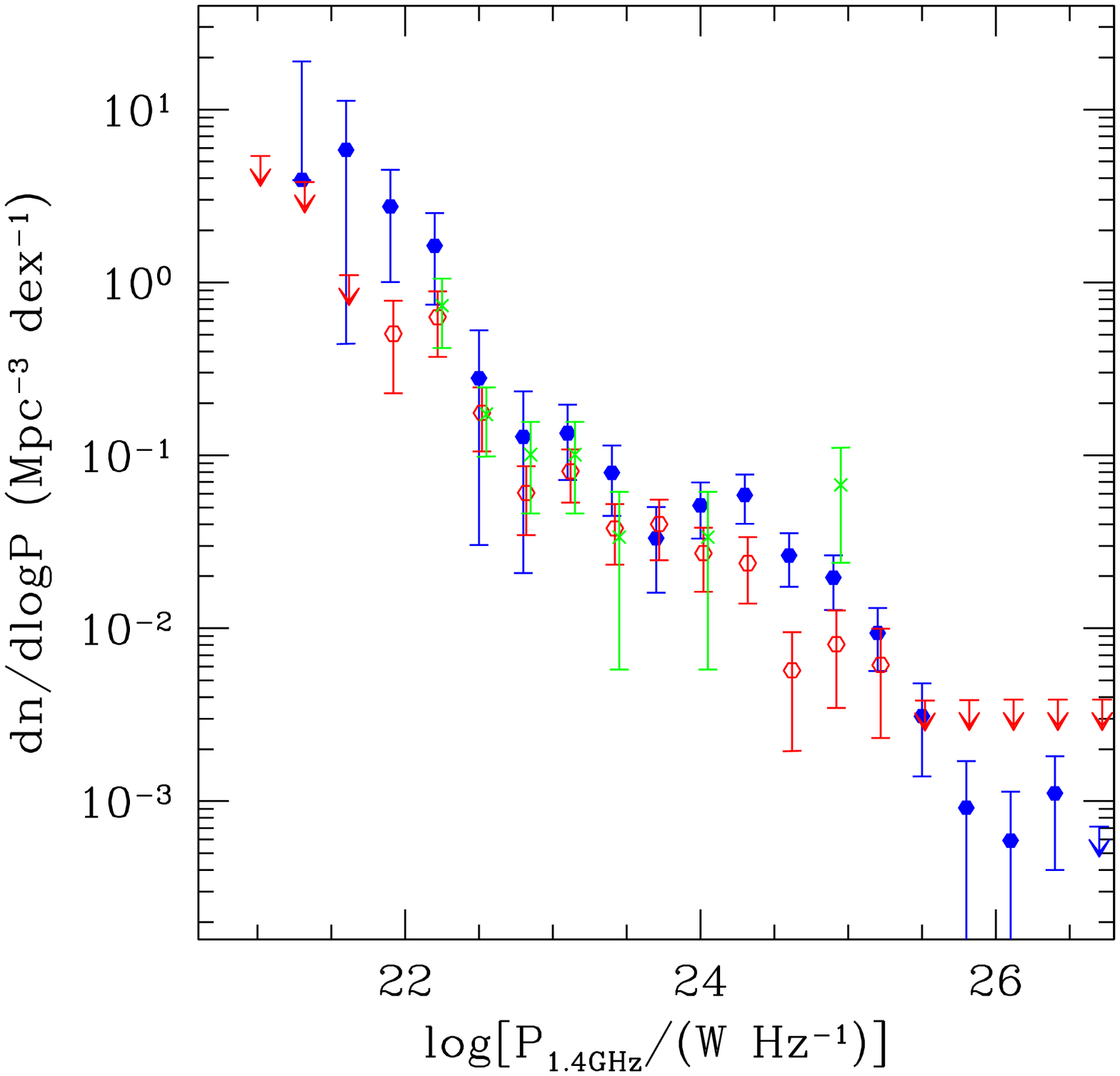}\end{minipage}
   \end{center}}
   {\myputfigure{f1.pdf}{0.1}{1.}{-20}{-5}}
   \figcaption{\label{fig:ra_lf2}
    Comparison of RLFs constructed from the NVSS-only (solid symbols)
    and XC (hollow symbols) sources. The crosses show the RLF from a smaller
    sample of those clusters that have redshift information for their member radio
    galaxies (based on data from \citealt{miller01b}).
     }
\end{inlinefigure}

Earlier surveys conclude that powerful radio sources are almost exclusively
found in luminous galaxies (\eg $M_R\le -21$; \citealt{sadler89,owen91}; LO96).
In particular, LO96 notice that all galaxies fainter than $M_R= -21$ that are
identified with radio sources stronger than the flux limit of their survey turn
out to be background objects. First, we note that $M_R=-21$ roughly corresponds
to $M_K=-23.6$ (for early-type galaxies), only $\sim 0.5$ mag fainter than the
characteristic magnitude of the luminosity function. The $K$-band data of our
XC sources (more powerful than \eg $\log P=24$) reach deeper than that limit
(\eg $M_K\ge -23$, although the sample size is smaller; see Fig
\ref{fig:ra_klf} above). It is not clear to us either what (optical) magnitude limit the
LO96 survey reaches, or the number of faint galaxies with redshift measurements
that are determined as background.

As another check of our RLFs, we use the data from \citet{miller01b}, who have
conducted an extensive redshift survey over 18 nearby ($z<0.033$) clusters and
provided the redshifts for a sample of 326 radio galaxies.  Eight of these
clusters are in our sample (so that we can estimate their masses), and we
cross-match their source list with the 2MASS catalog. 
The RLF from this dataset is
shown in Fig.~\ref{fig:ra_lf2} as crosses (for 20 galaxies with $M_K\ge -24$).
It shows good agreement with our XC RLF within the (rather large)
uncertainties.

\subsection{Robustness of Our Treatment of the Radio Luminosity Function}
\label{sec:robust}

The analysis presented above indicates that about 50\% the cluster radio
galaxies in our sample are associated with galaxies less luminous than
$M_K=-23$. Because this is a new result, we discuss some possibilities that may
explain the differences between the RLFs of the NVSS-only and XC sources. These
include: (1) differential breakdown of the statistical background subtraction
scheme, presumably due to the differences in the dominance of the background
(the total-to-background number ratio is $>7.6$ and 1.3 for the XC and NVSS-only
cases, respectively), (2) presence of radio sources that are resolved into
several components, (3) missing matches due to astrometric problems and (4)
gravitational lensing of background sources.

For the first possibility, we have examined (a) overall shape and amplitude of
the $\log N$--$\log S$; (b) variation in the background as a function of
declination $\delta$ or Galactic latitude $b$ (see \S\ref{sec:ra_data}); and
(c) the effects of sidelobes. 
We find that our $\log N$--$\log S$ has to be increased in amplitude by
$\gtrsim 10\%$ to account for the differences in RLFs, which is unlikely, for
it is in good agreement with that derived from the entire NVSS survey region
($\delta\ge-40$ deg). In addition, the mean surface density thus derived is in
excellent agreement with the value obtained by \citet{blake02} and \citet{blake04}, which
is derived after carefully masking out from the NVSS catalog the Galactic
plane, nearby bright extended sources, and sidelobes of very strong sources.
Secondly, we do not find any clear changes in amplitude for both NVSS-only and
XC RLFs as a function of either $b$ or $\delta$. Thirdly, to minimize the
effects due to sidelobes that are not cleaned from the NVSS catalog, we
construct RLFs from clusters that do not have sources with flux density larger
than 1 Jy. The RLFs are statistically indistinguishable.

Finally, the RLFs are not sensitive to our choice of the completeness limit.
For example, using 15 or 20 mJy as the limit does not change either the shape
or amplitude of the RLF for bins at $\log P \ge 23.5$.

Although there has been an estimate of the
fraction of NVSS sources that are components of the same source \citep[$\sim
7\%$,][]{blake04}, their effect on the RLF is not trivial to evaluate.  
%
To examine if the resolution of the NVSS affects the resulting RLF, we
construct RLFs both for radio-source only and for radio-2MASS cross-matched
sources,
using the data from the FIRST (Faint Images of the Radio Sky
at Twenty-centimeters, \citealt{becker95}) survey, which has a resolution of
$5\arcsec$. Because of the sky coverage of the survey, only 242 clusters are
available. We use only sources with $S\ge 10$ mJy. The FIRST-only RLF
is shown in Fig.~\ref{fig:ra_lf} as hollow squares, and overall it is in agreement 
with the NVSS-only RLF.
We also find that the XC RLF based on the FIRST survey is consistent with
that determined using the NVSS-2MASS XC sources. Integrating down to $\log P =
23$, the FIRST-2MASS XC sources has the space density that is $0.98\pm 0.23$ of that
from the NVSS XC sources.
Therefore, there is no evidence that the resolution of the
NVSS survey is causing problems.

Astrometry should not be a problem, as 2MASS extended sources have accuracy
good to $<0.5\arcsec$, and NVSS sources with $S\ge 10$ mJy are accurate to
$1-2\arcsec$, which are much smaller than our maximal separation $d_{max}=20\arcsec$ for cross-matching the two catalogs.

In the case of lensing, \citet{cooray99} estimates the number of expected
lensing events due to clusters more massive than $\sim 1.3\times 10^{15}
M_\odot$ over the sky, and find it to be $\sim 1500$ for sources with $S\ge
10\,\mu$Jy at 1.4 GHz. The lensing probability $p$ depends on both the number
density $n$ and mass $M$ of clusters ($p\sim n M^{4/3}$; \citealt{cooray99}).  Considering the
shape of the cluster mass function, and our adopted flux limit (10 mJy),
lensing events due to our clusters may not be larger than the prediction of
\citet{cooray99}.  \citet{phillips01} conduct a survey over 4150 deg$^2$ as an
attempt to detect lensed quasars by clusters of masses $>10^{13}-10^{14}
M_\odot$. None of their 38 candidates are found to be actual lensing systems.
Therefore, we conclude that the frequency of lensing events is quite small
(\citealt{cooray98,andernach98}; see also \citealt{boyce06}).

None of these alternatives seem to be able to offer a satisfactory explanation. 
Below we present some evidence that supports a population of AGN in low mass
galaxies.

\begin{inlinefigure}
   \ifthenelse{\equal{\figtype}{EPS}}{
   \begin{center}
   \epsfxsize=8.cm
   \begin{minipage}{\epsfxsize}\epsffile{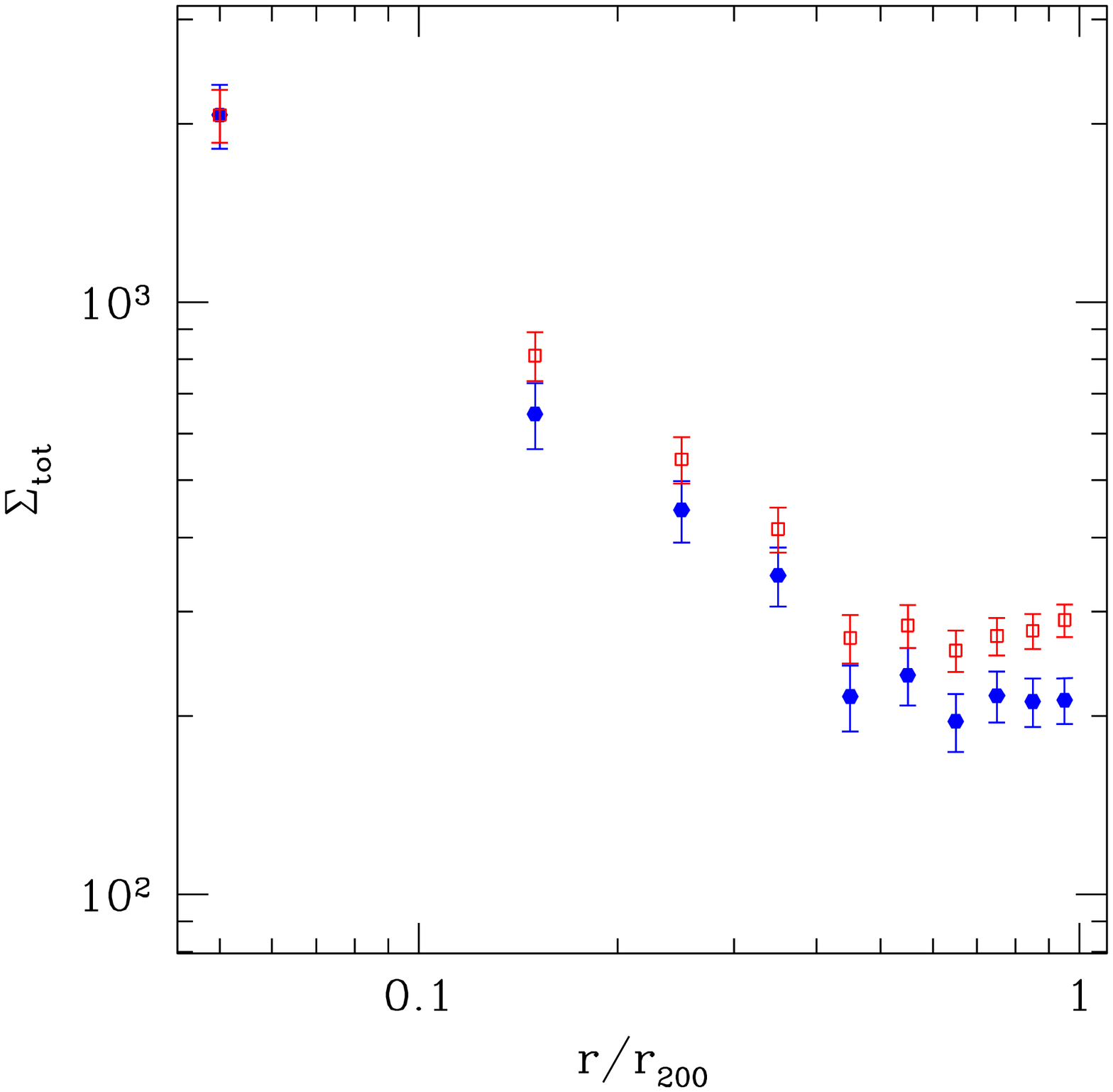}\end{minipage}
   \end{center}}
   {\myputfigure{f1.pdf}{0.1}{1.}{-20}{-5}}
   \figcaption{\label{fig:tot_nvss_xc}
       Total surface density profile (cluster plus background) for the NVSS--XC
    sources (the NVSS sources {\it without} a 2MASS counterpart within
    $d_{max}$ chosen for each cluster, see \S\ref{sec:ra_data}; hollow squares). The profile
    for all the NVSS sources is shown as solid points. Both are from sources
    with $\log P \ge 23$. The two profiles are arbitrarily normalized to match
    at the innermost bin. The fact that NVSS--XC sources are {\it clustered},
    together with the small expected effects from gravitational lensing
    amplification, indicate that they are very unlikely to be background
    sources.
     }
\end{inlinefigure}

\subsection{Evidence Supporting the Existence of Such Population}

We can estimate the RAF expected for the low mass cluster galaxies so that the
difference between the NVSS-only and XC RLFs can be accounted for. The basic
idea is to calculate, for galaxies whose radio luminosity is above a given
limit, the ratio between the numbers of galaxies more and less luminous than a
certain $K$-band magnitude (which we choose to be $M_K=-23$). For the KLF with the faint-end power-law slope
$\alpha=-0.9$, for galaxies within the magnitude range from $-23$ to $M_{lim}$,
where $M_{lim}=-19$, $-18$, and $-17$, the RAFs are 1.84\%, 1.51\%, and 1.30\%,
respectively. Therefore the active fraction of these low-mass cluster galaxies
does not need to be high given reasonable forms of the KLF.

Further support of our conclusion comes from the spatial distribution, as well
as the RLF of these sources.  We first examine the surface density profile for
the NVSS sources that do not have a 2MASS counterpart (hereafter referred to as
the NVSS--XC sources), and find that these sources are clustered 
(Fig.~\ref{fig:tot_nvss_xc}).  
If they were all background objects we would not expect them to be clustered
around the positions of known galaxy clusters.

Secondly, we construct the RLF for the NVSS--XC sources. For this purpose we
use the $\log N$--$\log S$ based on NVSS--XC sources from a region of 60 deg
radius centered at the NGP. We follow the same procedure as outlined in
\S\ref{sec:ra_rlf}, and the resulting RLF from all clusters is shown in
Fig.~\ref{fig:lf_nvss_xc}. Integrating the two RLFs down to $\log P = 23$ (24),
we find that the number density of NVSS--XC sources is $67\pm 18 \%$ ($55\pm
11\%$) of the NVSS-only sources, which is similar to the estimate
based on the NVSS-only v.s.~XC RLF comparison (\S\ref{sec:lowmass}).

Is there evidence from OIR observations for the existence of low mass cluster
galaxies that host radio-loud AGNs?
We have obtained $K$-band imaging data sufficient to probe to $M_K \approx -19$
(about five magnitudes fainter than $M_{K*}$) for a sample of 10 clusters at
$z<0.2$ (Lin et al.~2006, in preparation). Due to the field-of-view of the
observations ($20'\times 20'$, conducted with {\it FLAMINGOS} imager at KPNO
2.1m), we focus on the region enclosed by the radius $r_{2000}$ (within which
the mean overdensity is $2000 \rho_c$) for this sample of clusters.  There are
21 sources in the NVSS catalog that are within $r_{2000}$ in these clusters and
are stronger than 10 mJy; we expect 4.44 background sources based on the mean
surface density of the NVSS.

\begin{figure*}
\vspace{-5mm}
\centering
\includegraphics[width=11.2cm]{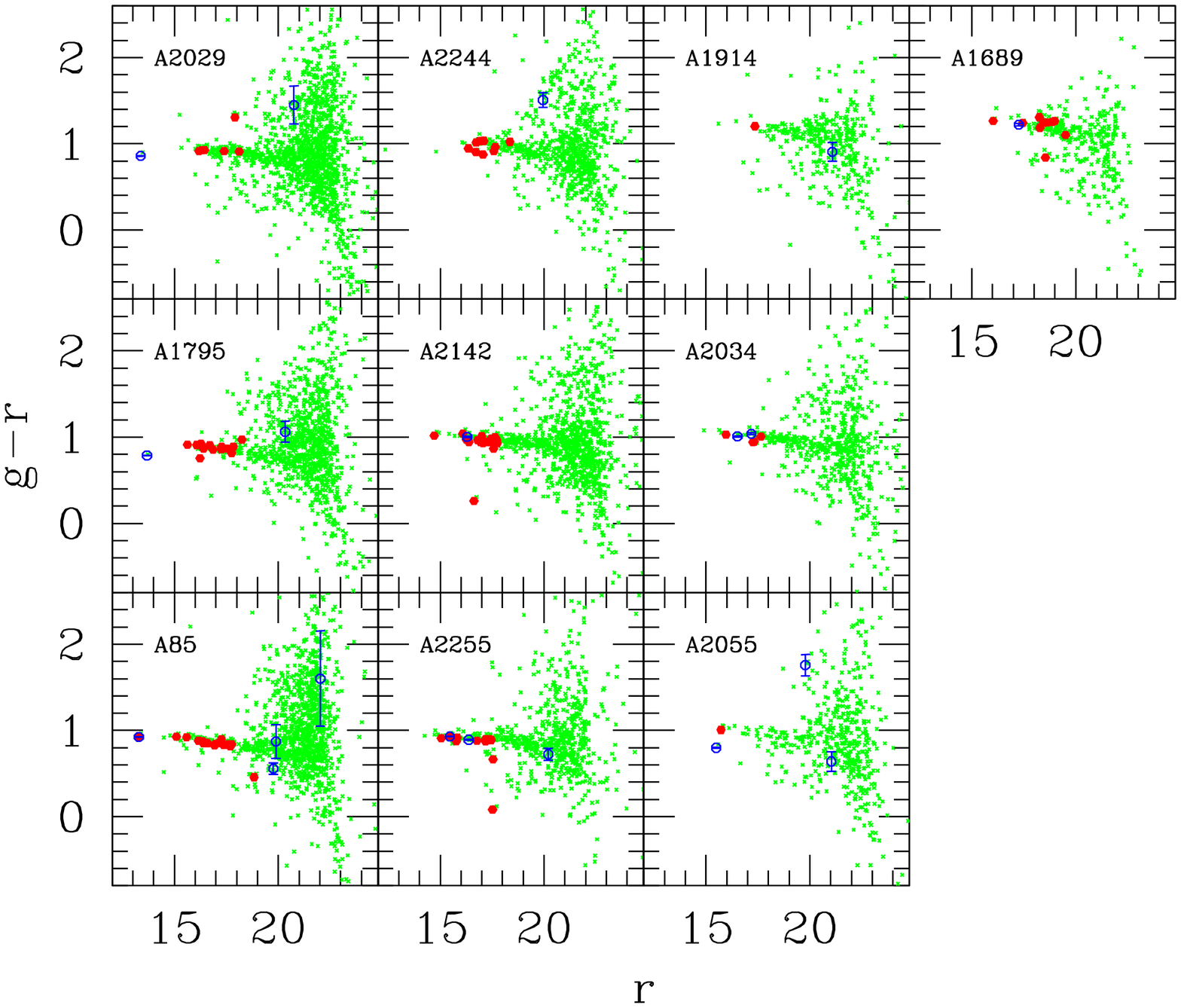}
\vspace{-10mm}
\figcaption{
   Color-Magnitude diagrams for a sample of clusters covered by SDSS, for which
we have obtained deep $K$-band imaging data. All SDSS extended sources
projected within $r_{2000}$ from cluster centers are shown as small points. For
a limited number of galaxies whose redshift is available (from SDSS and NED)
and is within 0.001 from the cluster redshift, they are displayed as large
solid points.  The NVSS sources with $K$-band and SDSS counterparts are shown
as hollow symbols with errorbars.  Among these sources, those that are faint
($r\gtrsim 20$) and lie along the cluster red-sequence are candidates of the
NVSS--XC sources whose optical counterpart is too faint to be detected by 
2MASS. 
\label{fig:cmd}}
\end{figure*}

Because all 10 clusters lie within the region covered by SDSS data release 4
(DR4), we seek optical and $K$-band counterparts of the NVSS sources. We find
matches in both the $K$-band and optical catalogs for all (21) but one source
(down to our completeness limit $K \sim 19.3$). In Fig.~\ref{fig:cmd} we show
the $g-r$ v.s.~$r$ color-magnitude diagrams for both the radio sources
(hollow symbols) and all galaxies (small points) projected within $r_{2000}$.
In all cases the red sequence of the cluster galaxies is apparent. As a
guidance for cluster membership we gather redshift information from both SDSS
and the NASA/IPAC Extragalactic Database (NED) and show the galaxies whose
redshift is within 0.001 from the cluster redshift as the large solid points.
It is clear that most of the optically bright ($r\lesssim 19$) radio sources are cluster
members.  Using the red sequence as a rough guide for the cluster membership
assignment, and taking into account the uncertainties in the photometry for the
faint sources, it is likely that $\approx 1/2$ of the optically faint
($r\gtrsim 20$) radio sources are members.  Therefore, we conclude that the
background estimates from both the NVSS $\log N$--$\log S$ and the optical
color of the XC sources give roughly consistent results (the former method
gives $16-17$ members, and the latter yields an estimate of $14-15$ members).

\begin{inlinefigure}
   \ifthenelse{\equal{\figtype}{EPS}}{
   \begin{center}
   \epsfxsize=8.cm
   \begin{minipage}{\epsfxsize}\epsffile{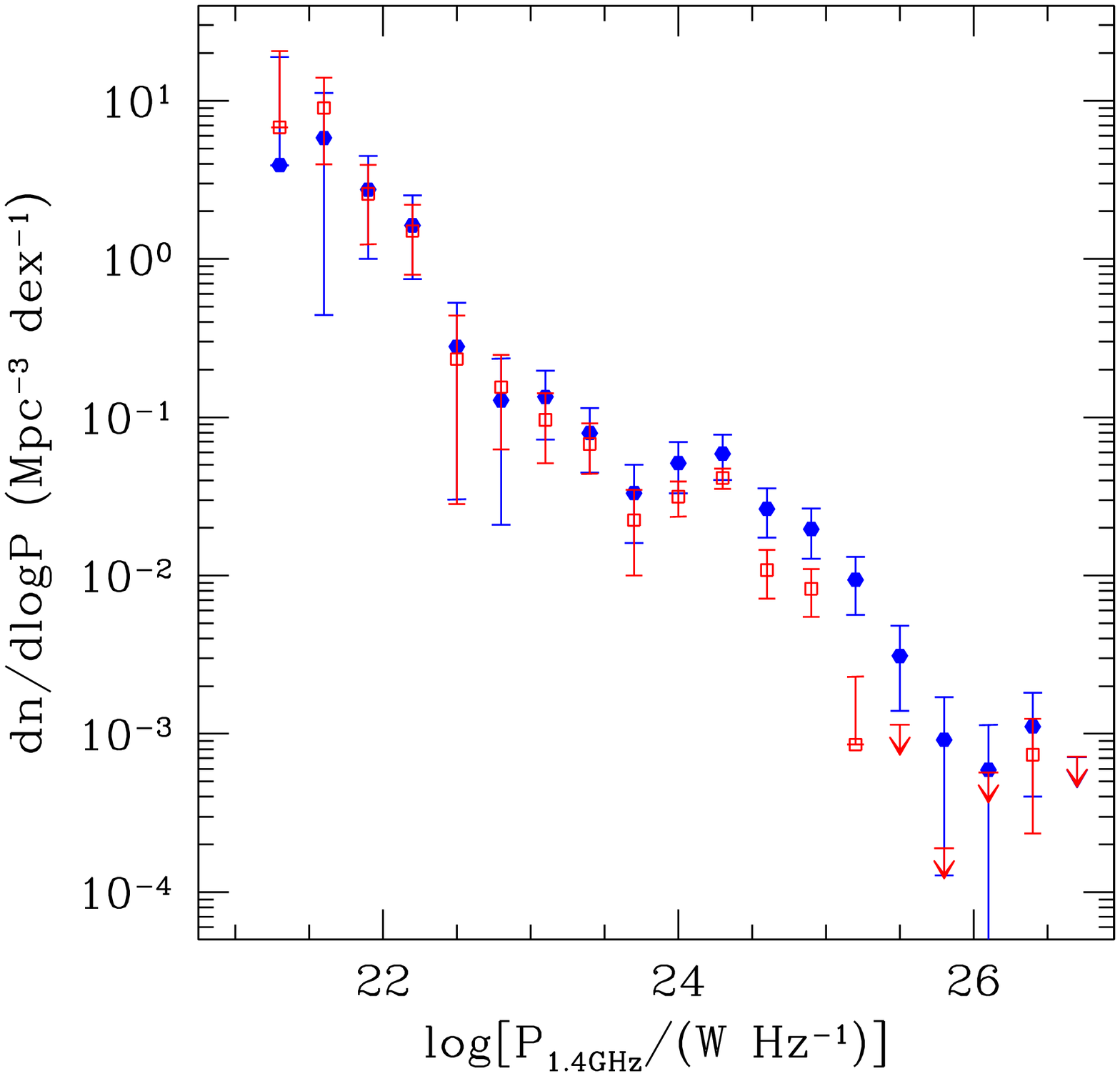}\end{minipage}
   \end{center}}
   {\myputfigure{f1.pdf}{0.1}{1.}{-20}{-5}}
   \figcaption{\label{fig:lf_nvss_xc}
    RLF for NVSS--XC sources (hollow squares). As a comparison, the RLF for
    all NVSS sources is shown as the solid points. The RLF of NVSS--XC sources
    appears to be similar to that of NVSS--only sources.
     }
\end{inlinefigure}

We examine the radio luminosity for candidate cluster member galaxies of
$r \sim 20-21$ in four clusters (A85, A2255, A2055, A1914), and find that 
$P = 4\times 10^{23} - 2\times 10^{24}$ W\,Hz$^{-1}$. These luminosities are
slightly smaller than that where the discrepancy between the NVSS-only and
XC RLFs is largest (\eg $\log P = 24-25$, see Fig.~\ref{fig:ra_lf2}). We expect,
however, if a similar analysis is carried out for a large, nearby cluster sample
(\eg 200 clusters), we will find low-mass radio galaxies whose radio luminosity
is greater than those found here.

\subsection{Seeking Spectroscopic Confirmation}

Conclusive support for the existence of the active low-mass cluster galaxies
would come from spectroscopy. From Fig.~\ref{fig:cmd} it is clear that the
majority of such sources, typically having $r<20$, would lie
below the limits of current large galaxy redshift surveys (e.g.~SDSS and
2dFGRS).  Nevertheless, we combine the three large public datasets (NVSS,
2MASS, SDSS DR3) to seek further evidence.
In Fig.~\ref{fig:ra_2df} we show the distribution of radio sources in the
$M_K$-$P$ space, based on a sample with measured redshifts from the SDSS
survey. More specifically, we cross-correlate our XC sample (not limited to
those within cluster fields) with the main galaxy sample in the SDSS DR3. There
are 1318 galaxies matched that are above the completeness limits of both 2MASS
and NVSS.

We caution that the majority of these galaxies would not be cluster galaxies.
More importantly, {\it we have not assessed the statistical completeness of the
sample}, as well as taking into account the variation of the volume probed at
different locations on the $M_K$-$P$ plane. The plot is to be used as merely a
suggestion for the form of the BLF; the concentration of data points on the
figure should not be taken to be proportional to the actual BLF.  Despite these
caveats, the figure reveals a couple of interesting properties of galaxies in
the local ($z\lesssim 0.2$) Universe. For massive galaxies (\eg $M_K\le -24$),
at a given $M_K$, their radio luminosities span a very wide range.  This is to
be expected, for at the moment we observe, the central SMBHs of galaxies must
be at various stages of their duty cycle.  Secondly, the maximum radio
luminosity the galaxies can have seems to correlate with their stellar mass,
which is in agreement with the findings of \citet{owen93} and LO96. This result
can be expected from the correlation between the masses of the bulge and the
central SMBH \citep{magorrian98}, and the classical idea of maximum luminosity
given the mass of the source (the Eddington limit; \eg \citealt{deyoung02}).

\begin{inlinefigure}
   \ifthenelse{\equal{\figtype}{EPS}}{
   \begin{center}
   \epsfxsize=8.cm
   \begin{minipage}{\epsfxsize}\epsffile{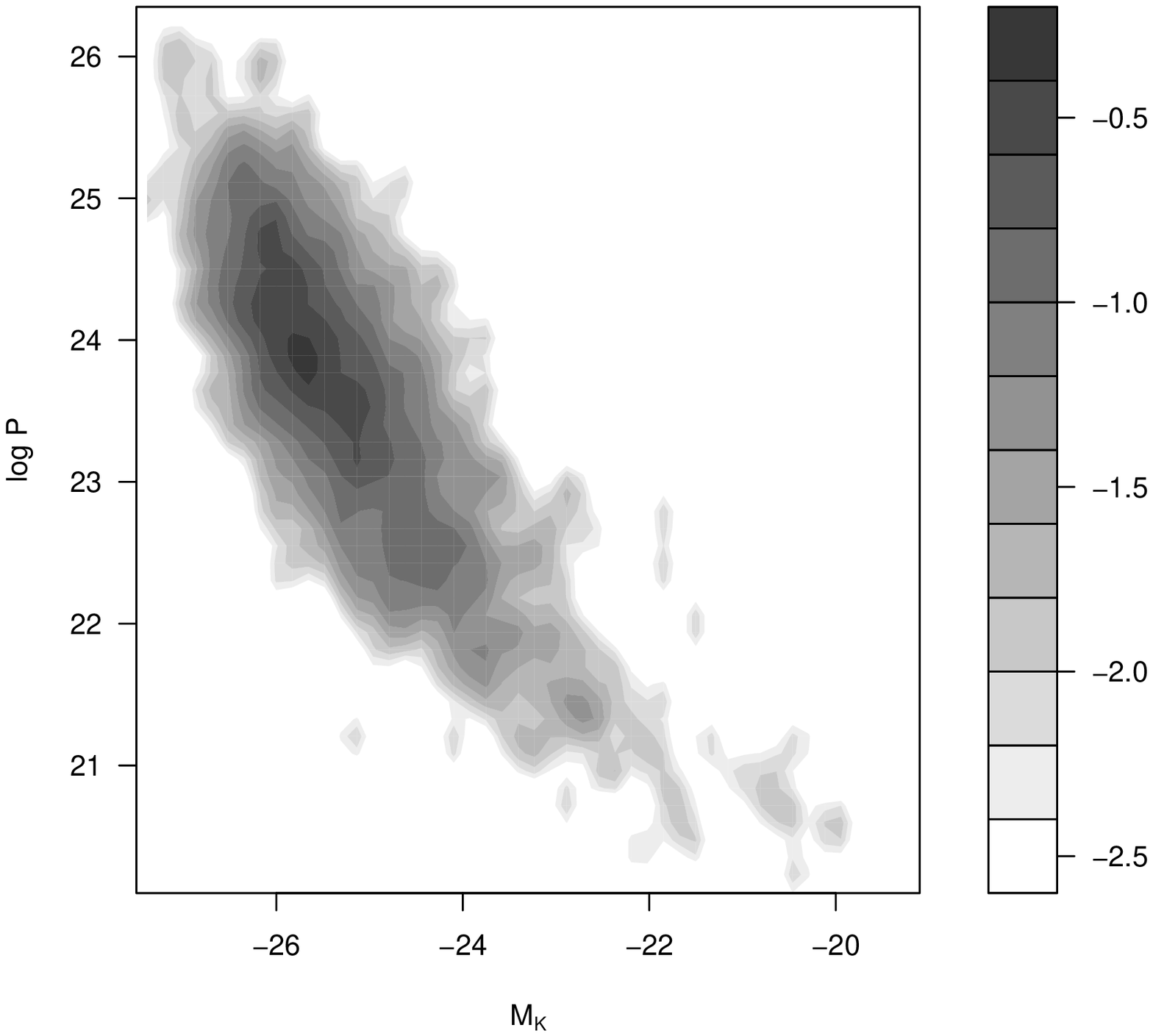}\end{minipage}
   \end{center}}
   {\myputfigure{f1.pdf}{0.1}{1.}{-20}{-5}}
   \figcaption{\label{fig:ra_2df}
    Correlation between 1.4 GHz and $K$-band luminosities for (predominantly
    field) galaxies at $z\lesssim 0.2$. We cross correlate our
    XC sample with the main galaxy sample from the SDSS DR3; in
    total there are 1318 matches that are above the completeness limits of
    2MASS and NVSS. The density is shown in logarithmic
    scale, where the units are arbitrary. We caution this
    sample is for illustrating the shape of the bivariate LF only,
    and is not meant for rigorous statistical analysis.
     }
\end{inlinefigure}

If we restrict the galaxies to lie within clusters, the resulting distribution
is similar, but with far fewer data points. This simply indicates that the SDSS
is not deep enough (and is not designed) to probe such a faint population.  We
next search for evidence from two large surveys targeting cluster galaxies: the
ESO Nearby Abell Cluster Survey (ENACS, \citealt{katgert96}) and the Cluster
and Infall Region Nearby Survey (CAIRNS, \citealt{rines03}).  By cross
correlating the NVSS catalog with these surveys, we do not find any faint
cluster galaxies (e.g.~$M_K>-23$) that are associated with strong NVSS sources.
We note that, however, for our purpose, the depth of these surveys is also not
sufficient.  For example, the completeness of the CAIRNS is well below 50\% at
about $M_K \sim -21$.  Moreover, our estimate is that only 1 to 2\% of these
faint galaxies are radio luminous, and so we require large numbers of faint
galaxy spectra to be assured of a single confirmed AGN.  Deep spectroscopy
extending to $M_K \sim -19$ is needed in a dozen or so clusters to examine the
low mass cluster radio galaxy population.

\subsection{Discussion}
\label{sec:disc_faint}

A recent comprehensive study of the properties of radio galaxies in the local
Universe has shown that the fraction of galaxies that host radio-loud AGNs is a
strong function of both the BH and stellar masses \citep[][hereafter
B05]{best05}.  To be more specific, they find that RAF $\propto M_{star}^{2.5}$
and RAF $\propto M_{BH}^{1.6}$, where $M_{BH}$ and $M_{star}$ are supermassive
black hole
and stellar masses, respectively.  The study also shows that the radio and
(optical) emission-line AGN activities are independent of each other. B05
conclude that the radio selection picks up the most massive SMBHs that
are believed to be largely dormant at present.

Can this strong dependence on stellar mass of the nuclear radio activity be
reconciled with our proposed population of radio-active low-mass cluster
galaxies? A possibility lies in the confining pressure that the ICM imposes on
cluster galaxies. For an elliptical galaxy surrounded by hot gas that
is isothermal with temperature $T$ and has the X--ray luminosity $L_X$, we
follow the simple arguments outlined in B05 and estimate the rate
that gas cools out to be 
$\dot{M}_{cool} \propto L_X / T$.  
Based on the observed scaling relation between $L_X$ and optical luminosity of
galaxies, the Faber-Jackson relation, and that between $M_{BH}$ and the
velocity dispersion $\sigma$, B05 argue that 
\[
\dot{M}_{cool} \propto \sigma^6 \propto M_{BH}^{3/2}.
\]
We note that because the galaxy sample B05 use is drawn from the
main galaxy sample of the SDSS, they do not probe to the low mass regime in
cluster galaxies.

\citet{finoguenov04} examine the effects of cluster environment have on the
$L_X$--$L_{opt} \sigma^2$ relation (where $L_{opt}$ is the optical luminosity of
galaxies) and find that it flattens in clusters with
respect to the field.  The inferred relation in cluster is $L_X \propto
T^{11/4}$, which leads to 
\[ 
\dot{M}_{cool} \propto \sigma^{7/2} 
\] 
(using $T \propto \sigma^2$ for an isothermal gas).  Therefore, if the
$M_{BH}-\sigma$ scaling remains the same, the cooling rate becomes only weakly
dependent on stellar or black hole mass.  Although we do not consider several
other factors such as the radiative accretion efficiency onto the central 
SMBH and the abundance of the gas that feeds the SMBH,
the change in the gas cooling rate suggests that a simple extrapolation of the
RAF from properties of high mass galaxies is unlikely to hold.
Furthermore, we note that the
RAF is a strong function of the radial distance from the cluster center.
There is clearly something special about being deeper within
the cluster potential well-- at least for the high mass galaxies, and so it is
reasonable that this physics is also at work for the low mass galaxies. 
More rigorous investigations, both theoretical and observational, are definitely
needed to further study this population.

\section{ICM Energy Injection from AGNs}
\label{sec:ra_heating}

We start by summarizing several observational results about the BCGs and the
central region of clusters. (1) The spatial distribution of radio-loud AGNs is
much more centrally concentrated than both the ICM and cluster galaxies
(\S\ref{sec:sdp}).  (2) About 80\% of the BCGs are found within $0.1 r_{200}$
from the cluster center \citep{lin04b}.  (3) BCGs are much more likely to host
AGNs of a given radio luminosity compared to less massive galaxies
(\S\ref{sec:ra_klf}). (4) BCGs have $2-3$ times higher likelihood to host AGNs
compared to the non-BCG galaxies of comparable luminosity.
(5) The BCGs in high mass clusters show a
$\sim 3$ times higher probability of hosting an AGN 
compared to those in lower mass clusters (\S\ref{sec:ra_klf}).

It is clear that the BCGs spend more time in the active phase than the other
cluster galaxies do.  This also implies this elevated AGN activity is linked to
the environment of BCGs; apparently the cluster core is a favorable place to
turn on an AGN \citep[e.g.][]{burns90,eilek04}.  This is most likely due to the
repeated mergers central galaxies experience over the assembly history of the
cluster \citep[e.g.][]{lin04b}, and/or interactions with the dense ICM
surrounding them. 
All the observations mentioned above favor the idea that radio activities of
central galaxies may be a plausible source to halt the radiative cooling of the
ICM toward the center
\citep[e.g.][]{binney95,david01,bruggen02,nath02,mcnamara05,croston05}.

With the recent finding that the radio luminosity at 1.4 GHz may be a rough
indicator of the mechanical luminosity $L_{mech}$ the AGN outflow (``bubbles''
or ``cavities'' seen in \xray images) has on the ICM \citep{birzan04}, we can
use the AGN RLF to estimate the amount of energy injected into the ICM by the
cluster AGN population, and compare it to the thermal energy content of the
ICM.  Following \citet{birzan04}, who find that $L_{mech}$ is of order $100 \nu
P_\nu$ where $\nu=1.4$ GHz, we assume that $L_{mech} = 400\eta \nu P_\nu$,
where $\eta$ is an efficiency coefficient, and the factor of 400 comes from our
further assumption that the gas contained in the AGN outflows is relativistic
and adiabatic \citep{birzan04}. With this expression, we can calculate the
total energy output from AGN $E_{AGN}$ by integrating the mechanical luminosity
over time from $z=0$ to $z=1$. A $\gamma=2.5$ power-law evolution in AGN
population is also assumed (see \S\ref{sec:ra_sz}).

The ICM thermal energy is estimated as $E_{th}=(3/2)kT_X f_{ICM}M_{200}/\mu
m_p$, where $f_{ICM}=0.12$ is the ICM mass fraction (assumed to be constant with
respect to cluster mass), $\mu=0.59$ is the mean
molecular weight, and $m_p$ is the proton mass. For a $M_{200}=10^{15} M_\odot$
cluster, we find that within the virial radius, $E_{AGN}/E_{th}=0.0047\eta$;
for a low mass cluster ($10^{14} M_\odot$), the ratio is $0.0204\eta$. Because
of the different mass dependences of $E_{AGN}$ and $E_{th}$, the AGN heating is
more relevant in low mass systems than in massive ones.  We can also express
the amount of AGN heating in terms of the energy per ICM particle, which would
be independent of cluster mass. We find that $E_{AGN}$ corresponds to
$0.059\eta$ keV per particle.

These numbers suggest that AGN mechanical heating is not a significant
contributor to the thermal energy of the ICM within the entire cluster virial
region.  However, while the ICM is rather extended, the AGN distribution is
very centrally peaked; this implies that the AGN would be more energetically
important toward the central parts of clusters. We use the measurements of the
gas density profile for a sample of 45 nearby clusters \citep{mohr99} to
determine the ICM thermal energy within $10\%$ of the virial radius.  We find
that within this radius, $E_{AGN}/E_{th}=0.011\eta$ and $0.046\eta$ for
$10^{15} M_\odot$ and $10^{14} M_\odot$ clusters, respectively. Cast
differently, the AGN energy is equivalent to $0.13\eta$ keV per particle. It is
apparent that, as long as the efficiency coefficient is not too small (\eg
$\eta\gtrsim 0.1$), the AGNs are an important source of heating of the ICM.
Based on the data of \citet{birzan04}, $\eta$ may range from $1/30$ to 20.
Another complication is that, $\eta$ may be strongly time-varying, which makes
the amount of heating extremely difficult to estimate.  Our crude calculation
is only meant to be an order of magnitude estimate of the role AGNs can play in
the thermodynamic history of clusters.

\section{Implications for SZE Surveys}
\label{sec:ra_sz}

Our results can be used to estimate the contamination of the cluster SZE
signal due to radio point sources. We address this issue in two steps: we first
transform the observed 1.4 GHz AGN RLF into the frequency at which an SZE
experiment operates; in the second step we use a simple Monte Carlo scheme to
estimate the fraction of clusters whose AGN flux is a significant fraction of
the SZE flux.

Given the frequency $\nu$ (in GHz) of an SZE experiment, we transform
our observed 1.4 GHz RLF to that frequency, via 
\beq
\label{eq:ra_tlf} 
{dn \over d\log P_\nu}\left( \log P_\nu \right) = \int
  {dn \over d\log P_{1.4}}
  \left( \alpha \log {1.4 \over \nu} +\log P_\nu \right)
  f(\alpha) d\alpha
\eeq
where $P_\nu$ and $P_{1.4}$ are the luminosities
at $\nu$ and 1.4 GHz, respectively,
which are related to each other by $P_{1.4} = (1.4/\nu)^\alpha
P_\nu$; $f(\alpha)$ is the
probability distribution of the spectral indices from cluster radio
sources, taken as the SID presented in
\S\ref{sec:ra_index} (see Fig.~\ref{fig:ra_alpha}). 
(We note in passing that \citealt{hlin02} use an expression very
similar to Eqn.~\ref{eq:ra_tlf} to extrapolate the LF to higher
frequencies.) To account for the possible steepening of the spectral
index at high frequencies (\eg $\nu \gtrsim 100$ GHz), the LFs at
frequencies greater than 90 GHz are obtained by extrapolating the 90
GHz RLF with a modified SID $\tilde{f}(\alpha)=f(\alpha+0.5)$.

In Fig.~\ref{fig:ra_fr} we show the estimated AGN RLFs at 15, 30, 90, and 150
GHz. The most apparent feature in these extrapolated RLFs is the rise of the
bright end, which is due to the non-negligible population that has a rising
spectrum ($\alpha>0$; see \S\ref{sec:ra_index}), and the steepness of the 
RLF. As the bright end will definitely contribute significantly to the fluxes
from radio sources, this points out the importance of an accurate model for the
SID of the radio sources over the frequency range of interest.

Our technique involves an extrapolation by as much as a factor of 100 in
frequency, and so this simple approach of using the 1.4~GHz to 4.85~GHz
spectral behavior together with a break at 100~GHz should be viewed as a crude
extrapolation-- especially at the highest frequencies.  Existing observations
tend to suggest a flattening of the spectral index for samples selected at
higher frequencies.  For example, a VLA survey finds that for sources (mainly
flat-spectrum, core-dominated QSOs) with $S\ge 0.1$ Jy, $\bar{\alpha}(8,90)$
can be described by a Gaussian whose mean and dispersion are $-0.37$ and 0.34,
respectively \citep{holdaway94}. At fainter flux limit ($\sim 20$ mJy), results
from the CBI experiment provide an estimate that $\bar{\alpha}(1.4,31)=-0.45$,
with maximum and minimum indexes being 0.5 and $-1.32$, respectively
\citep{mason03}. An extensive follow-up from 1.4 to 43 GHz of the 9C survey
finds that all but two sources in their sample of 176 show a falling or concave
spectrum \citep{bolton04}; about 20\% of their sources have spectra that peak
at or above 5 GHz. With sensitive (0.4 mJy) 1cm observations toward 56
clusters, \citet{cooray98} find $\bar{\alpha}(1.4,28.5)=-0.77\pm 0.06$ from 54
sources. 
A recent follow up study finds that $\bar{\alpha}(1.4,28.5)=-0.70$ from 87
sources \citep{coble06}. 
Finally, the {\it WMAP} satellite has conducted an all-sky survey of
bright radio sources ($\gtrsim 1$ Jy) at $41-94$ GHz, and found that the mean
index $\bar{\alpha} \approx 0$ with a scatter $\sim 0.3$ \citep{bennett03b}.
It is clear that reliable estimates of the cluster radio galaxy luminosity
function at high frequencies will have to await new observations.

\begin{inlinefigure}
   \ifthenelse{\equal{\figtype}{EPS}}{
   \begin{center}
   \epsfxsize=8.cm
   \begin{minipage}{\epsfxsize}\epsffile{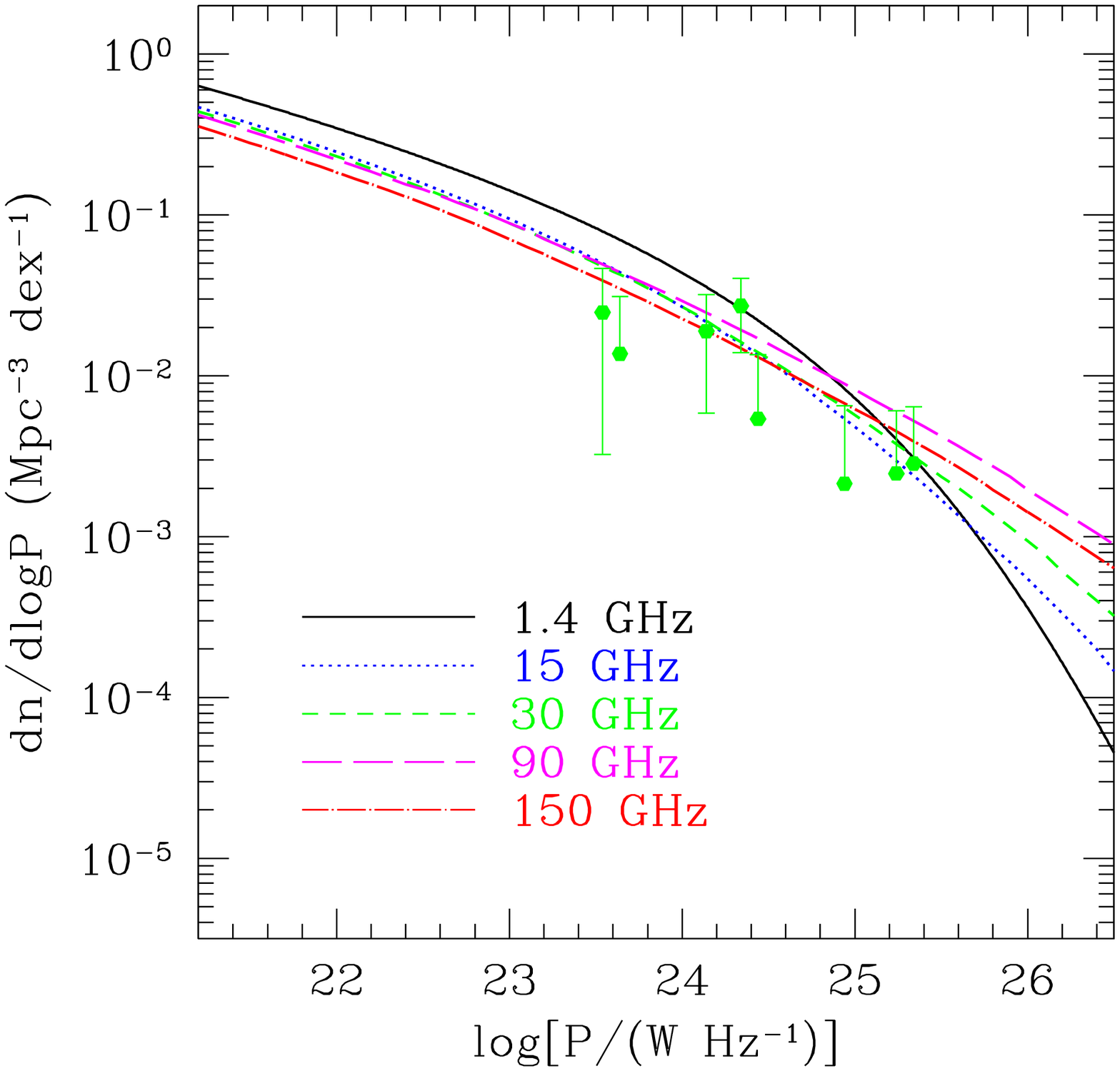}\end{minipage}
   \end{center}}
   {\myputfigure{f1.pdf}{0.1}{1.}{-20}{-5}}
   \figcaption{\label{fig:ra_fr}
    The AGN luminosity functions (within $r_{200}$) at several frequencies. We
    transform the 1.4 GHz cluster AGN RLF (solid line) to 15, 30, 90,
    and 150 GHz (dotted, short-dashed, long-dashed, and dot-dashed
    lines, respectively), using Eqn.~\ref{eq:ra_tlf}. Also shown as
    points is an estimate of the 30 GHz cluster radio source LF, 
    based on the data presented in \citet{coble06} and
    \citet{cooray98}. The background is estimated using the $\log N$--$\log S$
    of \citet{knox04}.
    The agreement between the points and the short-dashed line
    implies our method of extrapolating the RLF works well.
     }
\end{inlinefigure}

The data points with uncertainties in Fig.~\ref{fig:ra_fr}  show an estimate of
the cluster radio source LF within $r_{200}$ at 30~GHz, based on 68 clusters.
The data are taken from \citet{coble06} and \citet{cooray98}.
Taking into account the beam response as a function of radius, we estimate the
background using the $\log N$--$\log S$ at 30 GHz \citep[from][]{knox04}.
Despite a large degree of scatter, the match between the amplitudes of the data
points and the short-dashed curve suggests that our extrapolated RLF at 30~GHz
is in reasonable agreement with the available data.

\begin{table*}[htb]
\begin{center}
\begin{minipage}{0.65\textwidth}
\begin{center}
\caption{Luminosity Function Fits}
\label{ra_lffit}
\vspace{1mm}
\begin{tabular}{lllll}
\hline \hline 
frequency & $y$ & $b$ & $x$ & $w$ \\
\hline
\phn\phn1.4 & 37.97 & 2.40 & 25.80 & 0.78\\
\phn15      & 40.31 & 4.27 & 27.10 & 0.76\\
\phn30      & 40.60 & 3.91 & 27.89 & 0.81\\
\phn90      & 41.53 & 3.61 & 29.12 & 0.84\\
150         & 40.96 & 3.34 & 28.88 & 0.85\\
\hline
\end{tabular}
\end{center}
{\small
Note.-- See Eqn.~\ref{eq:ra_lffit} for the functional form used in the fit.
}
\end{minipage}
\end{center}
\end{table*}

For ease of comparison with other work, we have fit our extrapolated RLFs and
provide the fitting parameters in Table~\ref{ra_lffit}. The same functional
form as Eqn.~\ref{eq:ra_lffit} is used.

Given the mass and redshift of a cluster, and a model of AGN population
evolution (see below), the RLF can be integrated to give the total AGN flux. We
note, however, the flux thus obtained is usually dominated by the rare, very
luminous sources, and therefore the effects of contamination from AGNs would be
overestimated. To obtain a more realistic estimate, we employ a simple Monte
Carlo simulation scheme.  First, we integrate the RLF (from $10^{20}$ to
$10^{28}$ W$\,$Hz$^{-1}$) to obtain the expected total number of AGNs, $\langle
N \rangle$. Using $\langle N \rangle$ as the mean, we generate a Poisson random
number $N_p$, to account for cluster-to-cluster variations of the total AGN
number. We assign fluxes to the $N_p$ sources, using the RLF as the probability
distribution. The sum of the fluxes from these sources gives the total AGN flux
($S_{AGN}$) of the cluster, which can be compared with the cluster SZE flux
within the virial radius ($S_{SZE}$). We use the expression of the SZE
flux-cluster mass relation provided in \citet{majumdar04}. Repeating this
process $10^5$ times, 
%
%
we estimate the AGN contamination fraction by calculating
the fraction of clusters whose AGN flux is a significant proportion of their
SZE flux. Let us denote $S_{AGN}\ge s |S_{SZE}|$.
We regard a cluster as ``lost'' from a survey when $s=1$, and correspondingly 
define the ``lost cluster fraction'' (LCF) as the proportional of clusters whose AGN flux 
is equal to or greater than the SZE flux.

\begin{inlinefigure}
   \ifthenelse{\equal{\figtype}{EPS}}{
   \begin{center}
   \epsfxsize=8.cm
   \begin{minipage}{\epsfxsize}\epsffile{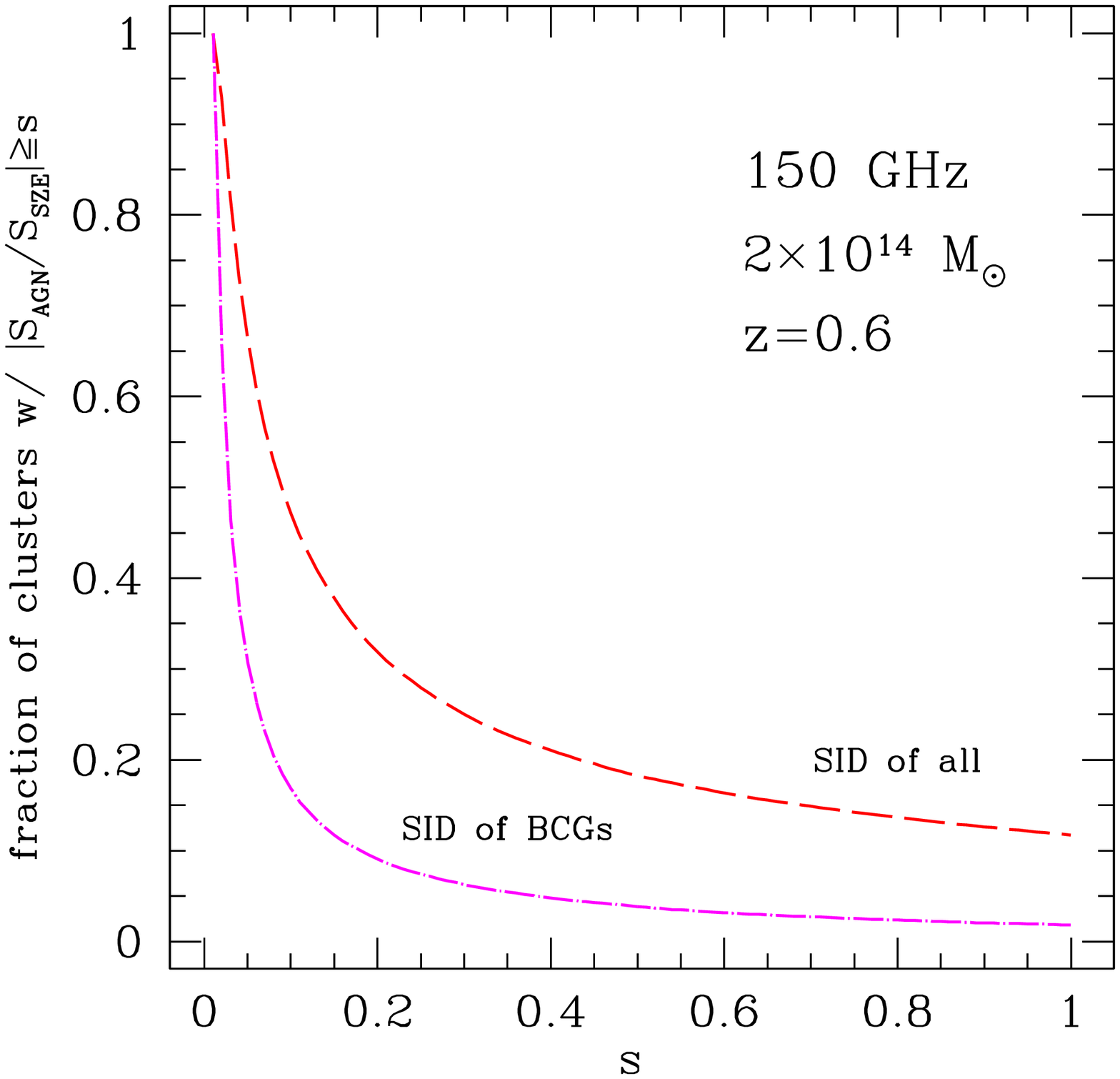}\end{minipage} 
   \end{center}}
   {\myputfigure{f1.pdf}{0.1}{1.}{-20}{-5}} 
   \figcaption{\label{fig:cont}
       The cumulative fraction of clusters whose AGN flux is greater or equal to
$s |S_{SZE}|$ as a function of $s$. We consider the case for clusters of mass
$2\times 10^{14} M_\odot$, at $z=0.6$, measured at 150 GHz. The dashed line is
the estimated fraction when the SID from all the sources for which we have flux
measurements at both 1.4 and 4.85 GHz are used to extrapolate the RLF, and the
dot-dashed curve is obtained when the SID of the BCGs is used. The lost cluster
fraction is the value of the curve at $s=1$ (about 12\% and 2\% for the two
cases). While the lower curve represents an extreme case, the upper curve
should be more representative of all the sources that contribute to AGN
contaminations.
     }
\end{inlinefigure}

For our modeling of the AGN population evolution, we consider two cases: a
no-evolution model (\ie the AGN RLF is independent of $z$), and a power-law
evolution model in which the number of AGN in a cluster of mass $M$ scales as
$(1+z)^\gamma$, but the shape of the RLF remains the same.  The true evolution
towards high redshifts is still under debate.  For example, an earlier study
suggests that the AGN RLF in intermediate redshift clusters is similar to that
in nearby clusters (\citealt{stocke99}; see also \citealt{perlman03,coble06}). However,
we note that the studies of \citeauthor{stocke99} and \citeauthor{coble06} 
focus on radio-loud AGNs
within a fixed metric radius over the redshift range of their cluster samples
($0.3<z<0.8$ and $0.1<z<1.0$), and therefore the AGN population may not be properly compared
(see also discussions in \citealt{branchesi06} for other possible
explanations).  Studies that favor a positive evolution also exist.  A deep
radio survey over the cluster MS1054-03 ($z=0.83$) finds that the radio-loud
AGN fraction is $\sim 4$ times higher than that in $z<0.1$ clusters found by
LO96 \citep{best02}.  Another determination of the RLF from a sample of 18
clusters at $0.3<z<0.8$ also concludes that the RLF in high-$z$ clusters has
higher amplitude than that of low-$z$ clusters \citep{branchesi06}.  Finally, a recent study shows
there is evidence for a $\sim 5$ times higher RAF in galaxies at $z\sim 1$
(M.~Gladders 2005, private communication).  Assuming this increase in RAF is
representative of cluster galaxies, we adopt $\gamma=2.5$ for the power-law
evolution model for the AGN.

As an example of the degree of contamination due to AGNs, we show in
Fig.~\ref{fig:cont} the cumulative fraction of clusters at $z=0.6$ whose AGN
flux is greater or equal to $s |S_{SZE}|$ as a function of $s$, at 150 GHz. The
chosen cluster mass is $2\times 10^{14} M_\odot$, close to the expected
detection limit of some future SZE surveys (e.g.~ACT, SPT). We consider the
$\gamma=2.5$ case here.  Two curves are shown; the dashed line is the estimated
contamination fraction when the SID from all the sources for which we have flux
measurements at both 1.4 and 4.85 GHz are used to extrapolate the RLF, and the
dot-dashed curve is obtained when the SID of the BCGs is used (hereafter the
all-SID and BCG-SID cases, respectively).  In \S\ref{sec:ra_index} we find
evidence that the BCGs, on average, may have a steeper spectral index compared
to non-BCG sources (the mean indices of the two populations are $-0.74\pm 0.05$
and $-0.47\pm 0.15$, respectively; the distribution about the mean is broad for
both populations, however). 
Therefore, the extrapolated 150 GHz RLF based on the SID of the BCGs has a
lower amplitude compared to the one constructed with the SID from all the
sources, and results in a lower contamination fraction. For example, the LCFs
(i.e.~the value of the curves at $s=1$) are 12\% and 1.8\% for the two cases.
The Figure also suggests that about 32\% (9\%) of clusters host AGNs whose
fluxes are at least 20\% of that of the SZE, for the all-SID (BCG-SID) case.

We do not attempt to model the difference in the SIDs in our Monte Carlo
simulations.  The differences between the dashed and dot-dashed curves in
Fig.~\ref{fig:cont} represent the uncertainty due to our lack of knowledge in
the spectral behavior of radio sources at high frequencies, with the BCG-SID
case being an extreme.  We expect a more sophisticated treatment will yield
estimates in the region bracketed by the curves. Below we will proceed with the
RLFs constructed with the SID from all the sources, as they should be more
representative of all the radio sources that may contaminate the SZE signal.

\begin{inlinefigure}
   \ifthenelse{\equal{\figtype}{EPS}}{
   \begin{center}
   \epsfxsize=8.cm
   \begin{minipage}{\epsfxsize}\epsffile{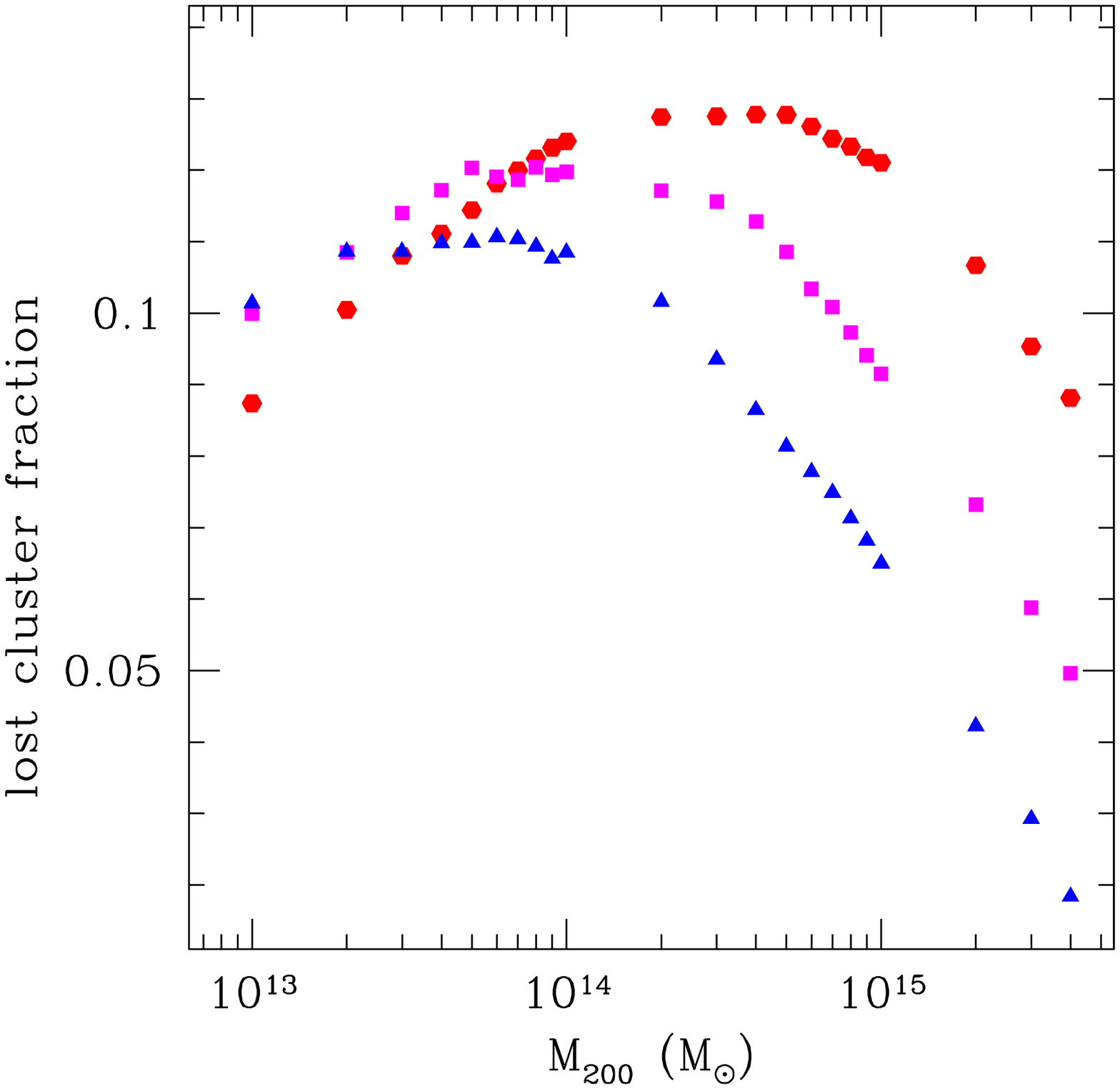}\end{minipage} 
   \end{center}}
   {\myputfigure{f1.pdf}{0.1}{1.}{-20}{-5}} 
   \figcaption{\label{fig:ra_fm}
    The 150 GHz lost cluster fraction (LCF), defined as the fraction of all
    clusters whose AGN flux is at least equal to their SZE flux. The
    circles, squares and triangles show the LCF at $z=0.1$, $0.6$
    and $1.1$, as a function of cluster mass. The forecast is based on the simple power-law
    extrapolation from the 1.4 GHz model that we have adopted (see details in text).
    We assume a power-law evolution of AGN population with redshift. See text for
    discussion on the shape of the LCF.
     }
\end{inlinefigure}

We note in passing that contamination due to super-luminous AGNs is very small
(e.g.~at 150 GHz, $\lesssim 1\%$ of all clusters are lost due to $\log P\ge 27$ sources); the main sources of contamination are the AGNs at $\log P=24-26$.

In Fig.~\ref{fig:ra_fm} we show the LCF at 150 GHz as a function of cluster
mass, estimated at three redshifts (circles: $z=0.1$, squares: $z=0.6$,
triangles: $z=1.1$). We only show the case for $\gamma=2.5$ (the no-evolution
LCF is just a factor of $(1+z)^\gamma$ smaller at each redshift). 
%
%
At each redshift, the LCF first increases with cluster mass, reaches a maximum
at masses $1-10\times 10^{14} M_\odot$, then decreases with mass. The latter
behavior can be simply understood because, in our modeling, the total number of
AGNs in a cluster simply scales with mass, while the SZE flux has a steeper
mass dependence ($\propto M^{1.68}$); therefore at the high mass end the SZE
flux wins over AGN. The decrease of the LCF toward the lower mass end, on the
other hand, is caused by the shape of the AGN RLF. For simplicity, consider a
cluster which hosts an AGN whose flux $S$ is equal to the cluster SZE flux.
The number of such sources per cluster is simply $\phi(P) V$, where $P$ is the
luminosity corresponding to $S$, and $V$ is the cluster volume. For a cluster
ten times less massive, its SZE flux is about 50 times weaker, and therefore
one AGN with flux of $S/50$ can contaminate the SZE signal. However, in the low
mass system the number of such sources is $\phi(P/50)V/10$. For our
extrapolated 150 GHz RLF, because of its mild slope (approximate power-law slope
$-0.5 \sim -0.6$ for $\log P\ge 23$), there are not enough AGNs in low mass
clusters to contaminate the SZE to the same level as in the higher mass
systems.

From the figure one also notices that for high mass clusters, the LCF decreases
with redshift, while the opposite happens for low mass ones; the LCF of
intermediate mass clusters shows little redshift dependence. This also can be
understood from the shape of the RLF. Consider a cluster at redshift $z$ whose
SZE flux is $S_{SZE}$. At that redshift the AGN luminosity that corresponds to
$S_{SZE}$ is $P_z$.
At a larger redshift $z'$, because the SZE flux is roughly constant with
respect to redshift, the contamination radio luminosity is larger than $P_z$ by
a factor of $g$ (which is simply the ratio of the luminosity distances and
$k$-corrections at the two redshifts); at the same time, the amplitude of the
RLF is increased by a factor of $h=((1+z')/(1+z))^\gamma$. Therefore, the
relative size between the slope of the RLF and the ratio $r=h / g$ would
indicate the redshift behavior of the LCF.  For low mass clusters, where the
SZE flux is small, the portion of the RLF of interest is toward the faint end,
whose slope is flatter than $r$, which means the increase of the AGN population
with respect to $z$ compensates the ever-increasing luminosity required to
contaminate the SZE toward higher-$z$.  For high mass clusters, because of the
steepness of the RLF, the $\gamma=2.5$ evolution is not fast enough to maintain
the level of contamination. Finally, for intermediate mass clusters, the
balance between the shape of the RLF and our adopted AGN evolution produces the
roughly redshift-independent LCF.

In Fig.~\ref{fig:ra_szf} we show the LCF for a $2\times 10^{14} M_\odot$
cluster as a function of redshift, at three frequencies of interest to some SZE
experiments (30, 90 and 150 GHz; \eg SZA, AMiBA, SPT, ACT). Only the
$\gamma=2.5$ case is shown. It is apparent that the radio point source
contamination is significantly reduced for experiments operating at higher
frequencies. This is due to the increase of the SZE signal toward higher
frequency, and also the tendency for cluster radio sources to have negative
spectral indices. We note, however, in our estimates no attempt is made to
model the removal of cluster radio point sources (either through
multi-frequency bolometric observations or through interferometric filtering).
This requires careful consideration of instrumental effects and observational
strategies, which is beyond the scope of this paper.

As we note earlier, two approximations have been used in our analysis (the
extrapolation of the RLF based on the spectral behavior of radio sources at 1.4
to 4.85 GHz, and the power-law redshift evolution of radio-loud AGN abundance in
clusters). Dedicated study of the spectral energy distribution of cluster radio
sources, based on a large sample of clusters that span a wide range in
redshift, would be ideal to provide the data needed to carry out better
forecasts for the next-generation mm-wave SZE surveys. Such a knowledge is
crucial, in that to extract precise cosmological information from a survey, one
needs to be able to control and model the systematics to few percent level
\citep[e.g.][]{majumdar04,lima05}.
Useful information about the model employed in the forecast may result from
a cross-correlation analysis of a well-defined cluster sample and WMAP maps
\citep[e.g.~][]{afshordi05,lieu06}.

\begin{inlinefigure}
   \ifthenelse{\equal{\figtype}{EPS}}{
   \begin{center}
   \epsfxsize=8.cm
   \begin{minipage}{\epsfxsize}\epsffile{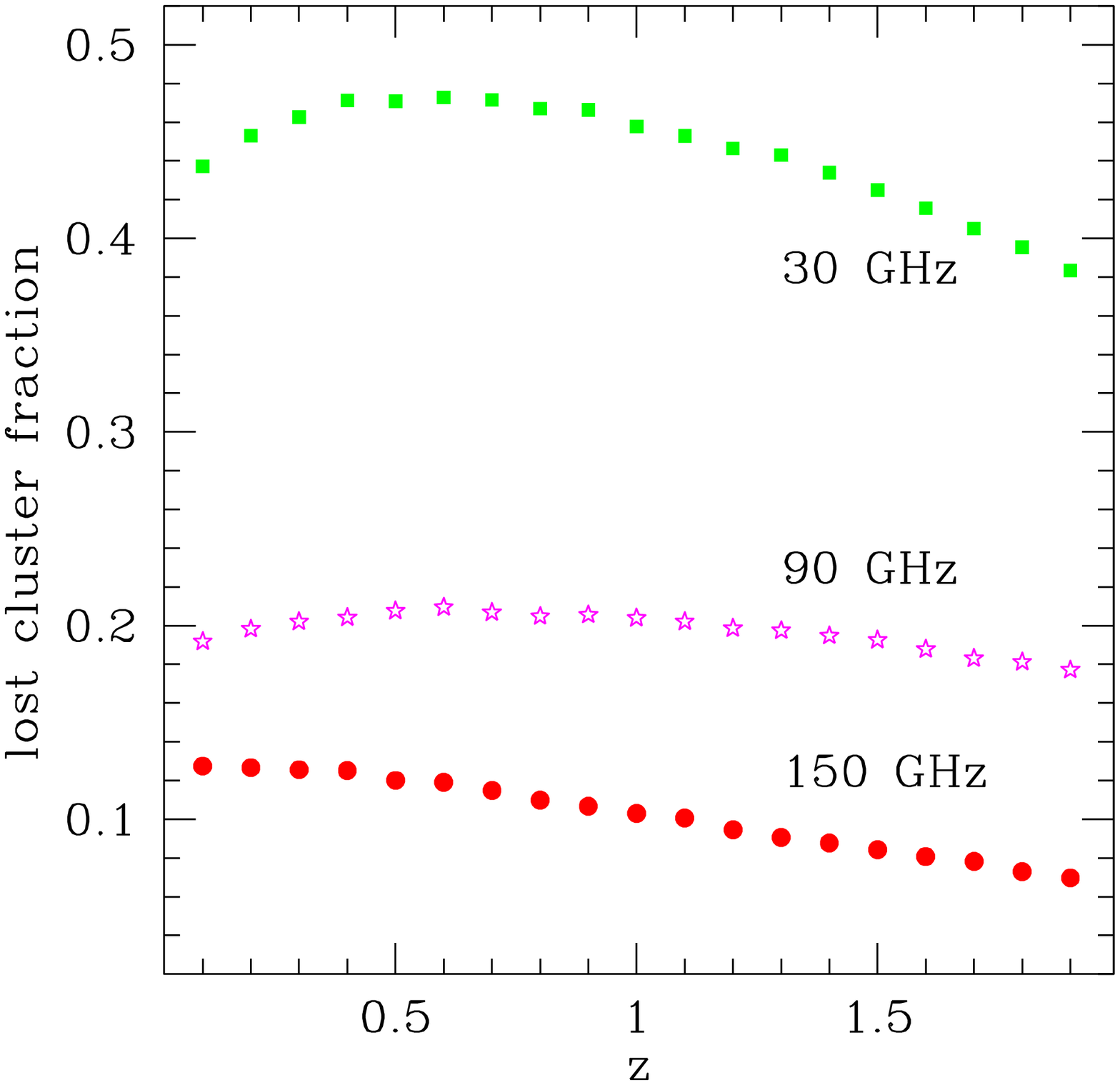}\end{minipage} 
   \end{center}}
   {\myputfigure{f1.pdf}{0.1}{1.}{-20}{-5}} 
   \figcaption{\label{fig:ra_szf}
       Lost cluster fraction for a
    $2\times 10^{14} M_\odot$ cluster at 30, 90, and 150 GHz (from top
    to bottom), assuming a power-law redshift evolution of AGNs. 
    For the simple power-law extrapolation model from 1.4 GHz that we have adopted (see details in text),  
    at 150 GHz the AGN contamination is $\sim 10\%$, and the rate
    increases as the frequency of a SZE experiment decreases.
     }
\end{inlinefigure}

\section{Summary}
\label{sec:ra_summary}

We have studied the nature of radio-loud AGNs from a sample of 573 \xray
selected clusters, using the public data from the NVSS and 2MASS surveys.
Without redshift information for individual radio sources, we perform a
statistical study of several ensemble properties of cluster radio sources,
paying particular attention to any cluster mass-related trends. In particular,
we investigate their radial distribution (in terms of the surface density), and
the 1.4 GHz and $K$-band luminosity functions. By cross-correlating the NVSS
and 2MASS catalogs, we study the fraction of galaxies that is active in the
radio, and estimate the duty cycle of radio-loud AGNs. 
We also discuss the implications of our findings for the next-generation
Sunyaev-Zel'dovich effect cluster surveys, as well as for the role of AGNs in
the heating of the intracluster medium.  Finally, we present evidence
supporting a population of low mass ($M_K\gtrsim -23$) cluster galaxies which
host radio-loud AGNs ($\log P\ge 23$,  $P$ in unit of W\,Hz$^{-1}$).  In this paper 
we set an (arbitrarily
chosen) division for high and low mass clusters at $\log M_{200} = 14.2$. Our
results can be summarized as follows:

\noindent 1.~We identify the brightest cluster galaxies (BCGs) in a statistical
sense for a subsample of 342 clusters, based on the expected $K$-band magnitude
and the location within clusters (\S\ref{sec:ra_bcg}). We find that the
$1.4-4.85$ GHz spectral index distribution (SID) of the BCGs is different from
that of non-BCG sources (\S\ref{sec:ra_index}); in general the BCGs have a
steeper spectrum.  The mean spectral index for BCGs is $\bar{\alpha}=-0.74\pm
0.05$, while that for non-BCGs is $\bar{\alpha}=-0.47\pm 0.15$.  About $14\%$
(17\%) of BCGs (non-BCGs) show a rising spectrum ($\alpha\ge 0$).

\noindent 2.~The radial distribution of radio-loud AGNs is very centrally
concentrated, and can be described by an NFW profile with concentration $c\sim
25$, when BCGs are excluded.  Inclusion of the BCGs increases the concentration
to $c \gtrsim 50$ (\S\ref{sec:sdp}; Table~\ref{ra_1}). More powerful radio
sources ($\log P > 24$) show a more concentrated distribution compared to the
weaker sources (\S\ref{sec:ra_dep}), which may be caused by the luminosity
segregation of galaxies as suggested by previous studies
\citep[e.g.][]{ledlow95}, or by the confining pressure due to the ICM. 
For $M_{K*}$ ($\approx -24$) or fainter galaxies, the
distribution can be described by an NFW profile with $c=4-5$; for the more
luminous galaxies, however, the radial profile has $c\gtrsim 9$
(\S\ref{sec:ra_2massprof}). 
The concentration of galaxy distribution presented here is higher than that
found in our previous study (LMS04); the main causes of the discrepancy are
the cluster sample selection (only clusters not strongly affected by the background
large scale structure are used) and the cluster mass
estimator. Our present analysis accounts for the variation in the background
number count due to the large scale structure better.
Finally, we find no indication for cluster
mass-dependence of the radial distribution of the radio sources
(\S\ref{sec:ra_dep}).

\noindent 3.~After accounting for the difference in the mean overdensities in
the field and in clusters ($200/\Omega_M$), the amplitude of the cluster radio
luminosity function (RLF) is still higher than that of the (scaled) field RLF
(\S\ref{sec:ra_rlf}). Between the luminosity interval $\log P = 24-27$, the
cluster AGN number density is $\sim 5700$ times higher than the field value
(see also \citealt{massardi04}), corresponding to a $5700/(200/\Omega_M)
\approx 6.8$ times higher probability for a galaxy to be radio-loud.  An
earlier study finds no difference between the field and cluster RLFs in terms
of the fraction of galaxies that is active in the radio \citep{ledlow96}; in
\S\ref{sec:ra_lo96} we discuss possible origins of the different conclusions,
and find that the comparison is non-trivial.

\noindent 4.~The RLFs for high and low mass clusters show marked difference at
the very bright end; the most luminous AGNs are found only in high mass
clusters (\S\ref{sec:ra_rlf}). The amplitude of the RLF for low mass clusters
is slightly higher than that of high mass clusters; 
limited by the poor statistics, we do not further quantify the difference.

\noindent 5.~We construct the $K$-band LFs for galaxies (based on 2MASS
sources) as well as for radio-loud AGNs (from the NVSS sources that have a
2MASS counterpart, which we refer to as the XC sources). By comparing the
number of galaxies more luminous than a $K$-band magnitude for the 2MASS and XC
sources (selected with a certain radio luminosity range), we infer the radio
active fraction (RAF) for cluster galaxies (\S\ref{sec:ra_klf}). When the
limiting radio ($K$-band) luminosity is increased (decreased), the RAF
decreases (see Table \ref{ra_5}). About $5\%$ of cluster galaxies more massive
than $M_{K*}$ host AGNs. Our data also suggest that the RAF in
massive clusters is a factor $\lesssim 3$ larger than that in low mass ones, with
differences mainly coming from luminous galaxies ($M_K\le M_{K*}$).
Because of the different radial distribution of radio sources and galaxies, we
caution that the RAF depends on the radius over which it is measured
(\S\ref{sec:ra_lo96}).

\noindent 6.~Special attention is paid to the RAF of BCGs (\S\ref{sec:ra_klf}).
Compared to the majority of galaxies (\eg $M_K\le M_{K*}$), the BCGs are 10
times more likely to host an AGN of typical radio luminosity ($\log P\ge 23$),
and the probability enhancement of having a $\log P\ge 25$ AGN is even higher.
From another point of view, compared to the cluster galaxies of comparable
$K$-band luminosity (but not centrally located in clusters), the BCGs still
have $2-3.6$ times larger probability of harboring AGNs.  This implies that the
central regions of clusters favor AGN activities. The BCG RAF shows a strong
cluster mass dependence: 36\% of the BCGs in high mass clusters have $\log P\ge
23$, while only 13\% of their counterparts in low mass clusters are as powerful
in the radio.

\noindent 7.~Using the fraction of galaxies that host radio loud AGNs as a
function of galaxy luminosity \citep{best05} and the stellar mass function
\citep{bell03} of field galaxy population, we compare the RAFs in clusters and
in the field, and find that the latter is $10-20\%$ of the former
(\S\ref{sec:ra_cffield}).

\noindent 8.~The RAF can be turned into an estimate of the typical lifetime
$t_r$ of radio sources (\S\ref{sec:duty}). For $M_{K*}$ galaxies formed at
$z=3$, $t_r \approx 5.7\times 10^8$ yr, which is consistent with the quasar
lifetime estimated from their spatial clustering (\citealt{croom05}).

\noindent 9.~By comparing the RLFs constructed from the XC sources
and from the sources
that are selected solely from the NVSS catalog, we infer that roughly $40-50\%$ of
the powerful cluster radio sources ($\log P\ge 23$) reside in low mass galaxies
(\eg $M_K\gtrsim -23$; see \S\ref{sec:lowmass}).  Support for the existence of
this population comes from 
({\it i}) their abundance as implied by the near-IR cluster LF, ({\it
ii}) the surface density profile and the RLF of the radio sources that do {\it
not} have a 2MASS counterpart; in particular we show that they are strongly
clustered. And finally ({\it iii}) their optical color being consistent with the
cluster red sequence.

\noindent 10.~It is very likely that the central region of clusters plays an important
role in fueling the central engine of radio loud AGNs; possible mechanisms include
the higher confining pressure of ICM, and the cooling instability (``cooling flow'')
towards cluster center.
Observations that strongly support this view include (1) the concentrated spatial distribution
of radio sources, irrespective of the stellar mass of their host galaxies, 
(2) the RAF of BCGs is $>2$ times
higher than that of galaxies of comparable stellar mass, and (3) 
the RAF of cluster galaxies is larger than their counterparts in the field, and the amplitude
of the cluster RLF is higher than that of the field RLF when the difference in the overdensities
are accounted for.

\noindent 11.~Because of their centrally concentrated spatial distribution, the
radio-loud AGNs are an important heating source for the intracluster medium,
especially within 10\% of the virial radius.  We estimate that AGNs can provide
$\sim 0.13\eta$~keV~per~particle near cluster center, where $\eta$ is a
coefficient relating mechanical energy input to radio emission luminosity
($1/20 < \eta < 20$) (\S\ref{sec:ra_heating}).

\noindent 12.~Using the SID derived at low frequencies (1.4 and 4.85 GHz), we
extrapolate our 1.4 GHz RLF to higher frequencies (\S\ref{sec:ra_sz}). Our
extrapolation at 30 GHz agrees reasonably well with existing data. We provide fits to the
RLFs at several frequencies in Table \ref{ra_lffit}. Assuming a power-law
evolution of the radio-loud AGN population, such that clusters of a given mass
have five times more AGN at $z=1$ than at $z=0$ (but no change in the shape of
the RLF), we model the contamination of cluster SZE signal by AGNs using a
Monte Carlo scheme. 
We define the 
``lost cluster fraction'' (LCF) as
the fraction of clusters whose AGN flux is greater or equal to their SZE flux,
$S_{AGN} \ge |S_{SZE}|$. At 150 GHz, at a
given redshift, the LCF first increases with mass, reaches a maximum of
$10-13\%$ at $1-10\times 10^{14} M_\odot$, then decreases with mass. 
For clusters more massive than
$10^{15}M_\odot$, the LCF decrease with redshift; low mass clusters
($M_{200}<10^{14} M_\odot$) show the opposite behavior. For clusters of
intermediate masses, the LCF decreases weakly with increasing redshift.  We
caution that our estimate is based on a large extrapolation in frequency for
the spectral behavior of AGNs, and on the cluster AGN abundance evolution that
is currently not well understood. 
Moreover, our estimates do not include the ability of cluster survey experiments 
to remove point source flux through angular or spectral techniques.

\acknowledgements

We are grateful to an anonymous referee for helpful comments and suggestions that 
have improved the paper.
We acknowledge Neal Miller, Tom Crawford, Andrey Kravtsov, Zoltan Haiman, Gil Holder, and
Alexis Finoguenov for useful suggestions and ideas.
We thank John Carlstrom and Kim Coble for
providing the 30 GHz point source data. YTL has benefitted
from discussions with Alastair Sanderson, Chris Blake, Ben Wandelt, Nelson Padilla,
T.-C.~Wei,
Y.-S.~Shiao, and I.H.  We acknowledge financial support from NSF Office of Polar
Programs award OPP-0130612 and NASA LTSA award
NAG5-11415.
YTL acknowledges support from the Princeton-Catolica Fellowship,
NSF PIRE grant OISE-0530095, and FONDAP-Andes.

This publication makes use of data products from the 2MASS, which is a joint
project of the University of Massachusetts and the IPAC/Caltech. 
This research has made use of the NED and BAX. 
This paper has made use of data from the SDSS and 2dFGRS surveys.  
Funding for the SDSS and SDSS-II has been provided by the Alfred P.~Sloan Foundation, the Participating Institutions, the National Science Foundation, the U.S.~Department of Energy, the National Aeronautics and Space Administration, the Japanese Monbukagakusho, and the Max Planck Society, and the Higher Education Funding Council for England.
The SDSS web
site is http://www.sdss.org/.  The 2dFGRS has been made possible by the
dedicated efforts of the staff of the AAO, both in creating the 2dF instrument
and in supporting it on the telescope.
This paper makes use of data products from the NVSS and FIRST surveys.

\appendix
\section{Systematics and Uncertainties in Surface Density Profile Fitting}

\subsection{Radio Sources}
\label{sec:appen}

Here we provide some details about the fitting of the surface density profile
of radio sources. First we argue for the use of a suitable cluster sample
to study the stacked profile, then we describe the mock observations used
to understand the systematics and uncertainties in the fitting procedure.

In general, to study the density profile of sources statistically (i.e.~without the
aid of spectroscopic redshifts), one has to determine the background either
globally (i.e.~from the $\log N$--$\log S$) or locally (e.g.~from an annulus
surrounding the cluster where the contribution of cluster sources is negligible).
At large radii, the estimates from both methods should agree, at least when
many different line-of-sights are considered (that is, when many cluster fields
are stacked). For example, for sources with $\log P\ge 23$ in our {\it RASS}
cluster sample, within $r_{out}=2 r_{200}$ the locally derived background level
is $\sim 5\%$ lower than that based on the $\log N$--$\log S$. With $r_{out}=5 r_{200}$,
the two estimates agree to $0.2\%$. This suggests that the {\it RASS} sample
is suitable for the study of the surface density profile of radio sources (this is 
not the case for 2MASS galaxies;  see \S\ref{sec:ra_2massprof}).

We employ mock catalogs to help determine the uncertainty in the best fit
profile (in particular that of the concentration, $\sigma_c$) and understand 
the systematics of the fitting procedure. Given a cluster sample 
(e.g.~the 188 clusters which allow the construction of surface density
profile for sources with $\log P\ge 23$ in \S\ref{sec:sdp}) 
and a concentration, for every cluster 
we generate cluster sources whose spatial distribution follows an NFW profile,
in addition to uniformly distributed background sources. 
To incorporate the presence of large scale structure in the vicinity of cluster fields,
for every real cluster we measure the surface densities of radio sources in annuli of radii
$2-3 r_{200}$ and $3-5 r_{200}$ and compare them with the predictions based on
the global number count, $\log N$--$\log S$. Averaging over all clusters, the
ratio of local number counts to the ones based on $\log N$--$\log S$ is almost
unity in both annuli, with dispersion about 0.2, indicating that (1) the presence of
the clusters do not affect the source count outside the virial region (an indication
that radio sources are quite centrally distributed), (2) the 
background is fairly uniform and, on average, $\log N$--$\log S$ gives good estimates
of the background level (which confirms the conclusion drawn in the previous
paragraph).
While the number of background sources in mock clusters is determined by the
limiting radio luminosity, the $\log N$--$\log S$, and the angular
extent of the cluster, we model the number
of cluster radio sources $N_r$ by assuming it is proportional to the number of 
galaxies $N_g$ in a cluster; using the $N_g$--$M_{200}$ relation obtained by
\citet[][LMS04]{lin04}, 
\beq
\label{eq:appen}
N_g = 37 (M_{200}/10^{14} M_\odot)^{0.85}\ \ \ \ \ {\rm (for }\ M_K\le -21\ {\rm galaxies)}, 
\eeq
we write $N_r = 37 \xi (M_{200}/10^{14} M_\odot)^{0.85}$. 
The actual number of sources assigned to a cluster is drawn from a Poisson
random number whose mean is $N_r$. We adjust
the value of $\xi$ until the total numbers of sources within $r_{200}$ and $5 r_{200}$
of the composite (stacked) cluster match well those in the observed composite
cluster. The value of $\xi$ depends on the limiting radio power of sources; for 
$\log P\ge 23$ sources we can construct realistic mock composite clusters by setting
$\xi=0.02$. After choosing a suitable $\xi$, we generate 50 mock composite clusters and fit
the resulting profiles; the standard deviation of the distribution of best-fit $c$ is then
used as $\sigma_c$ of the observed profile.

We further use these mock observations to study the effects of uncertainties
in the cluster mass (i.e.~$r_{200}$), cluster center, and position of individual radio sources
on the surface density profile. For each cluster in the sample, we perturb its mass
in logarithmic space
from the one inferred from the X-ray luminosity by a Gaussian random number with
a standard deviation of 0.2 (corresponding to $\sim 50\%$ fractional uncertainty in mass;
\S\ref{sec:ra_sample}). 
Because apertures of $3\arcmin$ radius are used to locate the centroid of 
extended sources from the raw {\it RASS} data in the {\it NORAS} and {\it REFLEX} 
surveys \citep{boehringer00,boehringer04b}, we expect the uncertainty in the center 
position to be $\lesssim 1\arcmin$. As our sources are stronger than 10 mJy, the
uncertainty in the radio source position is about $1\arcsec$ (\S\ref{sec:ra_data}).
We therefore randomly displace the cluster center and radio source positions
up to $1\arcmin$ and $1\arcsec$, respectively.

For objects with a very concentrated spatial distribution like radio sources, it is the
uncertainty in the cluster center that has the largest impact on the shape (concentration)
of the density profile. For a circularly symmetrical profile, false centers tend to smear 
the profile and lower the fitted concentration. The uncertainty in cluster mass translates
into a scaling of $r_{200}$; the concentration will be biased low (high) for clusters whose
masses are underestimated (overestimated). Finally, as long as there is no
systematic shift of the positions of the sources, the small positional uncertainty for NVSS
sources should not bias the value of concentration.

The effects of these uncertainties are quantified in Table~\ref{bias}. We use 50 mock composite
clusters, each is composed of 188 clusters, to study how well our fitting code can recover the input concentration $c_{{\rm in}}$ in the Monte Carlo simulations, where we (1) assume there is no 
uncertainty at all (the ``none'' entry), (2) vary
the cluster center up to $1\arcmin$ (``center''), (3) perturb the position of individual
sources up to $1\arcsec$ (``individual''), (4) change the cluster mass in logarithmic
space by a Gaussian with standard deviation of 0.2 (``mass''), and (5) combine all
three uncertainties (``combined''). Three input values of concentration are considered ($c_{{\rm in}}
= 20, 25, 30$). We record the mean and standard deviation of the 50 best-fit values of concentration 
as $\bar{c}_{{\rm fit}}$ in each of the entries. The number of mock observations is sufficiently large
that the mean and scatter of $c_{{\rm fit}}$ do not change much when we create more mock composite
clusters.

\begin{table*}[htb]
\begin{center}
\begin{minipage}{0.65\textwidth}
\begin{center}
\caption{Effects of Uncertainties in Cluster Properties on Concentration (Radio Sources)}
\label{bias}
\vspace{1mm}
\begin{tabular}{cccccc}
\hline \hline
& \multicolumn{5}{c}{$\bar{c}_{{\rm fit}}$}\\
\cline{2-6}
$c_{{\rm in}}$ & none & center & individual & mass & combined\\
\hline
30 &  $31\pm 9$ & $28\pm 11$ & $31\pm 10$ & $32\pm 21$ & $28\pm 8$\\
25 & $25\pm 10$ & $24\pm 7$ & $24\pm 6$ & $25\pm8$ & $24\pm 6$\\
20 & $21\pm 6$ & $18\pm 4$ & $20\pm 5$ & $20\pm 6$ & $19\pm 5$\\
\hline
\end{tabular}
\end{center}
{\small
Note.-- Results based on 50 mock composite clusters for sources more powerful
than $\log P = 23$. $c_{{\rm in}}$ is the input concentration in the Monte Carlo simulations,
$\bar{c}_{{\rm fit}}$ records the average and standard deviation based on 50 mock 
observations. See test for the meaning of different entries in the Table.
}
\end{minipage}
\end{center}
\end{table*}

From the Table we see that our fitting code is able to recover the input concentration without
any biases, when no uncertainty is included in the simulations (the second column in the Table).
Very interestingly, the uncertainties in cluster mass only introduce a small, positive bias in the
concentration.
This is likely because cluster masses are perturbed high and low in the log space with similar
probability; 
as long as the X-ray luminosity is a unbiased mass indicator over the mass range of
clusters in our sample, such modeling of uncertainty in cluster mass is justified.
More importantly, the effect of uncertainties in cluster center
is slightly offset by the combined
effect of uncertainties in cluster mass and position of radio sources, making the resulting profile
only a little bit biased towards lower ($4-7\%$) concentration.
We note, however, if we perturb the cluster center up to $2\arcmin$, the bias on $c$ will increase.
Including only the central position uncertainty, $\bar{c}_{{\rm fit}}/c_{{\rm in}} \approx 0.7$ for the
three $c_{{\rm in}}$ values considered. When all three kinds of uncertainties are present, the ratio
is still about 0.7.
Even though we could not rule out the possibility that some of the {\it RASS} clusters may have their
center offset from the true location by as large as $2\arcmin$, the proportion of such clusters should
be very small ($\lesssim 2\%$; \citealt{boehringer00}), and therefore our simulation would represent an extreme and very unlikely case.

In summary, these tests suggest that the concentration of the spatial distribution of radio sources in clusters
found in \S\ref{sec:sdp} should be representative of the true value, with a possibility of being biased low
by $\lesssim 7\%$.


\subsection{2MASS Galaxies}
\label{sec:appen2}

We also use mock observations to study the effects of uncertainties in cluster properties
(cluster center and mass) on the best-fit parameters of the galaxy surface density profile.
The procedure is similar to that outlined in \S\ref{sec:appen}, with some modifications that
are described below.

First, the number of cluster galaxies is based on Eq.~\ref{eq:appen},
$N_g = 37 \xi' (M_{200}/10^{14} M_\odot)^{0.85}$, where $\xi'$ serves to scale the galaxy number
to that appropriate for the chosen $K$-band magnitude limit $M_{K,lim}$. 
The $K$-band luminosity function of cluster galaxies determined in LMS04
is used to set $\xi'$. Second, the local-to-global number count ratios in the two annuli around each
cluster ($2-3 r_{200}$ and $3-5 r_{200}$) show larger variation compared to the case of radio
sources. Averaging over all clusters, the ratios in the inner and outer annuli are both greater than
unity, with large dispersions. To better understand this, we have selected a subsample 
from all the {\it RASS} clusters which are not closer to each other than $6 r_{200}$, and repeat the
local-to-global number count comparison. With this subsample, the effects on the local background
counts due to nearby clusters in projection should be 
much reduced (although there would still be clusters
whose X-ray flux is below the limits of {\it NORAS} and {\it REFLEX} surveys present in the annulus
regions). The ratios are $1.4\pm 0.7$ and
$1.2\pm 0.4$ for the innter and outer annuli, respectively. In addition to variations in the 
large scale galaxy distribution, this also implies that the galaxies have a broad spatial distribution
whose effect may be seen beyond $3 r_{200}$! In fact, using a smaller sample of clusters which
are isolated from each other by at least $8 r_{200}$, we find that the local-to-global number count
ratio in the annulus of radii $5-8 r_{200}$ is $1.0\pm 0.3$. This exercise suggests that we should
examine the spatial distribution of galaxies using data out to at least e.g.~$5 r_{200}$.
To include field-to-field variations in the background galaxy counts, we incorporate the
local-to-global number count ratio from the real clusters in the mock cluster observations.

Combining our model of number of cluster galaxies,  the $\log N$--$\log S$
from the 2MASS all-sky data release, and the local background variation based on
real clusters,
we are able to generate mock composite clusters whose
numbers of galaxies within $r_{200}$ and $5 r_{200}$ match those in the observed ones well.
We further include uncertainties in the cluster center and mass in the mock clusters, by
randomly displacing the cluster center up to $1\arcmin$, and perturbing the cluster mass
in logarithmic space by a Gaussian random number with a standard deviation of 0.2.
After creating an ensemble of composite clusters, we fit the surface density 
profiles and use the standard deviation
of the resulting best-fit concentrations as the uncertainty of the concentration for the observed
composite cluster.

\begin{table*}[htb]
\begin{center}
\begin{minipage}{0.65\textwidth}
\begin{center}
\caption{Effects of Uncertainties in Cluster Properties on Concentration (Galaxies)}
\label{bias2}
\vspace{1mm}
\begin{tabular}{ccccc}
\hline \hline
& \multicolumn{4}{c}{$\bar{c}_{{\rm fit}}$}\\
\cline{2-5}
$c_{{\rm in}}$ & none & center & mass & combined\\
\hline
5.0 &  $5.0\pm 0.5$ & $4.5\pm 0.4$ & $5.0\pm 0.6$ & $4.9\pm 0.5$ \\
4.0 & $4.0\pm 0.4$ & $3.6\pm 0.3$ & $4.0\pm 0.4$ & $3.9\pm 0.4$ \\
3.0 & $3.0\pm 0.3$ & $2.8\pm 0.2$ & $3.0\pm 0.3$ & $2.9\pm 0.3$ \\
\hline
\end{tabular}
\end{center}
{\small
Note.-- Results based on 100 mock composite clusters for galaxies more luminous
than $M_K=-24.5$. $c_{{\rm in}}$ is the input concentration in the Monte Carlo simulations,
$\bar{c}_{{\rm fit}}$ records the average and standard deviation based on 100 mock 
observations. See text for the meaning of different entries in the Table.
}
\end{minipage}
\end{center}
\end{table*}

We record in Table~\ref{bias2} the effects of the uncertainties in cluster center and mass on
the concentration. We construct 100 mock composite
clusters with galaxies more luminous than $M_K=-24.5$ (slightly more luminous than $M_{K*}$), 
with three values of input concentration
($c_{{\rm in}}=3,4,5$). The second column shows the recovered concentration
$\bar{c}_{{\rm fit}}$, which is the
mean and standard deviation of the distribution of the best-fit $c$ from 100 mock observations,
when no uncertainties are included.
The third and fourth columns are the results when the cluster center and mass are perturbed,
respectively, and the last column shows the result when both uncertainties are present.
Our code can recover the true concentration without any bias when the mass and cluster center
are perfectly known. Any error in cluster center determination translates into a bias in the
concentration. The uncertainty in cluster mass, on the other hand, does not produce any bias, as
discussed in \S\ref{sec:appen}.
This test shows that the concentration we measure would be very close to the true value,
with a potential bias of $\lesssim 4\%$ toward lower value.

\bibliographystyle{apj}
\bibliography{cosmology,refs}

\end{document}